\documentclass[a4paper,fleqn]{tmp-dc}

\usepackage[authoryear,longnamesfirst]{natbib}
\usepackage[edges]{forest}
\usepackage{graphicx}
\usepackage{circuitikz}
\usepackage{booktabs}
\usepackage{dsfont}

\def\tsc#1{\csdef{#1}{\textsc{\lowercase{#1}}\xspace}}
\tsc{WGM}
\tsc{QE}

\newcommand{\partb}[1]{\frac{\partial #1}{\partial \beta}}
\newcommand{\partm}[1]{\frac{\partial #1}{\partial m_{s,t}}}
\newcommand{\partr}[1]{\frac{\partial #1}{\partial r}}
\newcommand{\partrh}[1]{\frac{\partial #1}{\partial \rho}}
\begin{document}
\let\WriteBookmarks\relax
\def\floatpagepagefraction{1}
\def\textpagefraction{.001}

\shorttitle{Bankruptcies intermittency}    

\shortauthors{F.P.Pijpers, D.Diemer}  

\title [mode = title]{Intermittency in firm bankruptcies from register data}  




\author[1,2]{F.P. Pijpers}[orcid=0000-0001-7572-9435]

\cormark[1]


\ead{f.p.pijpers@uva.nl}


\credit{Conceptualization, Methodology}

\affiliation[1]{organization={Statistics Netherlands},
            addressline={Henri Faasdreef 312}, 
            city={The Hague},
%          citysep={}, % Uncomment if no comma needed between city and postcode
            postcode={2492 JP}, 
            country={Netherlands}}

\author[2]{D. Diemer}%[]

% Footnote of the second author

% Email id of the second author
\ead{dimdiemer@gmail.com}


\credit{Methodology, Data cleaning, Software}

\affiliation[2]{organization={Korteweg-de Vries Institute, University of Amsterdam},
            addressline={Science Park 105-107}, 
            city={Amsterdam},
%          citysep={}, % Uncomment if no comma needed between city and postcode
            postcode={1098 XG}, 
            country={Netherlands}}

\cortext[1]{Corresponding author}



\begin{abstract}
By analysing the clustering of bankruptcy activity from administrative register data in the Netherlands over time, the intermittency of this process is investigated. Using methodology from survival analysis a statistical characterization of bursts is obtained: short active periods alternating with quieter periods. Using Statistics Netherlands microdata on Dutch business bankruptcies declared on court hearing-days from 2009 to 2024, combined with input-output tables from the Dutch national accounts, it is studied whether this clustering can be explained by differences between sectors, years, and inter-sectoral exposure.
Sector differences explain most of the heterogeneity in the data, while year effects matter to a lesser extent and the inter-sectoral exposure variables constructed from the input-output tables further improve model fit.
\end{abstract}


%\item Registered bankruptcies of firms in the Netherlands show clear signs of clustering: burst episodes of high activity separated by relatively quiet %epochs.
%\item A modelling framework is developed that separates burst occurrence from burst size in bankruptcy data.
%\item The framework is used to quantify how these components vary across sectors and years from a complete administrative register of Dutch firm %bankruptcies.
%\item By adding exposure measures based on the input-output tables, it examines how inter-sectoral connections are associated with the burst onset %probabilities and the conditional size of bankruptcy counts over different timescales.

%\nocite{*}

% Keywords
\maketitle

% Main text
\section{Introduction}\label{intro}\noindent
Modern economies are highly interconnected sociotechnical systems in which firms, institutions, and technical infrastructures are linked through supply chains and financial linkages. Financial stress in one part of the economy can propagate through supply chains, financial relations and other forms of interdependence, potentially resulting in firm failures, bankruptcies, elsewhere. 

The Dutch national statistical institute, Statistics Netherlands (SN, also known under the Dutch abbreviation: CBS), publishes tables concerning bankruptcies as part of its regular output on economic indicators. The underlying dataset, which is used for this research, consists of information on bankruptcies declared by Dutch courts, including the date of the court decision and the sector classification, from which bankruptcy counts are constructed indexed by economic sectors and hearing-days. This administrative dataset has the properties of individual firms that were declared bankrupt, and is combined with publicly available input-output tables from the Dutch national accounts. These tables are used to construct exposure measures in order to study inter-sectoral dependencies at different timescales.

One feature of bankruptcy data is that bankruptcies are not evenly distributed over time: even within a given sector, calm periods may be followed by short episodes in which multiple bankruptcies occur on several consecutive hearing-days, a property known as \textit{intermittency}. 

Earlier work at SN shows that the aggregate bankruptcy process over 1981-2023 exhibits features of intermittency. Specifically, \cite{lopes_cardozo_internship_2024} show that a Weibull model fitted to the residual component of the aggregate bankruptcy time series yields a shape parameter below one, which suggests clustering of bankruptcies rather than a memoryless arrival process. Building upon this result, this thesis develops a framework that decomposes intermittency into burst onset and burst size components that can both be related to observed covariates.

The analysis is divided into two parts. First, the \textit{burst onset analysis} (BOA) studies the probability that a new burst starts after inactivity. This probability is formulated as a discrete-time hazard in a proportional hazards model and estimated using a Bernoulli generalised linear model with a complementary log-log link and a Weibull baseline. 
Second, the \textit{burst size analysis} (BSA) studies the size of bankruptcy counts on active hearing-days. Because the data contain many zero-count days and the positive-count days are overdispersed, this component is modelled using a zero-truncated negative binomial structure. These two analyses allow us to distinguish between \textit{when} bursts start after a period of inactivity and the \textit{size} of activity once a sector is active, yielding a more informative description of bankruptcy dynamics than a single count model would provide.

\subsection{Intermittency}\noindent
Intermittency is a property of a dynamical system in which a process alternates between phases of relatively low activity and active episodes at seemingly irregular times. Within dynamical systems theory, intermittency been studied extensively and has been formalised in several ways \citep{ott_blowout_1994, pomeau_intermittent_1980}. This concept is not restricted to purely theoretical systems, however. In fact, many economic and social processes exhibit similar alternations of quiet and active periods, and can therefore be studied from the same perspective.

A related notion is \textit{self-organised criticality} (SOC). In SOC, a system evolves towards a critical state in which relatively small events can trigger large cascades of events, often referred to as avalanches. In economic networks, avalanches are often discussed in relation to financial structures such as supply chains, where limited buffers may increase local stress and amplify shocks. Recent work on timeliness criticality and critical fragility formalises how efficiency pressures can move sociotechnical systems closer to a critical state \citep{moran_timeliness_2024, moran_critical_2025}. The SOC perspective offers one interpretation under which firm failures may increase financial stress on economically connected firms through supply chains.
Intermittency and SOC are distinct concepts, but they share a common feature: both involve extended periods of calm followed by bursts of activity associated with the slow build-up of pressure and the depletion of buffers. It has been shown that Dutch bankruptcies occur in bursts of activity \citep{lopes_cardozo_internship_2024}, and related work shows that bankruptcy cascades tend to exhibit SOC-like behaviour \citep{hong_power_2007}. Hong, Lee \& Lee specifically analyse Korean firm bankruptcies, but argue that the mechanisms are not intrinsically Korea-specific. These findings support the idea that bankruptcies should not be treated as time-independent events, but rather as a process in which clustering in time is present.

\section{Burst model and empirical estimators}\noindent
The formal description of intermittency, starts with a discrete-time counting process indexed by hearing-days. Let $Y_{s,t}$ be the number of bankruptcies observed in sector $s$ (e.g. agriculture, transport) on hearing-day $t$. Define the 
\begin{equation}
\label{eq:buacind}
A_{s,t}\equiv\mathds{1}{\{Y_{s,t}>0\}},
\end{equation}
which equals one if at least one bankruptcy is recorded in sector $s$ on hearing-day $t$. Also define the \textit{burst onset indicator}:
\begin{equation}
\label{eq:boaind}
B_{s,t} \equiv\mathds{1} {\{A_{s,t}=1,\, A_{s,t-1}=0\}}, \qquad t\geq 1,
\end{equation}
and set $A_{s,0}\equiv 0$. Thus, $B_{s,t}=1$ only at the first hearing-day of a burst. 

Define the \textit{activation probability} and \textit{conditional positive mean} by
\begin{eqnarray}
\pi_{s,t}&\equiv&\mathbb{P}(A_{s,t}=1)\nonumber\\
\mu_{s,t}^+ &\equiv& \mathbb{E}[Y_{s,t}\mid A_{s,t}=1].
\end{eqnarray}
$\mathbb{E}[Y_{s,t}]$ is the expected number of bankruptcies in sector $s$ on hearing-day $t$. Conditioning on whether the sector is active gives, for each $(s,t)$,
\begin{eqnarray}
\mathbb{E}[Y_{s,t}] &=& \mathbb{P}(A_{s,t}=1)\mathbb{E}[Y_{s,t}\mid A_{s,t}=1] + \nonumber\\
&&\qquad\mathbb{P}(A_{s,t}=0)\underbrace{\mathbb{E}[Y_{s,t}\mid A_{s,t}=0]}_{= \,0} \nonumber\\
&=& \mathbb{P}(A_{s,t}=1)\mathbb{E}[Y_{s,t}\mid A_{s,t}=1] \nonumber\\
&=&\pi_{s,t} \mu_{s,t}^+.
\end{eqnarray}
Since hearing-days are not necessarily equally spaced in calendar time, the gaps between consecutive hearing-days are used. Let $\Delta_t$ be the calendar-time gap between hearing-days $t-1$ and $t$ (for $t\geq 2$, as the gap preceding the first hearing-day is ambiguous), and define the \textit{total calendar time up to hearing-day} $T$ as
\begin{equation}
\mathcal{T}_T \equiv \sum_{t=2}^T\Delta_t.
\end{equation}
Define the \textit{average bankruptcy rate} in sector $s$, measured in calendar time $R^{\text{emp}, 1}_{s}$, as
\begin{equation}
R^{\text{emp}, 1}_{s} \equiv \frac{1}{\mathcal{T}_T} \mathbb{E}\left[\sum_{t=2}^T Y_{s,t}\right].
\end{equation}
Then the first decomposition of intermittency is obtained by applying $\mathbb{E}[Y_{s,t}] = \pi_{s,t}\mu_{s,t}^+$:
\begin{equation}
R^{\text{emp}, 1}_{s} \equiv \frac{1}{\mathcal{T}_T}\sum_{t=2}^T\pi_{s,t}\mu_{s,t}^+.
\end{equation}

The first decomposition distinguishes between an activation component and a conditional size component. To obtain a second decomposition, consider the observed bursts in the window $t=2,\dots, T$. For a sector $s$, let $j=1,\dots, N_{s,T}$ index the bursts in this window and let $M_{s,j}$ be the total number of bankruptcies in burst $j$. Then the total number of bankruptcies over $t=2,\dots,T$ can be written as 
\begin{equation}
\sum_{t=2}^T Y_{s,t} = \sum_{j=1}^{N_{s,T}} M_{s,j}.\label{eq:burstsize}
\end{equation}
Taking expectations and conditioning on $N_{s,T}$ yields
\begin{eqnarray}
\frac{1}{\mathcal{T}_T}\mathbb{E}\left[\sum_{t=2}^T Y_{s,t}\right] &=& \frac{1}{\mathcal{T}_T}\mathbb{E}\left[\sum_{j=1}^{N_{s,T}}M_{s,j}\right] \nonumber\\
&=& \frac{1}{\mathcal{T}_T}\mathbb{E}\left[\mathbb{E}\left[ \sum_{j=1}^{N_{s,T}}M_{s,j}\Bigg| N_{s,T}\right]\right] \nonumber\\
&\overset{(*)}{=}& \frac{1}{\mathcal{T}_T}\mathbb{E}[N_{s,T}\, \mathbb{E}[M_s]] \nonumber\\
&=&\mathbb{E}[M_s]\frac{\mathbb{E}[N_{s,T}]}{\mathcal{T}_T},
\end{eqnarray}
where $(*)$ holds under the assumption that all burst sizes $M_{s,j}$ are i.i.d. with $\mathbb{E}[M_{s,j}] = \mathbb{E}[M_s]$ and that $M_{s,j}$ is independent of $N_{s,T}$ for all $j$. 

Under these assumptions on burst sizes, the second decomposition of the average bankruptcy rate is:
\begin{equation}
R^{\text{emp}, 2}_{s}\equiv \mathbb{E}[M_s]\frac{\mathbb{E}[N_{s,T}]}{\mathcal{T}_T}.
\end{equation}

Thus, the calendar-time bankruptcy rate can be decomposed in two ways: the first in terms of an activation component and a conditional size component, and the second as the product of an average burst size and the average burst frequency. Substituting empirical estimators for each component yields expressions $R^{\text{emp}, 1}_{s}$ and $R^{\text{emp}, 2}_{s}$, and by replacing the empirical components for fitted components from the BOA and BSA, the model-based estimators are obtained $R^{\text{est}, 1}_{s}$ and $R^{\text{est}, 2}_{s}$.

\subsection{Empirical estimators and their fitted counterparts}\noindent
The empirical estimator $R_s^{\hspace{0.5pt}\text{emp}}$ corresponding to the average bankruptcy rate in sector $s$ is defined as
\begin{equation}
R_s^{\hspace{0.5pt}\text{emp}} \equiv \frac{1}{\mathcal{T}_T} \sum_{t=2}^T Y_{s,t}.
\end{equation}

\paragraph{Decomposition 1.} 
For the first decomposition, define the \textit{empirical activation rate} and the \textit{empirical conditional positive mean} as
\begin{eqnarray}
A^\text{\hspace{0.5pt}emp}_s &\equiv& \frac{1}{\mathcal{T}_{T}}\sum_{t=2}^T A_{s,t} \nonumber\\ 
\mu_{s}^{+,\,\text{emp}} &\equiv& \frac{\sum_{t=2}^T Y_{s,t}}{\sum_{t=2}^T A_{s,t}}.
\end{eqnarray}

Multiplying these gives the empirical form of the first decomposition:
\begin{eqnarray}
R^{\text{emp}, 1}_{s} &\equiv& A^\text{\hspace{0.5pt}emp}_s{\mu_s^{+,\,\text{emp}}} \nonumber\\
&=& \left(\frac{1}{\mathcal{T}_{T}}\sum_{t=2}^T A_{s,t}\right)\left(\frac{\sum_{t=2}^T Y_{s,t}}{\sum_{t=2}^T A_{s,t}}\right) \nonumber\\
&=& \frac{1}{\mathcal{T}_{T}} \sum_{t=2}^T Y_{s,t} .
\end{eqnarray}
In the fitted counterpart of Decomposition 1, the empirical conditional mean ${\mu_s^{+,\,\text{emp}}}$ is replaced by the estimated conditional mean from the burst size analysis, while the activity component is kept fixed. Since the BSA produces fitted values $\hat{\mu}_{s,t}^+$, define the \textit{estimated conditional mean} over all active hearing-days as
\begin{equation}
\hat{\mu}_s^+ \equiv \frac{\sum_{t=2}^T A_{s,t}\hspace{0.5pt}\hat{\mu}_{s,t}^+}{\sum_{t=2}^T A_{s,t}}.
\end{equation}

The fitted version corresponding to the first decomposition is
\begin{equation}
R^{\text{est}, 1}_{s} \equiv A^\text{\hspace{0.5pt}emp}_s\hat{\mu}_s^{+}.
\end{equation}

\paragraph{Decomposition 2.} For the second decomposition, define the \textit{empirical burst size} and the \textit{empirical burst onset rate} as
\begin{eqnarray}
M_s^\text{\hspace{0.5pt}emp} &\equiv& \frac{1}{N_{s,T}}\sum_{j=1}^{N_{s,T}} M_{s,j} \nonumber\\ \lambda_s^\text{\hspace{0.5pt}emp} &\equiv& \frac{N_{s,T}}{\mathcal{T}_T},
\end{eqnarray}
where $N_{s,T}$ indexes the bursts in $t=2,\dots, T$ and $M_{s,j}$ is the size of burst $j$. 

This yields the empirical version of the second decomposition, using (\ref{eq:burstsize}):
\begin{eqnarray}
R^{\text{emp}, 2}_{s} &\equiv& M_s^\text{\hspace{0.5pt}emp}\lambda_s^\text{\hspace{0.5pt}emp} \nonumber\\
&=& \Bigg(\frac{1}{N_{s,T}}\sum_{j=1}^{N_{s,T}} M_{s,j} \Bigg)\Bigg(\frac{N_{s,T}}{\mathcal{T}_T}\Bigg) \nonumber\\
&=& \frac{1}{\mathcal{T}_T}\sum_{t=2}^T Y_{s,t} .
\end{eqnarray}

Define the set of hearing-days on which a burst can start as
\begin{equation}
\mathcal{I}_s \equiv \{t\in \{2,\dots, T\}\mid A_{s,t-1}=0\}.
\end{equation}
The empirical onset rate component is replaced by the burst onset rate from the burst onset analysis:
\begin{equation}
\hat{\lambda}_s \equiv \frac{1}{\mathcal{T}_T}\sum_{t\in\mathcal{I}_s} \hat{p}_{s,t},
\end{equation}
where $\hat{p}_{s,t}$ is the estimate of the burst onset probability $p_{s,t} = \mathbb{P}(B_{s,t}=1\mid A_{s,t-1}=0,\, w_{s,t-1})$ produced in the BOA. Furthermore, it is possible to write 
\begin{equation}
\mathbb{E}[N_{s,T}] = \sum_{t\in \mathcal{I}_s}\mathbb{E}[B_{s,t}] \approx \sum_{t\in \mathcal{I}_s}\hat{p}_{s,t},
\end{equation}
therefore
\begin{equation}
\hat{\lambda}_s \approx \frac{\mathbb{E}[N_{s,T}]}{\mathcal{T}_{T}}.
\end{equation}
The fitted counterpart of the second decomposition is thus
\begin{equation}
R^{\text{est}, 2}_{s} \equiv M_s^\text{\hspace{0.5pt}emp}\hat{\lambda}_s.
\end{equation}
Since both empirical decompositions are identical  $R^{\text{emp}, 1}_{s} =$  $R^{\text{emp}, 2}_{s}$, differences can only arise when empirical components are replaced by fitted components. Therefore, differences between the empirical and fitted decompositions reflect how well the BOA and BSA components estimate the empirical quantities. At the same time, it is reasonable to expect the estimated decompositions to be closer to the empirical decompositions when the model is refined to include more covariates, since a model with more degrees of freedom is able to better reproduce the data. 

Beyond reproducing the empirical estimates, the fitted versions of the decompositions are useful in the sense that they reveal \textit{why} the estimate changes: whether it is due to changes in the frequency of bursts or due to changes in the burst size during an active period. This is reflected in the fact that Decomposition 1 isolates the BSA component, keeping the empirical activity rate fixed, and Decomposition 2 isolates the BOA component, while keeping the empirical burst size fixed.

\subsection{Covariates}\noindent
The burst onset and size analyses it is also important to account for covariates, i.e. to model the probability
\begin{equation}
p_{s,t} \equiv \mathbb{P}(B_{s,t}=1\mid A_{s,t-1}=0,\, w_{s,t-1}),
\end{equation}
where $w_{s,t-1}$ is a vector of covariates. This burst onset probability describes the frequency of bursts and how it varies across years, sectors, and exposure variables. The burst size analysis models the conditional positive mean
\begin{equation}
\mu_{s,t}^+ \equiv \mathbb{E}[Y_{s,t}\mid A_{s,t}=1,\,z_{s,t}],
\end{equation}
where $z_{s,t}$ is a vector of covariates. Together, the BOA and BSA yield the fitted quantities $\hat{p}_{s,t}$ and $\hat{\mu}_{s,t}^+$ that are used in the two decompositions.

\subsection{Model structure}\noindent
A nested modelling approach is used. First, a baseline model (model $\text{BOA}_0$) is fitted, which serves as a comparison and yields estimates of the shape and scale parameters for the Weibull baseline. The baseline model is then expanded by adding the years and sectors separately as covariates (models $\text{BOA}_Y$ and $\text{BOA}_S$) to investigate changes over time and across sectors. 
Finally, both sets of covariates are included in an additive model ($\text{BOA}_{SY}$). This structure allows assessing which covariates improve model fit the most at each step while yielding parameter estimates that describe the residual process. Because the models are nested, likelihood-ratio tests can provide evidence for or against including certain covariates in a model.

\begin{figure}
\centering
\includegraphics[width=\linewidth]{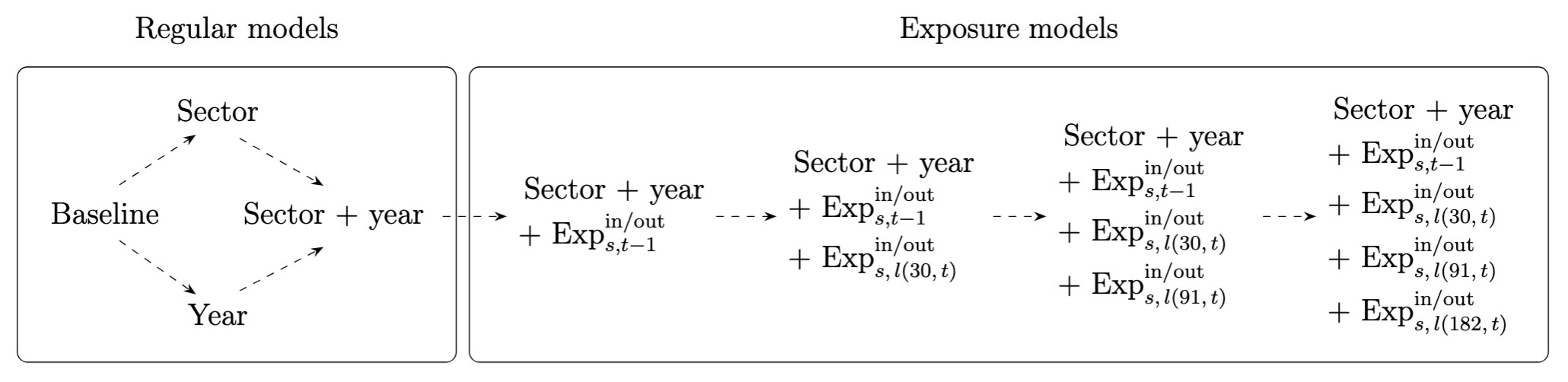}
\caption{Model hierarchy in the analysis, labelled by their covariates. See \ref{subsec:exposurevars} for the definition of the lags and the lag indices.}
\label{fig:model_hierarchy}
\end{figure}

The model that best fits the data based on likelihood-ratio tests is augmented with exposure variables. The exposure variables are added sequentially, in order of increasing lag length. The hierarchy is given in fig. \ref{fig:model_hierarchy}. The exposure models, like the regular models, are compared using likelihood-ratio tests. For brevity, model information and estimates are reported only for the regular models, as well as for the best-fitting exposure model.

For each model, the optimiser is provided with the starting values $\mu=\nu = 0$, so that $k=\exp(\mu)=1$ and $\lambda = \exp(\nu) = 1$. Gamma is initialised as the zero vector. 

After the parameters are estimated, confidence intervals are constructed in the usual asymptotic sense, but on the log scale: starting with the MLE approximation $\hat{\theta} = (\hat{\gamma},\, \log\hat{k},\, \log\hat{\lambda})$ using the inverse Hessian of the negative log-likelihood to compute standard errors. Then the confidence interval is constructed: for example, for $\log \hat{k}$ as
\begin{equation}
\text{CI}(\log\hat{k})\equiv [\log\hat{k} \pm q_{1-\alpha/2}{\rm SE}(\log\hat{k})],
\end{equation}
where $q_{1-\alpha/2}$ is the $(1-\alpha/2)$-quantile of the ${\rm Normal}(0,1)$ distribution (here $\alpha=0.05$). Finally, transform back by exponentiating the endpoints: $\text{CI}(\hat{k}) = \exp(\text{CI}(\log \hat{k}))$,
and analogously for $\hat\lambda$.

\subsubsection{Comparing models}\noindent
For each model, the usual Akaike and Bayesian information criteria (AIC and BIC) are calculated.
In the summary, the difference in AIC and BIC of a given model from the best model is presented.

Whereas the AIC and BIC indicate the goodness of fit, a useful comparison tool to assess the significance of the covariates is the likelihood-ratio test. This test is only available for nested models, i.e. when one is a restricted version of another. The simpler model is taken as the null hypothesis and the model with the added covariates as the alternative. 

Furthermore, the number of L-BFGS gradient iterations and the maximum gradient difference are reported. The L-BFGS gradient iteration count is the number of gradient evaluations required in the L-BFGS algorithm before convergence, and the maximum gradient difference is the maximum difference between the analytical gradient and the numerical gradient, computed using the finite-differences method.

Lastly, a Wald test is applied to assess whether the shape parameter $k$ equals one against the alternative $k<1$ in each model, using the Wald statistic $z_k$, defined on the scale of $\mu$:
\begin{equation}
z_k = \frac{\hat{\mu} - 0}{\hat{\sigma}_\mu}.
\end{equation}
when the corresponding $p$-value $p=\Phi(z_k)$ is small, this gives evidence in favour of rejecting the null hypothesis.

\section{Data}\noindent
The microdata consist of administrative records on bankruptcies pronounced by Dutch courts, where each row corresponds to a publication or update in a case and contains variables such as the publication date, the publication type, the sector classification of the firm and an anonymous business identifier (BEID).

\subsection{Bankruptcy microdata}\noindent
The primary dataset is ``Adjudicated bankruptcies pronounced by Dutch courts, by type of bankrupt entity, economic activity and region (Bankruptcies)''. A description is available on the SN website \citep{cbs_uitgesproken_2026}.\footnote{In Dutch: ``Uitgesproken faillissementen door Nederlandse rechtbanken naar type gefailleerde, activiteit en regio (Faillissementen)''} From this dataset the date of the court decision is used, the legal personality (sole proprietorship or business) and the sector classification according to the SBI 2008.

The microdata are filtered and aggregated to create counts of bankruptcies $y_{s,t}$ for a given sector and hearing-day. As no input-output table was available yet for the year 2025, no exposure variables can be created for post-2024 data. For this reason, the dataset is right-truncated to include data up to and including 31-12-2024. This means that the analysed period is 1 January 2009 through 31 December 2024.

Because bankruptcies are pronounced on hearing-days, it is necessary to filter on the publication type and sort by date so that the publication date corresponds to the court hearing date at which the bankruptcy is finalised. For administrative reasons, the same bankruptcy case appears three times in the dataset: at the start of the process, when a draft is published and finally at the definitive publication. Only the rows with ``definitive'' as publication status are retained. 
\begin{figure}
\includegraphics[width=1.0\linewidth]{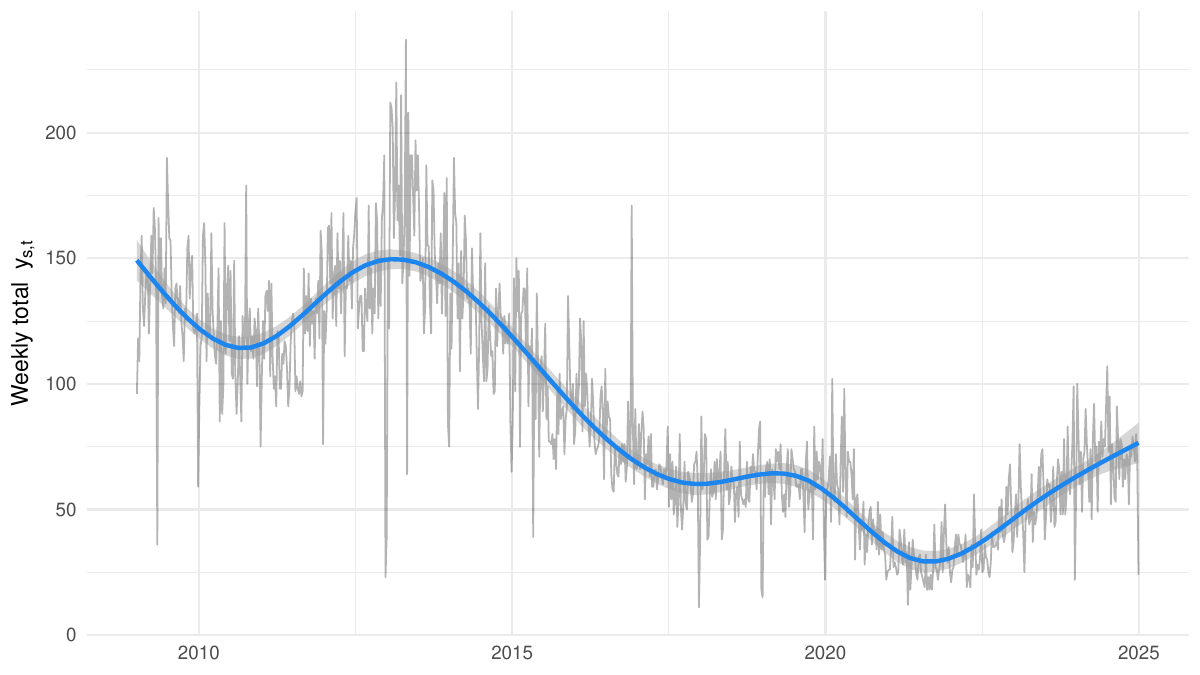}
\caption{Weekly total bankruptcy counts across all sectors, with a smoothed conditional mean overlaid in blue.}
\label{fig:WeeklyTotals}
\end{figure}

The sample is further restricted by filtering on legal entity. Sole proprietorships (single-person firms) are excluded for two reasons: first, they are unevenly distributed over the sectors, as starting a solo business in construction of farm equipment is comparatively harder than starting a solo study guidance practice, for example. Second, because such one-person firms operate under different laws and tax systems than other businesses it is inappropriate to combine them into one class. 

A key feature of the data is that multiple bankruptcies can be declared on the same hearing-day. Hearing-days are predominantly Tuesdays \citep{CBS_faillissementen_2026}, but other weekdays and even some weekend days have been recorded in the data. This structure allows constructing a discrete-time counting process, where arrivals correspond to counts of bankruptcies on hearing-days. Let $t=1,\dots, T$ index the sequence of hearing dates in the sample. Then for each sector $s$, define the \textit{observed bankruptcy count}:
\begin{eqnarray}
y_{s,t} &\equiv& \#\{\text{bankruptcies in sector } s \nonumber\\
&&\qquad\text{ declared on hearing-day } t\}.
\end{eqnarray}
The resulting dataset is a grid of sectors and hearing-days in which $y_{s,t}$ denotes the number of bankruptcies for a given choice of $s,t$ and equals 0 on hearing-days without bankruptcies. The bankruptcy data consist of 408,434 rows, and after filtering 259,920 remain. Using the hearing-day calendar, four variables are constructed that are central to the analysis:
\begin{itemize}
\item the \textit{activation indicator} $A_{s,t}$,
\item the \textit{burst onset indicator} $B_{s,t}$, 
\item the time between consecutive hearing-days $\Delta_t$ and 
\item the elapsed time since the end of the last burst $R_{s,t-1}$. 
\end{itemize}
The details of the construction of $\Delta_t$ and $R_{s,t-1}$ are given in appendix \ref{app:RDelCon}. A `burst' is understood to be a sequence of consecutive active hearing-days within a given sector.
The sectors are encoded and identified by a Standard Business Index (SBI), a hierarchical code which can have up to 5 digits. 
Using only leading digits automatically implies a logical aggregation of sectoral business activity. In this paper only the leading two digits are used. In appendix \ref{subsec:SBIcodes} further details of this international standard coding are provided.

\subsection{Omission of financial variables}\noindent
In general, if the BEID is available for multiple datasets, these datasets can be linked in order to construct a dataset with multiple variables (e.g. turnover in the last quarter, year, or other financial variables). While such variables may be useful in principle, they are omitted here for the following reasons. First, many firm closures are not bankruptcies: a firm may merge with another firm, restructure or cease operations for other reasons. Second, struggling firms can strategically choose to cease operations without declaring bankruptcy, for example to avoid incurring judicial costs. This incomplete coverage introduces undesirable bias. Lastly, linking the financial variables to the bankruptcy data reduces the number of observations substantially due to incomplete coverage. 

\subsection{Cross-exposure between sectors and epochs}\label{subsec:exposurevars}\noindent
For the majority of Dutch firms it is not known which firms
are their suppliers or users, and therefore the exposure of
each firm to their supply chain has to be estimated through proxies. At a sectoral level, there is a routine publication of SN, describing at a macroscopic level the interconnectedness of the Dutch economy through input-output tables (I/O tables).  
The I/O-tables contain a square block of intermediate transactions between sectors. Further details are described in appendix \ref{sec:IOtables}.

Essentially this square block of intermediate transactions is a weighted, directed, adjacency matrix where the rows and columns are the sectors, and each cell quantifies the value of goods and services exchanged between a pair of sectors. 
That block can thus be extracted and transformed into weight matrices that capture inter-sectoral exposure. Because the industry classification in the I/O-tables is coarser than the sector classification in the bankruptcy data, some sectors are grouped. Specifically, the groups 13\,-\,15, 37\,-\,39, 59\,-\,60, 87\,-\,88 and 90\,-\,92. 
Both the I/O-block and the bankruptcy counts are aggregated accordingly. 

Let $s_1,\dots, s_n$ be the sectors in a given I/O-table, and let $Z(y)\in \mathbb{R}_{+}^{n\times n}$ be the extracted I/O-block for year $y$. The entry $Z_{ij}$ corresponds to the intermediate sales from sector $s_i$ to $s_j$ in year $y$. $Z(y)$ is used to construct two matrices: an input matrix $W^\text{in}(y)$, corresponding to exposure of a sector to its suppliers, and an output matrix $W^\text{out}(y)$, corresponding to exposure to buyers. 

To obtain interpretable exposure measures, the matrices are normalised, so that, for example, a value of $0.32$ corresponds to 32\% of the total exposure in that row or column. Define the input weights by column-normalising:
\begin{equation}
W^\text{in}_{ij}(y) = \frac{Z_{ij}(y)}{\sum_{k} Z_{kj}(y)},
\end{equation}
so that $W_{ij}^\text{in}$ represents the share of sector $s_j$'s inputs from sector $s_i$. Similarly, define the output weights by row-normalising:
\begin{equation}
W^\text{out}_{ij}(y) = \frac{Z_{ij}(y)}{\sum_{k} Z_{ik}(y)}.
\end{equation}
In both cases, if the normaliser is zero, the corresponding weights are set to zero. 

A small number of entries in the I/O-tables are negative. These values are unusual and typically not bookkeeping errors, but rather a product of how the national accounts separate the flow of transactions during a given period from the amount of stock held at a point in time: the stock level of inventories is not shown, rather, only the change in inventories. Thus, a negative value corresponds to withdrawals exceeding deposits during that time. This can happen for multiple reasons, e.g. when a business knows or expects the price of a certain product to increase in the future, leading to stockpiling of that product. After the price increase the firm can take advantage of that buffer, while buying less, thereby leading to a negative entry in the I/O-table. For consistency, these values are set to zero.

Finally, to restrict the analysis to purely inter-sectoral exposures, the exposures of sectors to themselves are removed by setting the diagonal of the weight matrices to zero:
\begin{eqnarray}
W_{ii}^\text{in}(y) &=& 0 \nonumber\\ 
W_{ii}^\text{out}(y) &=& 0,
\end{eqnarray} 
for all $i$. Explicit exposure covariates are constructed using the burst onsets. Let $\vec{B}_t = (B_{1,t},\dots,B_{n,t})^\top$ be the vector of burst onset indicators across sectors for hearing-day $t$. Let $y(t)$ be the calendar year corresponding to hearing-day $t$. Define the \textit{input and output exposure variables}
\begin{eqnarray}
\text{Exp}_{s,t}^\text{in} &\equiv&  \sum_{i} W_{i,s}^\text{in}(y(t)) B_{i,t} \nonumber\\
\text{Exp}_{s,t}^\text{out} &\equiv& \sum_{j} W_{s,j}^\text{out}(y(t))B_{j,t}.
\end{eqnarray}
Since $B_{i,t}\in \{0,1\}$ and the weights $W_{i,s}$ are normalised, these exposure variables can be interpreted as weighted measures of burst onsets in connected sectors. A high value of $\text{Exp}_{s,t}^\text{in}$ indicates that sectors supplying to $s$ have experienced burst onsets recently, while a high value of $\text{Exp}_{s,t}^\text{out}$ indicates that buyers from $s$ experienced burst onsets. 

It is often useful to use lagged versions of the exposure variables, as a shock in one sector may need time to propagate to another sector. The speed at which these shocks propagate depends on factors such as the structure of the supply chain the firms are embedded in or the firms' financial liquidity. Multiple lags are therefore considered in the analysis. Let $T(t)$ denote the calendar date of hearing-day $t$. For $d\in \mathbb{N}$ calendar days, define the \textit{calendar-time lag index}
\begin{equation}
l(d, t) = \max\{ u <t \mid T(u)\leq T(t)-d\},
\end{equation}
which denotes the most recent hearing-day $u$ that is at least $d$ calendar days before hearing-day $t$. For $k$ in hearing-days, simply write $t-k$ for the $k$-th previous hearing-day, provided it exists. Because hearing-days are irregularly spaced in calendar time, a calendar-time lag does not correspond one-to-one with a fixed number of hearing-days in the past. Use is made of four lag lengths, namely $l(30, t)$, $l(91, t)$ and $l(182, t)$ denoting (approximately) the monthly, quarterly and half-year lags, and the index $t-1$ is used for the previous hearing-day lag. As an example, a model that includes exposure variables for two lag lengths, namely one hearing-day and one month, has a model formula of the form
\begin{equation}
\sim (...) {}+{}\text{Exp}^{\text{in}}_{s,t-1} + \text{Exp}^{\text{out}}_{s,t-1} + \text{Exp}^{\text{in}}_{s,\,l(30,\, t)} + \text{Exp}^{\text{out}}_{s,\,l(30,\, t)}.
\end{equation}

The exposure variables across different lags, together with the other covariates, allow assessing the extent of inter-sectoral dependence in the bankruptcy process.

\section{Methodology}\noindent
\subsection{Burst onset}\noindent
The burst onset probability is:
\begin{equation}
p_{s,t}\equiv\mathbb{P}(B_{s,t}=1\mid A_{s,t-1}=0,\, w_{s,t-1}).
\end{equation}
Given a sector $s$, it is the probability that a burst starts on hearing-day $t$, conditional on inactivity on hearing-day $t-1$ and on covariates $w_{s,t-1}$ known at $t-1$. Because the bankruptcy process can be interpreted as a discrete-time counting process, the burst onset probability can be expressed as a \textit{discrete-time hazard}: the probability that a burst occurs in the next hearing-day interval, given that it has not occurred up to the start of that interval. 

Hazard functions arise naturally in survival analysis, which is the area of statistical data analysis that focuses on predicting and modelling time-to-event data, where this time is known as the \textit{survival time}. It originates in biomedical settings, where survival time indicates the time until death or recovery. Whenever the goal is to study the effect of covariates on survival/event times, a standard approach is the Cox proportional hazards model \citep{peng_estimating_2003, carroll_use_2003, klein_survival_2005}. Introduced by Cox in 1972, the PH model has been applied in a wide range of fields: examples include estimating the time until failure of a piece of equipment \citep{kumar_proportional_1994, diamoutene_survival_2016}, the length of U.N. peacekeeping missions \citep{box-steffensmeier_event_2004} and predicting the time from the founding of a business to bankruptcy \citep{Shumway01}. 

While the methods are well known it is useful to repeat some of the notation and definitions here, for clarity.
Let $T$ be a non-negative continuous random variable denoting an event time and let $f$ and $F$ be the PDF and CDF respectively of $T$. The \textit{survival function} $S(t)$ of $T$ is defined as
\begin{equation}
S(t) \equiv 1-F(t) = \mathbb{P}(T>t).
\end{equation}
The \textit{hazard function} $h(t)$ is defined as
\begin{equation}
h(t) \equiv \lim_{\delta\downarrow 0} \frac{\mathbb{P}(t\leq T < t+\delta \mid T\geq t)}{\delta} = \cfrac{f(t)}{S(t)}.
\end{equation}
Furthermore, the \textit{cumulative hazard function} is defined as the integral of the hazard function:
\begin{equation}
H(t) \equiv \int_0^t h(u){\rm d} u,\quad t>0.
\end{equation}
An important identity that relates the cumulative hazard function and survival function is
\begin{equation}
S(t) = \exp(-H(t)).\label{eq:fundrel}
\end{equation}
When modelling firm bankruptcies using survival theory, it is reasonable to assume that the survival time of a firm depends on multiple factors, such as revenue, liquidity or competitiveness within a sector. It is therefore desirable to introduce such variables as covariates into a model for survival time. 

Let $w$ be a vector of covariates. In the PH model, the hazard $h$ has the form
\begin{equation}
h(t\mid w) = h_0(t)\exp \big(w^\top \gamma\big),
\end{equation}
where $h_0(t)$ is known as the \textit{baseline hazard} and $\gamma$ is the coefficient vector corresponding to $w$. Note that the hazard reduces to the baseline hazard when no covariates are present. Thus, the baseline hazard contains all information that is not explained by the covariates.

The term \textit{Cox proportional hazards model} is often reserved for models where $h_0$ is left unspecified. If a baseline hazard is chosen, the model becomes fully parametric, and the preferred term is \textit{parametric proportional hazards model}. In what follows a Weibull baseline is chosen, as the goal is to assess whether a signature of intermittency is present in the data after conditioning on the covariates. For simplicity, the parametric proportional hazards model is referred to as the PH model, or PH framework. The PH model enjoys the \textit{proportionality property}, which means that additive changes to the covariates act multiplicatively on the hazard.

It is straightforward to show that the discrete-time hazard over $(R,\,R+\Delta]$ conditioned on the covariates can be written as
\begin{eqnarray}
p(R,\,\Delta \mid w) &&= 1-\exp(-[H(R+\Delta\mid w)-H(R\mid w)])\nonumber\\
&&= 1-\exp(- [\exp(w^\top \gamma)H_0(R+\Delta) \nonumber\\
&&\qquad-\exp(w^\top \gamma)H_0(R)])\nonumber\\
&&= 1-\exp(-\Lambda_0(R,\,\Delta)\exp(w^\top\gamma)).\label{eq:PHform}
\end{eqnarray}
in which the \textit{cumulative baseline hazard increment} is defined as
\begin{equation}
\Lambda_0(R,\,\Delta) \equiv H_0(R+\Delta) - H_0(R).  
\end{equation}
Given $\Delta_t$ and $R_{s,t-1}$, for a pair $(s,t)\in\mathbb{I}$, the burst onset probability $p_{s,t}$ can be written as a discrete-time hazard in the PH model over the interval $(R_{s,t-1}, \,R_{s,t-1}+\Delta_t]$ using (\ref{eq:PHform}):
\begin{eqnarray}
p_{s,t} &=& \mathbb{P}(B_{s,t}=1\mid A_{s,t-1}=0,w_{s,t-1})\, \\
 &=& 1-\exp\left(-\Lambda_0(R_{s,t-1},\Delta_t)\exp(w_{s,t-1}^\top \gamma)\right),\nonumber\label{eq:pst}
\end{eqnarray}
where $w_{s,t-1}\in \mathbb{R}^p$ is the vector of covariates, $\gamma\in \mathbb{R}^p$ is the corresponding vector of coefficients, 
Once a baseline hazard is chosen, the model becomes fully parametric. $p_{s,t}$ can be viewed as a Bernoulli success probability for each pair $(s,t)\in \mathbb{I}$ and the model parameters and the covariate coefficients estimated by maximising the likelihood over the full set of observations. With a suitable link function, this yields a Bernoulli generalised linear model with success probability $p_{s,t}$.

\subsubsection{Choice of the link function}\noindent
A known limitation of discrete-time models is that the inference can depend on the (often arbitrary) interval length \citep{allison_discrete-time_1982}. The cloglog link avoids this issue, as the estimates produced by the cloglog link within the PH framework are typically less sensitive to the length of the interval than those obtained from other common link functions. Moreover, it is the natural link for discrete-time hazards. This result follows from the stronger property that a continuous-time proportional hazards model corresponds to a Bernoulli generalised linear model (GLM) with a complementary log-log link \citep{prentice_regression_1978, cox_regression_1972, holford_life_1976}.
\begin{equation}
g(x) = \log[-\log(1-x)].
\end{equation}
Applying $g$ to the onset probability $p_{s,t}$ in (\ref{eq:pst}) yields the linear predictor in the GLM:
\begin{eqnarray}
{\rm cloglog}(p_{s,t}) &=& \log[-\log(1-p_{s,t})]\\
&=& \log\left[\Lambda_0(R_{s,t-1},\,\Delta_t)\exp(w_{s,t-1}^\top\gamma)\right] \nonumber\\
&=& \log\Lambda_0(R_{s,t-1}, \Delta_t) + w_{s,t-1}^\top \gamma.
\nonumber
\end{eqnarray}

A Weibull baseline is used as it generalises the exponential distribution, making it more versatile within the PH framework. In particular, the Weibull distribution with shape parameter $k$ and scale parameter $\lambda$ allows the hazard to vary with time through $k$. The shape parameter can be interpreted as follows:
\begin{itemize}
\item[$k<$] 1: the hazard decreases with time, corresponding to clustering of activity after a burst onset. 
\item[$k>$] 1: the hazard increases with time, corresponding to dispersion rather than clustering.
\item[$k=$] 1: The Weibull distribution reduces to the exponential distribution. Then, by the memoryless property, the probability of a new burst onset does not depend on the elapsed time since the previous burst ended.
\end{itemize}
In the context of intermittency, a shape parameter estimate of $\hat{k}<1$ would suggest clustering in the data, as the probability of a new burst is higher shortly after a recent burst.

Assume that the waiting time until the next burst onset is Weibull-distributed with shape parameter $k>0$ and scale parameter $\lambda > 0$. For $t\geq 0$, the CDF is 
\begin{equation}
F_0(t) = 1-\exp{(-(t/\lambda)^k}). 
\end{equation}
The corresponding survival function, baseline hazard and cumulative hazard are
\begin{eqnarray}
S_0(t) &=& \exp(-(t/\lambda)^k), \nonumber\\ 
h_0(t) &=& \frac{k}{\lambda}\Big(\frac{t}{\lambda}\Big)^{k-1}, \nonumber\\ 
H_0(t) &=& \Big(\frac{t}{\lambda}\Big)^k.
\end{eqnarray}
Hence, the cumulative baseline hazard increment over $(R_{s,t-1}$, $R_{s,t-1}+\Delta_t]$ becomes
\begin{eqnarray}
\Lambda_0(R_{s,t-1},\Delta_t) &=& H_0(R_{s,t-1}+\Delta_t)-H_0(R_{s,t-1}) \nonumber\\
&=& \Big(\frac{R_{s,t-1}+\Delta_t}{\lambda}\Big)^{\hspace{-1pt}k} - \Big(\frac{R_{s,t-1}}{\lambda}\Big)^{\hspace{-1pt}k} \nonumber\\
&=& \frac{(R_{s,t-1}+\Delta_t)^k - R_{s,t-1}^{k}}{\lambda^k}.
\end{eqnarray}
Before conditioning on covariates, the baseline behaviour is expected to be broadly consistent with the earlier work of \cite{lopes_cardozo_internship_2024}. Through error-term analysis, a Weibull model is fit to the inter-arrival times of the bankruptcy renewal process and an estimate is reported of $\hat{k} = 0.45$ for the shape parameter. Although the \cite{lopes_cardozo_internship_2024} dataset spans 1981 to 2024 and includes bankruptcies of sole proprietorships as well as of firms, it is plausible that the underlying dynamics are comparable. At a minimum, the results support a shape parameter $k<1$ indicating intermittent behaviour.

\subsection{Burst size}\noindent
A description of intermittency is incomplete if it is based solely on the probability that a burst starts after a quiet period, as this ignores variability in the activity once activity starts. 

\begin{figure}
    \centering
    \includegraphics[width=0.48\textwidth]{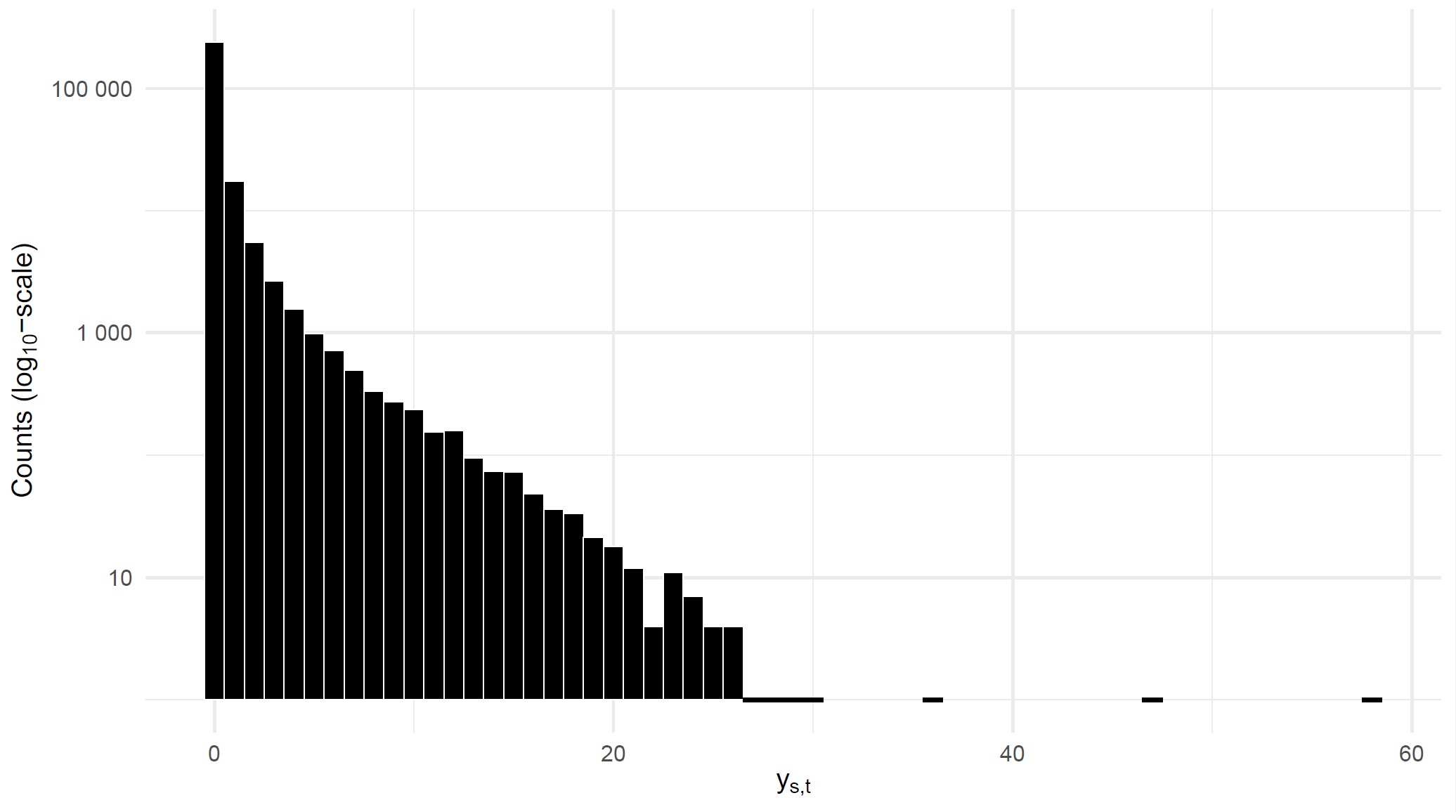}
    \caption{Histogram of the counts $y_{s,t}$ with frequency shown on the $\log_{10}$-scale. From this figure it becomes clear that the majority of the observations are zero counts.}%
    \label{fig:counts}%
\end{figure}

In the data, the proportion of active hearing-days is about 10\%, so zero-count days (i.e. $t$ where $A_{s,t}=0$) are prevalent. A histogram of the observations $y_{s,t}$ with a $\log_{10}$-scaled y-axis is given in \ref{fig:counts}. The distribution is strongly concentrated near zero, with a long, thin right tail over the positive counts. At the same time, only a handful of outliers are larger than 20, indicating the presence of overdispersion.

Fitting a Poisson or negative binomial distribution to the observed counts would assign the majority of the probability mass to the zero counts, thereby underestimating the true distribution of the positive counts. To circumvent this, the domain of the counts is split into two sets $\{Y_{s,t}\mid A_{s,t}=0\}$ and $\{Y_{s,t}\mid A_{s,t}=1\}$, covering both zero and non-zero counts. These sets are then modelled separately. Specifically, the zero counts are assigned a fixed probability, whereas a \textit{zero-truncated} distribution is fit to the positive counts. This construction is an example of a \textit{hurdle model}, named for the concept that the zero counts, like hurdles, must be overcome to reach the positive probability mass. The conditional distribution $Y_{s,t}\mid Y_{s,t}>0$ is estimated, as a component in describing the size of a burst. Suppose $Y$ models the counts in a discrete process. A hurdle model for $Y$ is then formulated as
\begin{equation}
\mathbb{P}(Y=0) = 1-\pi,\qquad\mathbb{P}(Y=y) = \pi f^+(y)
\end{equation}
for $y\in \{1,2,\dots\}$, where $\pi$ is an activation probability and $f^+$ is the PMF of a distribution that is truncated at zero. $f^+$ can readily be obtained from the regular PMF $f$ using the following identity:
\begin{eqnarray}
f^+(y) &=& \mathbb{P}(Y=y\mid Y>0) = \frac{\mathbb{P}(Y=y)}{\mathbb{P}(Y>0)} \nonumber\\
&=& \frac{f(y)}{1-f(0)}.\label{eq:truncidentity}
\end{eqnarray}

While there are several options, here the negative binomial distribution 
is used.\cite{lopes_cardozo_internship_2024} show that the bankruptcy data are overdispersed, or in other words that ${\rm Var}(Y)>\mathbb{E}[Y]$. If $Y\sim {\rm Poisson}(\mu)$ then ${\rm Var}(Y)=\mathbb{E}[Y]=\mu$, which implies that a Poisson model would be misspecified, providing an argument against using a Poisson distribution. 

%%%

\begin{table*}[t]
\centering
%\resizebox{\columnwidth}{!}{%
\begin{tabular}{lccccc}
& \multicolumn{4}{c}{Regular models} & \multicolumn{1}{c}{Exposure model} \\
\cmidrule(lr){2-5}\cmidrule(lr){6-6}
& \makecell{\textbf{Baseline} \\ {$(\text{{BOA}}_0)$}} & \makecell{\textbf{Year}\\{$(\text{BOA}_Y)$}} & \makecell{\textbf{Sector}\\{$(\text{BOA}_S)$}} & \makecell{\textbf{Sector + year}\\{$(\text{BOA}_{SY})$}} & \makecell{\textbf{Full exposure}\\{$(\text{BOA}_{SY}\text{E}_4)$}} \\
\hline
AIC                & 157112.9 & 156460.6 & 141154.9   & 138699.5   & 104375.4\\
BIC                & 157133.5 & 156636.1 & 141929.1   & 139628.6   & 105387.0\\
$\Delta\text{AIC}_j$        & 52737.5  & 52085.2  & 36779.5    & 34324.1    & 0\\
$\Delta\text{BIC}_j$        & 51746.5  & 51249.1  & 36542.1    & 34241.6    & 0\\
\hline
$\hat{k}$          & 0.7042  & 0.7188    & 1.090      & 1.179      & 0.9813\\ 
$\hat{\lambda}$    & 10.29   & 10.83     & 29.16      & 30.66      & 137.2\\
$\hat{p}_\text{avg}$& 0.1094 & 0.1093    & 0.1069     & 0.1065     & 0.1055\\
\hline
Constant term          & $-1.641$& $-1.712$  & $-3.675$   & $-4.035$   & $-4.829$\\
Smallest $\hat{\gamma}$ & $-$     & $-0.3751$ & $-5.552$   & $-6.001$   & $-4.600$\\
Largest $\hat{\gamma}$ & $-$     & 0.2387    & 2.662      & 2.863      & 3.030\\
\hline
\end{tabular}
%}
\caption{Model characteristics for the burst onset analysis models.}
\label{tab:onsetsummary}
\end{table*}

%%%
The advantages of the negative binomial distribution when compared to the Poisson distribution are evident from the properties of its variance:
for $X\sim {\rm NB}(r,m)$:
\begin{equation}
\mathbb{E}[X] = m \qquad\text{and}\qquad {\rm Var}(X) = m + \frac{m^2}{r}.
\label{eq:ZTNBEandV}
\end{equation}
When $r\to \infty$, ${\rm Var}(X) = m$ and the NB distribution reduces to a Poisson distribution. Therefore, the value of $r$ describes the overdispersion of $X$ relative to a Poisson distribution, as lower values of $r$ indicate a larger deviation from the Poisson distribution. Because the bankruptcy data are overdispersed, the negative binomial distribution as a basis for truncation is preferred.

\subsubsection{Zero-truncated negative binomial distribution}\noindent
Let $Y$ follow a negative binomial (NB) distribution with size $r>0$ and success probability $q\in (0,1)$. The boundary values $q=0$ and $q=1$ that exist in the regular NB model yield a degenerate distribution in the ZTNB model, hence these are excluded. 
In regression settings it is often more convenient to reparametrise the NB distribution in terms of its mean $m=r(1-q)/q$, rather than $q$, since $m$ can be linked directly to covariates through a linear predictor. The PMF of $Y$ in terms of the size parameter $r$ and mean $m$ is
\begin{equation}
\mathbb{P}(Y=y) = \frac{\Gamma(r+y)}{y!\hspace{1pt} \Gamma(r)}\left(\frac{r}{r+m}\right)^r\left(\frac{m}{r+m}\right)^y,
\end{equation}
so the probability mass at zero is
\begin{equation}
\mathbb{P}(Y=0) = \left(\frac{r}{r+m}\right)^r.
\end{equation}
Substituting these results into (\ref{eq:truncidentity}) yields the PMF of the zero-truncated negative binomial (ZTNB) distribution:
\begin{eqnarray}
f_\text{NB}^+(y;\hspace{1pt}r,\hspace{0.5pt}m) &=& \frac{\displaystyle\frac{\Gamma(r+y)}{y!\hspace{1pt}\Gamma(r)}
\left(\frac{r}{r+m}\right)^{r}
\left(\frac{m}{r+m}\right)^{y}}
{\displaystyle 1-\left(\frac{r}{r+m}\right)^{r}} \nonumber\\
&&\qquad y\in \{1,2,\dots\}.\label{eq:ZTNBPMF}
\end{eqnarray}
Thus, the mean of the ZTNB is
\begin{eqnarray}
\mathbb{E}[Y\mid Y>0] &=& \frac{\mathbb{E}[Y]}{\mathbb{P}(Y>0)}= \frac{\mathbb{E}[Y]}{1-\mathbb{P}(Y=0)} \nonumber\\
&=& \cfrac{m}{1-\Big(\cfrac{r}{r+m}\Big)^r}\,.\label{eq:meanZTNB}
\end{eqnarray}
This conditional mean is the central object in the BSA. The conditional size of an active hearing-day is modelled by fitting a ZTNB model for the positive counts, conditional on both activity and covariates $z_{s,t}$ with corresponding vector of coefficients $\beta$:
\begin{equation}
(Y_{s,t} \mid A_{s,t}=1, \,z_{s,t})\sim {\rm ZTNB}(r, m_{s,t}(\beta)).
\end{equation}
The mean $m_{s,t}(\beta)$ of the NB distribution depends on the covariate vector $\beta$ because it varies with the covariates $z_{s,t}$, depending on the link used in the estimation procedure. Using the truncated PMF in (\ref{eq:ZTNBPMF}), the unconditional distribution of the counts $Y_{s,t}$ can be written as a hurdle model:
\begin{eqnarray}
\mathbb{P}(Y_{s,t}=0) &=& 1-\pi_{s,t}\\
\mathbb{P}(Y_{s,t}=y) &=& \pi_{s,t}\hspace{0.5pt}f_\text{NB}^+(y; \hspace{1pt}r, \hspace{0.5pt}m_{s,t}(\beta)) \quad\text{for } y > 0, \nonumber
\end{eqnarray}
where $\pi_{s,t}=\mathbb{P}(A_{s,t}=1)$ is the activation probability. Conditional on activation, the distribution of $Y_{s,t}$ is governed by the ZTNB component through the mean parameter $m_{s,t}$ and size parameter $r$. In particular, using (\ref{eq:meanZTNB}), the conditional positive mean on a hearing-day $t$ in sector $s$ can be written as
\begin{equation}
\mu_{s,t}^+ \equiv \mathbb{E}[Y_{s,t}\mid A_{s,t}=1,\, z_{s,t}] = \frac{m_{s,t}(\beta)}{1-\left(\frac{r}{r\hspace{0.5pt}+\hspace{0.5pt}m_{s,t}(\beta)}\right)^r}.
\end{equation}

\section{Results}\noindent
\subsection{Burst onset}\noindent
The baseline model $\text{BOA}_0$ contains no covariates. 
Thus, the probability is governed purely by the Weibull structure in the baseline hazard. The baseline model serves as a reference for the other models. 

The baseline model is extended to include the calendar years as covariates: 
$\text{BOA}_Y$. Under the proportional hazards interpretation, an estimated coefficient $\hat{\gamma}_y$ multiplies the baseline hazard by $\exp(\hat{\gamma}_y)$ for a year $y$ where the sum of the $\hat{\gamma}_y$ is constrained to be $=$ to ensure model identification.

The estimates of the year effects are shown in table \ref{tab:BOAyeareffects}. 
\begin{table}
\centering
\setlength{\tabcolsep}{3pt}
\resizebox{\columnwidth}{!}{%
    \begin{tabular}{r|cccc}
 \textbf{Year} & 2009 & 2010 & 2011 & 2012 \\
  \textbf{Coefficient} & 0.2150 & 0.1609 & 0.1416 & 0.1900 \\
  \hline
  \textbf{Year} & {\bf 2013} & 2014 & 2015 & 2016\\
  \textbf{Coefficient} & {\bf 0.2387} & 0.1526 & 0.07004 & 0.03115\\
  \hline
  \textbf{Year} & 2017 & 2018 & 2019 & 2020 \\
  \textbf{Coefficient} & $-$0.07189 & $-$0.08977 & $-$0.1062 & $-$0.1543 \\
  \hline
  \textbf{Year} & {\bf 2021} & 2022 & 2023 & 2024\\
  \textbf{Coefficient} & {\bf $-$ 0.3751} & $-$0.2988 & $-$0.1046 & $6.700\cdot 10^{-4}$\\
  \end{tabular}
}
\caption{Estimated year effects in the BOA year model ($\text{BOA}_Y$), the highest and lowest values are marked in bold. Note that the estimate for 2024 is implied by the sum-to-zero constraint: $\gamma_{2024} = -\sum_{y=2009}^{2023}\hat{\gamma}_y$ }\label{tab:BOAyeareffects}
\end{table} 
The smallest and largest year effects correspond to $2021$ and $2013$ respectively. Their multiplicative effect on the hazard ranges from 
\begin{equation}
\exp(-0.3751) \approx 0.6873 \quad  \text{to} \quad \exp(0.2387) \approx 1.270.
\end{equation}

Next, the sector effects are included as covariates. The shape parameter estimate is larger than $1$, which does not lie within the 95\% confidence interval. The Wald test results in $p\approx 1-2.6\cdot 10^{-66}$), giving no evidence for $k<1$. The constant estimate is $-\hat{k}\log\hat{\lambda} \approx -3.675$. The estimated sector effects are given in table \ref{tab:BOAseceffects} (see appendix \ref{app:appendixOnsetTable}). The smallest and largest sector effects are
\begin{equation}
\min_{s\in \mathcal{S}}\hat{\gamma}_s =\hat{\gamma}_{19} = -5.552\qquad \text{and}\qquad \max_{s\in \mathcal{S}}\hat{\gamma}_s =\hat{\gamma}_{64} = 2.662,
\end{equation}
so that the multiplicative effect on the hazard ranges from 
\begin{equation}
\exp(-5.552) \approx 0.003881\quad \text{to}
\quad \exp(2.662) \approx 14.32.
\end{equation}
The average onset probability estimate is $\hat{p}_{\text{avg}} = 0.1069$. Relative to the baseline model, the AIC decreases by $15958.0$ and the BIC by $15204.4$, indicating a considerable improvement in fit.

For the final regular model, both the sector and year covariates are included. Again, the Wald test yields $p \approx 1-6.100\cdot 10^{-223}$, giving no evidence for $k<1$. The constant estimate is $-\hat{k}\log\hat{\lambda} \approx -4.035$. The estimated effects for both sector and year are given in table \ref{tab:BOAsecyeareffects}. Across all covariates, the smallest and largest estimates are both sectors estimates:
\begin{equation}
\min_{s\in\mathcal{S}}\hat{\gamma}_s=\hat{\gamma}_{19} = -6.001\quad\text{and} \quad \max_{s\in\mathcal{S}}\hat{\gamma}_s =\hat{\gamma}_{64} = 2.863,
\end{equation}
For the year effects specifically, the smallest and largest estimates are
\begin{equation}
\min_{y\in\mathcal{Y}}\hat{\gamma}_y =\hat{\gamma}_{2021} = -0.7376\ \ \text{and}\ \ \max_{y\in\mathcal{Y}}\hat{\gamma}_y =\hat{\gamma}_{2013} = 0.4513.
\end{equation}
Thus, the multiplicative effect of the sector covariates on the baseline hazard ranges from 
\begin{equation}
\exp(-6.001) \approx 0.002477 \quad \text{to}\quad \exp(2.863) \approx 17.52,
\end{equation}
and the multiplicative effect of the year covariates ranges from
\begin{equation}
\exp(-0.7376)\approx 0.4783 \quad \text{to}\quad \exp(0.4513)\approx 1.570.
\end{equation}
Compared to the sector model, the AIC decreases by $2455.4$ and the BIC by $2300.6$. The average onset probability estimate is $\hat{p}_{\text{avg}} = 0.1065$.

\subsubsection{Best regular model}\noindent
The four regular models are compared using the AIC, BIC and the likelihood-ratio tests, all of which are reported in table \ref{tab:onsetLRs}. The results show strong evidence in favour of including both sector and year effects. $\text{BOA}_{SY}$ is taken as the preferred model and augmented with the exposure variables in the next stage.
\begin{table*}[h]
\centering
\begin{tabular}{ll|lll|ll}
\multicolumn{1}{l}{\textbf{Model 1}} & \multicolumn{1}{l}{\textbf{Model 2}} & \multicolumn{1}{|c}{\textbf{df}} & \multicolumn{1}{c}{\textbf{LR}} & \multicolumn{1}{c|}{{$p$}\textbf{\hspace{1pt}-value}}& {$\Delta_{\rm AIC}$} & {$\Delta_{\rm BIC}$}\\
\hline
Baseline & Year          & 15 & 682.318 & $1.089\cdot 10^{-135}$  & 652.3 & 497.5\\
Baseline & Sector        & 73 & 16104.0 & $<10^{-3398}$ & 15958.0 & 15204.4\\
Year&Sector + Year       & 73 & 17907.1 & $<10^{-3788}$ & 17761.1 & 17007.5\\
Sector&Sector + Year     & 15 & 2485.41 & $<10^{-521}$  & 2455.4 & 2300.6\\
\hline
\end{tabular}
\caption{Likelihood-ratio tests and AIC/BIC differences for the burst onset models.}
\label{tab:onsetLRs}
\end{table*}

\subsection{Exposure models}\noindent
The exposure covariates are added sequentially to $\text{BOA}_{SY}$, in accordance with fig.\ref{fig:model_hierarchy}. For each new model, one input and one output exposure variable is added with lags of increasing length to capture the exposure effects at different timescales. Specifically, one short lag of one hearing-day and three longer lags, defined as the first hearing-day at least one month, one quarter, and two quarters in the past. 
This leads to the following exposure models:
\begin{itemize}
    \item $\text{BOA}_{SY}\text{E}_1\colon\,\,\, \text{BOA}_{SY} + {\rm Exp}^\text{in/out}_{s,t-1}$
    \item $\text{BOA}_{SY}\text{E}_2\colon\,\,\, \text{BOA}_{SY}+{\rm Exp}^\text{in/out}_{s,t-1} + {\rm Exp}^\text{in/out}_{s,l(30,t)}$
    \item $\text{BOA}_{SY}\text{E}_3\colon\,\,\, \text{BOA}_{SY}+{\rm Exp}^\text{in/out}_{s,t-1} + {\rm Exp}^\text{in/out}_{s,l(30,t)} + {\rm Exp}^\text{in/out}_{s,l(91,t)}$
    \item $\text{BOA}_{SY}\text{E}_4\colon\,\,\, \text{BOA}_{SY}+{\rm Exp}^\text{in/out}_{s,t-1} + {\rm Exp}^\text{in/out}_{s,l(30,t)} + {\rm Exp}^\text{in/out}_{s,l(91,t)} + {\rm Exp}^\text{in/out}_{s,l(182,t)}$.
\end{itemize}
Each added lag yields two new estimates $\gamma^{(\text{lag})}_{\text{in}}$ and $\gamma^{(\text{lag})}_{\text{out}}$ that represent the in-exposure and out-exposure at that timescale. The models are again compared using AIC/BIC and LR tests. The results are given in table \ref{tab:onsetexposureLRs}.

\begin{table*}[h]
\centering
\begin{tabular}{ll|lll|ll}
\multicolumn{1}{l}{\textbf{Model 1}} & \multicolumn{1}{l|}{\textbf{Model 2}} & \multicolumn{1}{c}{\textbf{df}} & \multicolumn{1}{c}{\textbf{LR}} & \multicolumn{1}{c|}{{$p$}\textbf{\hspace{1pt}-value}} & {$\Delta_{\text{AIC}}$} & {$\Delta_{\text{BIC}}$}\\
\hline
$\text{BOA}_{SY}$           & $\text{BOA}_{SY}\text{E}_1$  & 2 & 205.243 &  $2.704\cdot 10^{-45}$  &201.2   &180.8\\
$\text{BOA}_{SY}\text{E}_1$ & $\text{BOA}_{SY}\text{E}_2$  & 2 & 656.812 & $2.372\cdot 10^{-143}$ &652.8   &632.2\\
$\text{BOA}_{SY}\text{E}_2$ & $\text{BOA}_{SY}\text{E}_3$  & 2 & 30018.6 & $<10^{-6517} $         &30014.6   &29994.0\\
$\text{BOA}_{SY}\text{E}_3$ & $\text{BOA}_{SY}\text{E}_4$  & 2 & 3459.50 & $<10^{-750} $          &3455.5   &3434.8\\
\hline
\end{tabular}
\caption{Likelihood-ratio tests for the nested exposure models in the burst onset analysis.}
\label{tab:onsetexposureLRs}
\end{table*}
Across these models, the richest model, i.e. $\text{BOA}_{SY}\text{E}_4$, is preferred: it has the lowest AIC and BIC, and the LR tests favour the expanded model at each step. For these reasons, this model is treated as the final burst onset model. Its model characteristics are given in table \ref{tab:BOAexp4effects}.

\begin{table}[h]
\centering
%\resizebox{\columnwidth}{!}{%
\scalebox{0.95}{
\begin{tabular}{l|cccc}
$\hat{\gamma}$& $t-1$ & $l(30,t)$ & $l(91,t)$ & $l(182,t)$\\
\hline
in & 0.6821 & 0.2696 & 1.898 & 1.699\\
out & 0.3155 & $-$0.04931 & 1.229 & 0.7865\\
\end{tabular}
}

\par

\scalebox{0.95}{
\begin{tabular}{l|cccc}
$\exp(\hat{\gamma})$& $t-1$ & $l(30,t)$ & $l(91,t)$ & $l(182,t)$\\
\hline
in & 1.978 & 1.309 & 6.675 & 5.47 \\ 
out & 1.371 & 0.9519 & 3.418 & 2.196 \\
\end{tabular}
}
%}
\caption{Estimated effects of the lagged exposure variables $\hat{\gamma}$ in the $\text{BOA}_{SY}\text{E}_4$ model (top) with exponentiated forms $\exp(\hat{\gamma})$ (bottom).}
\label{tab:BOAexp4effects}
\end{table}
The average onset probability estimate is $\hat{p}_{\text{avg}} = 0.1055$, corresponding to an average burst onset probability of $10.6\%$, conditional on the previous hearing-day being inactive.

\begin{table}[ht]
\centering
%\resizebox{\columnwidth}{!}{%
\scalebox{0.95}{
\begin{tabular}{l|cccc}
$\hat{\beta}$ & $t-1$ & $l(30,t)$ & $l(91,t)$ & $l(182,t)$\\
\hline
in & $-$0.5175 & $-$0.6582 & 0.7841 & 0.7852 \\
out & $-$0.2765 & $-$0.1043 & 0.4306 & 0.2329 \\
\end{tabular}
}
\par

\scalebox{0.95}{
\begin{tabular}{l|cccc}
$\exp(\hat{\beta})$ & $t-1$ & $l(30,t)$ & $l(91,t)$ & $l(182,t)$\\
\hline
in  & 0.5960 & 0.5178 & 2.190 & 2.193 \\
out & 0.7584 & 0.9010 & 1.538 & 1.262 \\
\end{tabular}
}
\caption{Estimated effects of the lagged exposure variables $\hat{\beta}$ in the $\text{BSA}_{SY}\text{E}_4$ model (top) with corresponding multiplicative effects $\exp(\hat{\beta})$ (bottom).}
\label{tab:BSAexp4effects}
\end{table}

The optimisation routine converged in each model under the standard stopping criteria. As expected, the larger models required more L-BFGS iterations to converge, likely due to the higher-dimensional space to optimise over. A summary of each model is given in table \ref{tab:onsetsummary}. To show the increase in fit quality, the difference in AIC and BIC between a model $j$ and the final model is given by $\Delta\text{AIC}$ and $\Delta\text{BIC}$, defined as
\begin{equation}
\Delta \text{AIC}_j = \text{AIC}_j - \text{AIC}_{\text{BOA}_{SY}\text{E}_4},
\end{equation}
and similarly for $\Delta \text{BIC}_j$.

\begin{figure}
\centering
\includegraphics[width=0.95\linewidth]{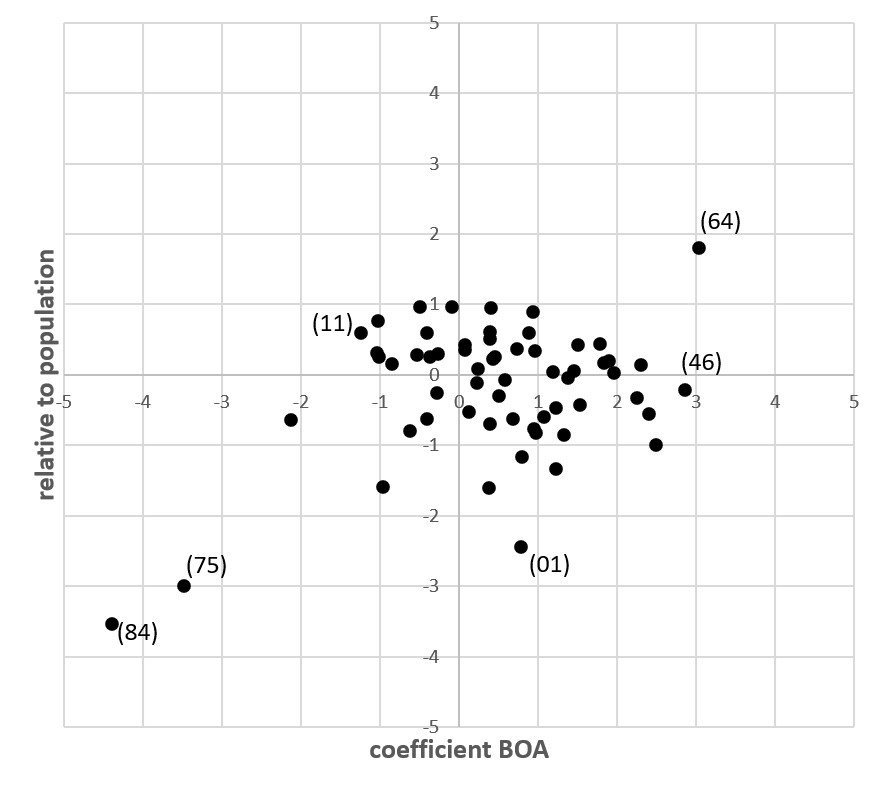}
\caption{The difference between the coefficients of best $\text{BOA}$ model and the logarithm of the fraction of firms in that sector is plotted against the $\text{BOA}$ coefficients of that model, listed in table \ref{tab:BOAsecyearexp4effects} }
\label{fig:BOA_relative}
\end{figure}

In addition to likelihood increase for the absolute numbers of bankruptcies per sector, it is also of interest to look at these likelihoods relative to the sizes of the sectors in terms of the number of companies that are active within it. Taking the difference between the coefficients determined in the best model for BOA or BSA and the logarithm of the fraction of the firm numbers compared to the total population of firms, produces precisely this relative effect. In Fig. \ref{fig:BOA_relative} 
these modified coefficients relative to the population are plotted against the original coefficients for the best BOA model.
Some of the sectors at the periphery of the point clouds are indicated with the numbers between brackets.
From this figure it is clear that once corrected for the size of the sector, in terms of numbers of active firms within it. The ordering of the largest and smallest likelihood ratios does change somewhat, but it is also clear that there is considerable spread. Differences in bankruptcy hazards between sectors are not purely a consequence of the sector size.

\subsection{Burst size}\noindent
Since the burst size model is fitted to the conditional distribution $Y_{s,t}\mid A_{s,t}=1$, the sample must first be restricted to active hearing-days. The data are filtered to remove the observations for which $A_{s,t}=0$. The same process is followed as in the burst onset analysis, with the same nested structure. The AIC and BIC, convergence diagnostics, estimated model parameters with corresponding 95\% confidence intervals and the covariate coefficients are calculated. In addition, the estimated ZTNB mean and its underlying NB mean are shown as a reference, to show the effect of the zero truncation of the distribution on the mean.

Table \ref{tab:BSAsummary} summarises the main statistics and estimates across the regular models, as well as the best exposure model. As in the burst onset analysis, the differences in AIC and BIC between the current model and the best model are given as $\Delta\text{AIC}$ and $\Delta\text{BIC}$.

\begin{figure}
\centering
\includegraphics[width=0.9\linewidth]{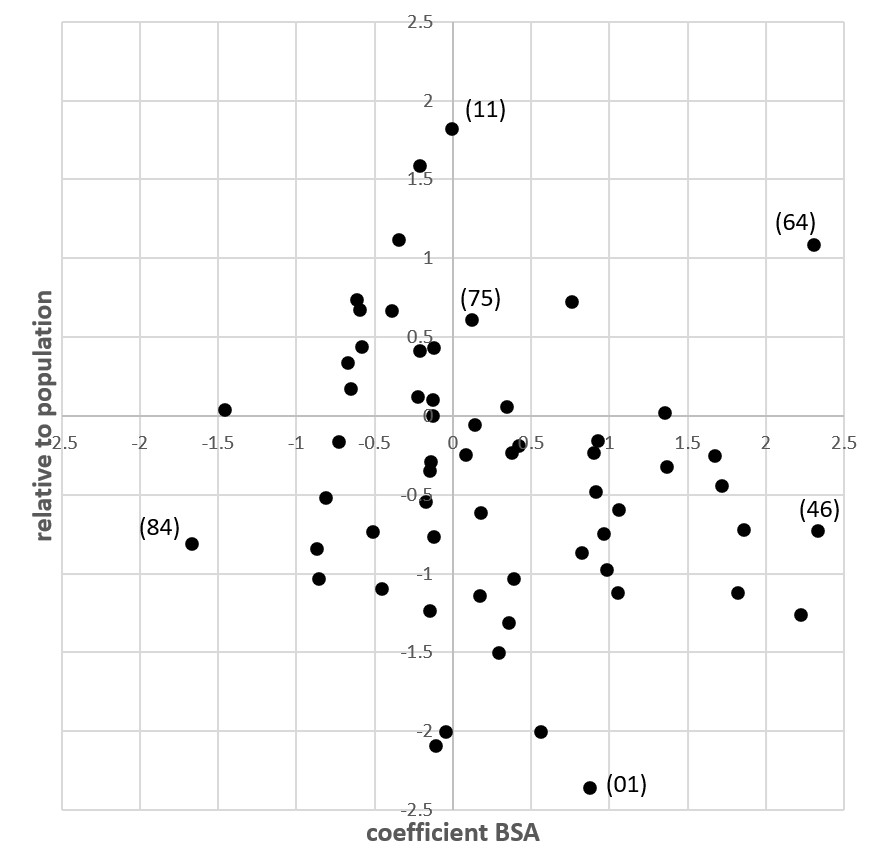}
\caption{The difference between the coefficients of best $\text{BSA}$ model and the logarithm of the fraction of firms in that sector is plotted against the $\text{BSA}$ coefficients of that model, listed in table \ref{tab:BSAsecyearexp4effects} }
\label{fig:BSA_relative}
\end{figure}

For the BSA also it is of interest to look at these likelihoods relative to the sizes of the sectors in terms of the number of companies that are active within it. In Fig. \ref{fig:BSA_relative} 
these modified coefficients relative to the population show are plotted against the original coefficients for the best BSA model.
Some of the sectors at the periphery of the point clouds are indicated with the numbers between brackets.
Again, the ordering of the largest and smallest likelihood ratios does change somewhat, but it is clear that there is considerable spread.

\begin{table*}[h]
\centering
%\resizebox{\columnwidth}{!}{%
\begin{tabular}{lccccc}
& \multicolumn{4}{c}{Regular models} & \multicolumn{1}{c}{Exposure model} \\
\cmidrule(lr){2-5}\cmidrule(lr){6-6}
& \makecell{\textbf{Baseline} \\ {$(\text{{BSA}}_0)$}} & \makecell{\textbf{Year}\\{$(\text{BSA}_Y)$}} & \makecell{\textbf{Sector}\\{$(\text{BSA}_S)$}} & \makecell{\textbf{Sector + year}\\{$(\text{BSA}_{SY})$}} & \makecell{\textbf{Full exposure}\\{$(\text{BSA}_{SY}\text{E}_4)$}} \\
\hline
AIC                & 95159.8 & 93882.3 & 87435.5 & 85521.4 & 80607.6\\
BIC                & 95176.4 & 94023.9 & 88060.6 & 86271.6 & 81424.4\\
$\Delta\text{AIC}_j$        & 14552.2 & 13274.7 & 6827.9  & 4913.8  & 0\\
$\Delta\text{BIC}_j$        & 13752.0 & 12599.5 & 6636.2  & 4847.2  & 0\\\hline
$\hat{r}$          & $2.707\cdot10^{-6}$ & $4.637\cdot10^{-7}$ & $0.3002$ & $0.5213$ & 1.994 \\
$\hat{\mu}^+$      & 2.387    & 2.387 & 2.386 & 2.396 & 2.372 \\
$\hat{\mu}$        & $9.978\cdot 10^{-6}$ & $1.735\cdot10^{-6}$ & 0.7783 & 1.078 & 1.602 \\
\hline
Intercept          & $-11.52$ & $-13.35$ & $-0.6960$ & $-0.4444$ &  $-0.09640$\\
Smallest $\hat{\beta}$  & $-$      & $-1.057$ & $-4.611$ & $-4.455$  &  $-3.711$\\
Largest $\hat{\beta}$  & $-$      & 0.6694   & 2.527    & 2.486     & 2.331\\
\hline
\end{tabular}
%}
\caption{Model characteristics for the burst size analysis models.}
\label{tab:BSAsummary}
\end{table*}

\subsection{Burst onset models for separate sectors} \label{app:AppendixSepSectors}
\begin{figure}
\centering
\includegraphics[width=0.9\linewidth]{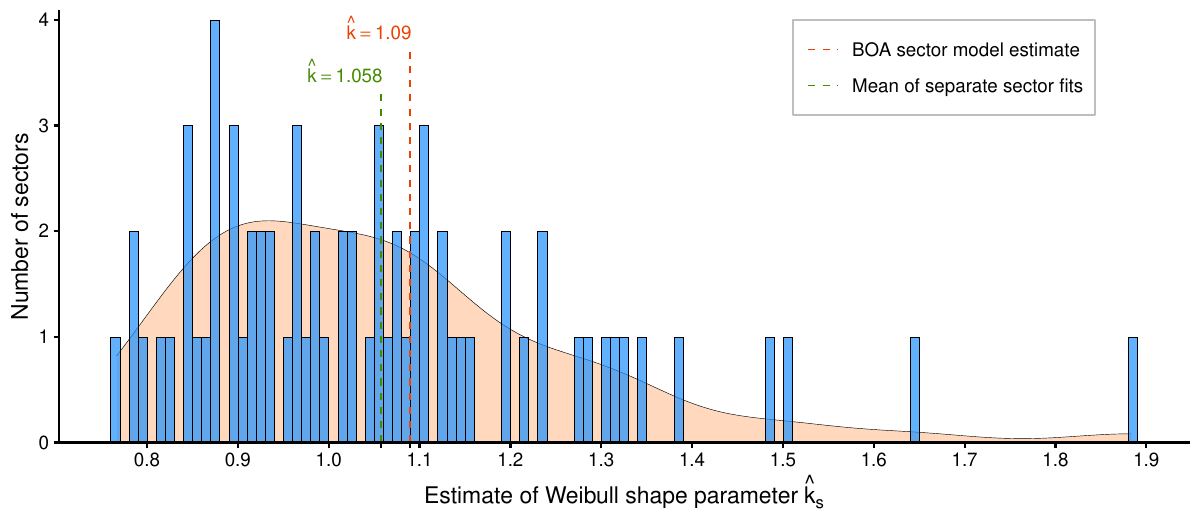}
\caption{Histogram and kernel density estimate of estimates $\hat{k}_s$ corresponding to the separate sector models. The shape parameter estimate in the $\text{BOA}_S$ model is annotated in red, the average of the separate sector estimates in green.}
\label{fig:Hist_sectors}
\end{figure}

\noindent
To investigate whether the Weibull shape parameter estimate $\hat{k} = 1.090$ from the $\text{BOA}_S$ model is a result of variation between the sectors, a model for each sector is created by restricting the bankruptcy data to only observations from a given sector, then fitting a baseline model to this sector-specific data. This yields estimates $\hat{k}_s$ for each $s\in\mathcal{S}$. \ref{fig:Hist_sectors} presents a histogram overlaid on top of an estimated kernel density plot of these estimates. This results in 
\begin{equation}
|\mathcal{S}|^{-1}\sum_{s\in\mathcal{S}} \hat{k}_s = 1.058\approx 1.090.
\end{equation}
The largest and smallest estimates of $\hat{k}$ correspond to sectors $36$ and $03$, with $\hat{k}_{36} = 1.883$ and $\hat{k}_{03} = 0.7665$ respectively. The five largest and smallest shape parameter estimates are given in \ref{tab:topbotsecs} below. For completeness, no sectors are left out based on a low burst count. Non-finite estimates are ignored, which can happen if there are two or less burst onsets in the data (this occurs in three sectors).

\begin{table}
\centering
\scalebox{0.95}{
%\resizebox{0.9\textwidth}{!}{%
\begin{tabular}{c|c|c}
\textbf{Sector} & \textbf{Estimate }{$\hat{k}$} & \textbf{\#Burst onsets} \\
\hline
36 & 1.883 & 4 \\
56 & 1.643 & 866 \\
47 & 1.506 & 911 \\
43 & 1.487 & 878 \\
49 & 1.382 & 775 \\
\end{tabular}
}

\par

\scalebox{0.95}{
\begin{tabular}{c|c|c}
\textbf{Sector} & \textbf{Estimate }{$\hat{k}$} & \textbf{\#Burst onsets} \\
\hline
03 & 0.7665 &  14 \\
21 & 0.7834 &  20 \\
65 & 0.7841 &  16 \\
11 & 0.7991 &  67 \\
01 & 0.8115 &  348 \\
\end{tabular}
}
\caption{The five largest (left) and five smallest (right) Weibull shape parameter estimates ($\hat{k}$) with the number of burst onsets denoted.}
\label{tab:topbotsecs}
\end{table}

\section{Discussion}\noindent
\subsection{The burst onset process}\noindent
In the baseline and year models $\text{BOA}_0$ and $\text{BOA}_Y$, the shape parameter estimates $\hat{k} \approx 0.7042$ and $\hat{k} \approx 0.7188$ are indicative of (residual) intermittent behaviour. 
Notably, this effect seems to vanish or even flip when sector effects are introduced: in the sector and sector + year models $\text{BOA}_S$ and $\text{BOA}_{SY}$, the estimates $\hat{k}=1.090$ and $\hat{k}=1.179$ imply that the hazard is increasing, and hence that $p_{s,t}$ increases with $R_{s,t-1}$, the time since the end of the previous burst. 
In the best model $\text{BOA}_{SY}\text{E}_4$, the estimate of $k$ moves back below one, with $\hat{k}=0.9813\approx 1$. This suggests that, once all covariates are included, the baseline hazard is close to exponential, and the residual process is close to memoryless, although the Wald test still points to $k<1$ with $p$-value $\approx 0.003$.

These results suggest that this effect is mostly explained by between-sector heterogeneity in the data, with a smaller emphasis on between-year differences, further explained by the effects across different lags. Therefore, when sectors, years and exposure effects are accounted for, the intermittency signature for the intervals between {\it burst onsets} largely disappears and the residual process becomes approximately memoryless.

These observations are consistent with the conclusions of \cite{lopes_cardozo_internship_2024}, i.e. that the pooled process shows intermittent behaviour. It also should be noted that in \cite{lopes_cardozo_internship_2024} the bankruptcy counts are not pooled into bursts, as is done here. By describing the bankruptcy counts as bursts, with a stochastic process describing the total count accumulated during each burst as separate from the onset timing process of such bursts, the statistics also change. The process of {\it burst onsets} can therefore look closer to memoryless even though the individual bankruptcy count process does show intermittency; the aggregation of counts into bursts acts to suppress an intermittency signature for the burst onset timing.

\subsection{Average conditional burst size}\noindent
Because the BSA models the distribution of the positive hearing-day counts, the quantity of interest is the average number of bankruptcies on an active hearing-day, conditional on the covariates:
\[
\mu_{s,t}^+ = \mathbb{E}[Y_{s,t}\mid A_{s,t}=1, z_{s,t}] = \cfrac{m_{s,t}}{1-(\frac{r}{r+m_{s,t}})^r},
\]
where $m_{s,t} = m_{s,t}(\beta)$ is the mean of the underlying NB distribution. Across all models, the estimated average conditional mean $\hat{\mu}^+$ is stable, ranging from 2.372 to 2.387. It is close to the empirical mean
\[
\overline{Y_{s,t}\mid A_{s,t}=1} = 2.387.
\]
Table~\ref{tab:BSAavgs} in appendix~\ref{app:appendixBSATable} compares the empirical and estimated conditional means per sector. While $\hat{\mu}^+$ remains stable, the average underlying NB mean $\hat{\mu}$ changes substantially, ranging from $9.978\cdot 10^{-6}$ through $1.602$. This is to be expected, as the untruncated mean is influenced by the large proportion of zero-count days. This effect is reduced when sector and year effects are included, and especially when exposure variables are added.

\subsubsection{Interpretation of the size parameter}\noindent
In the ${\rm ZTNB}(r,m)$ parametrisation, the size parameter $r$ describes the deviation from the Poisson distribution, as the NB distribution converges to the Poisson distribution when $r\to \infty$ (cf Eq.~(\ref{eq:ZTNBEandV}) ). The {\it index of dispersion} quantifies this deviation. It is defined as the ratio between the variance and mean of the distribution of $Y_{s,t}$, which is equal to one for a Poisson distribution. For the underlying NB distribution with size parameter $r$ and mean $m_{s,t}$ :
\[
{\rm Var}(Y_{s,t} \mid z_{s,t}) = m_{s,t} + \frac{m_{s,t}^2}{r}, \] 
so 
\[\frac{{\rm Var}(Y_{s,t}\mid z_{s,t})}{\mathbb{E}[Y_{s,t}\mid z_{s,t}]} = 1 + \frac{m_{s,t}}{r}.
\]
In the best model $\text{BSA}_{SY}\text{E}_4$: $\hat{r} = 1.994$ with 95\% CI $[1.8615,\, 2.1358]$. The sample average of the NB mean is $\hat{\mu} = 1.602$, which means that the index of dispersion for the underlying NB model is
\[
1+\frac{\hat{\mu}}{\hat{r}} \approx 1 + \frac{1.602}{1.994} \approx 1.80.
\]
As this is larger than one, it shows that the process is overdispersed and remains more variable than Poisson, even in the best-fitting model, which shows that the NB component underlying the zero-truncated distribution is preferable to Poisson in the best model. The earlier models also show that the covariates are crucial in identifying $r$. In the $\text{BSA}_0$ and $\text{BSA}_Y$ models, $\hat{r}$ is extremely small and the corresponding confidence interval is extremely wide. 
$r$ becomes well-identified once sector effects are included ($\hat{r}=0.3002$ in $\text{BSA}_S$ and $\hat{r}=0.5213$ in $\text{BSA}_{SY}$ with small confidence intervals). In the full $\text{BSA}_{SY}\text{E}_4$ model the estimate increases to $\hat{r} = 1.994$, with 95\% CI [1.8615, 2.1358]. 
This indicates that overdispersion is reduced when conditioned on sectors, or sectors and years, reaffirming that there is heterogeneity across sectors in the data.

\subsubsection{Exposure effects}\noindent
Table \ref{tab:BSAexp4effects} reports the estimated input and output exposure coefficients corresponding to the four lag lengths, along with their multiplicative effect on $m_{s,t}$. 
The estimates suggest that higher exposure is associated with smaller expected burst sizes at the short lags (one hearing-day and one month), but with larger expected burst sizes at longer lags (one to two quarters), which illustrates that the relation between exposure and burst size is different across timescales. As in the burst onset analysis, it is more informative to interpret the changes in exposure on a smaller scale than a full unit change, as this is not a frequent occurrence. 
For instance, a 0.1 increase in the exposure-in covariate at lag $t-1$ corresponds to a factor of $\exp(-0.5175\cdot 0.1)\approx 0.95$, whereas at the one-quarter lag it corresponds to $\exp(0.7841\cdot 0.1)\approx 1.08$. 

The asymmetry between observed exposure-in and expo\-sure-out effects in the burst onset models is replicated in the burst size models: at the longer lags (one and two quarters), exposure-in has larger positive coefficients than exposure-out, whereas at the shorter two lags (one hearing-day and one month) the opposite is true.
Although the asymmetry is less pronounced: the ratio between the exponentiated exposure-in and exposure-out effects across lags are approximately 0.89, 0.63, 1.34 and 1.58 respectively.

\section{Conclusions}\noindent
The Netherlands is a developed, open economy, currently the fifth largest in the European Union. As such, its firm demographics is a good starting point for studying the processes of sectoral interdependencies and exposure, and hence resilience of developed economies. It has been argued that delays in (scheduled) supply chains can escalate to a catastrophic pile-up \citep{moran_timeliness_2024, moran_critical_2025}. 
For firms, a large enough such delay in their supply or demand can even cause financial liquidity problems and bankruptcy. One empirical approach to determining whether the economy is
operating close to a critical transition is to look at time series of bankruptcies for which there is complete data at individual firm level available at the Dutch national statistical institute: Statistics Netherlands.

The process over time of firm bankruptcies in the Netherlands is decomposed into 
the onset of bursts of bankruptcy activity and their conditional size on active hearing-days. A survival-type analysis shows how these quantities vary across years, sectors, and inter-sectoral exposure. This framework provides a useful decomposition of intermittency within bankruptcy dynamics. 

Adding sector effects yields the largest improvement in fit in both the burst onset and burst size analyses. Secondary improvements are made by adding year effects, and fit is improved further when exposure variables are added. This indicates that heterogeneity within the bankruptcy data is attributable to differences across sectors, with smaller effects across years and additional variation explained by inter-sectoral exposure.

On average across sectors, the fitted version of the first decomposition differs from the empirical bankruptcy rate by about $0.62\%$, whereas the fitted version of the second decomposition differs by about $4.36\%$. Since the first decomposition replaces the empirical burst size component, while the second decomposition replaces the empirical burst onset component, this suggests that the remaining difference between the fitted components and the empirical bankruptcy rate comes mostly from the burst onset component, rather than from the conditional burst size component. Effectively, this indicates that the size of activity on active hearing-days is easier to recover from the data than the starting time of new bursts.

In the best BOA model, the Weibull baseline shape parameter estimate $k=0.9813$ is close to one after conditioning on years, sectors, and exposure covariates. This suggests that the residual process of {\it burst onsets} is close to memoryless, although evidence in favour of $k<1$ still remains. In the best BSA model, the size parameter estimate $\hat{r} = 1.994$ is well identified and indicates that the underlying distribution remains overdispersed relative to a Poisson distribution, while the estimated conditional positive mean remains stable around $\hat{\mu}^+ = 2.372$. Together, these results suggest that much of the heterogeneity in the bankruptcy data is explained by sector differences and by exposure to linked sectors across different timescales.

The addition of the I/O-based exposure variables improves fit in all BOA and BSA models. The estimated exposure effects differ across timescales, especially at longer lags. 
In the BOA, longer lags correspond to large and mostly positive effects, whereas in the BSA the shorter lags are negative and the longer lags positive. These characteristics are consistent with propagation mechanisms of shocks that operate over timescales in the quarterly to half-year range.

A natural interpretation of the exposure coefficients is that they capture effects that are associated with propagation of shocks through financial pathways, ie. payments for supply and demand. For input exposure, failure of suppliers can cause input shortages or operational problems, increasing stress in connected sectors. For output exposure, difficulties among major customers may result in delayed payments or decreasing demand, again increasing risk of bankruptcy. 
It is feasible for such effects to manifest at different timescales, as a build-up of stress is often gradual. These interpretations fit the broader idea that shocks can be amplified in connected systems. 

\medskip\noindent
Data availability: Much of the analysis has been carried out within the secure computing environment of Statistics Netherlands, since it is based on microdata from an administrative register of individual bankruptcies and therefore subject to disclosure control regulation. Further information on gaining access to such data for external researchers can be found at {\tt https://www.cbs.nl/en-gb/our-services/ customised -services-microdata/microdata-conducting-your -own-research}. The data on firm demographics used to produce figs. \ref{fig:BOA_relative} and \ref{fig:BSA_relative} is available as open data via {\tt https:// opendata.cbs.nl/}. Time series data of monthly aggregated bankruptcies, used in appendix \ref{app:appendixHawkes}, can also be found as open data via this site.

\appendix

\section{Hawkes process analysis}\label{app:appendixHawkes}\noindent
A standard model for self-exciting or mutually exciting point processes is the Hawkes process \citep{Hawkes1970,Hawkes1971}.
In this model there is a primary excitation of events, but then each such event can itself trigger more events, resulting in cascades. In one dimension, the model is a counting
process for events $N(t)$ with an intensity $\lambda(t)$ given by a constant term $\mu$ and a self-exciting term which is a function of the event history, leading to:
\begin{equation}
    \lambda(t) = \mu + \int\limits_0^\infty \phi(\tau) N(t-\tau){\rm d}\tau
\end{equation}
The kernel $\phi$ expresses how strongly, and how far back in time, past events enhance the likelihood of a new event occurring in an interval around time $t$. Preserving causality requires that $\phi\equiv 0$ for $\tau<0$. The integral of this Hawkes kernel is the branching ratio $r_b$, and it determines the behaviour of the model. An $r_b>1$ implies that each event itself triggers more than 1 new event: a runaway non-stationary process. $r_b<1$ leads to a stationary process, and evidently $r_b=1$ is a critical value. 
If $\phi=0$ for all $\tau$ then the process reduces to the usual memoryless Poisson process.

As is demonstrated in \cite{HardBou2014} it is possible to estimate 
this branching ratio from a time series of event counts in a fairly straightforward manner, by accumulating counts over time-windows with width $W$. In that paper it is shown that a good estimator $\langle r_b\rangle$ for that branching ratio is:
\begin{equation}
\langle r_b(W)\rangle = 1-\sqrt{\frac{\mu_W}{\sigma^2_W}}
\end{equation}
in which:
\begin{eqnarray}
    \mu_W &\equiv& \frac{W}{T} N_T = \frac{1}{m}\sum_{i}^{m} N_W(i)
    \nonumber\\
    \sigma^2_W &\equiv& \frac{1}{m-1}\sum_{i}^{m} \left[N_W(i)-\mu_W\right]^2
\end{eqnarray}
where   $T$ is the total length of the available time series
$N_W(i)$ is the event count within window $i$, and $m$ is the total number of windows of length $W$ that covers $T$.

This analysis can in principle be applied per sector, but for the present it is done only on the collected bankruptcies for firms in the Netherlands across all sectors. Table \ref{HawkesTab} shows
the results for three versions: the original time series of bankruptcies, but also two detrended time series, filtered to remove economic cycles with periods $>1\ {\rm yr}$. This detrending can be done by subtraction of a smooth trend (linear) or instead by dividing, i.e. taking the logarithm and then subtracting a smooth trend (logarithmic). The reason for doing so
is that the original time series has variations which might be exogenous; driven by the economic cycle, and thus unduly influence the determination of the branching ratio. A case can be made for either the linear or the logarithmic de-trending. However, there is very little difference between the three as is clear from table \ref{HawkesTab}.

\begin{table}[h]
\begin{tabular}{rrrrr}
&& \multicolumn{3}{c}{$r_b(W)$} \\
&& direct & \multicolumn{2}{c}{de-trended}\\
$W$ & m && {linear} & logarithm \\
\hline
2& 264& 0.9259 & 0.8727 & 0.8976 \\
3& 176& 0.9515 & 0.9212 & 0.9382 \\
4& 132& 0.9617 & 0.9376 & 0.9518 \\
5& 105& 0.9681 & 0.9491 & 0.9600 \\
6& 88& 0.9720 & 0.9554 & 0.9651 \\
8& 66& 0.9768 & 0.9629 & 0.9711 \\
10& 52& 0.9800 & 0.9681 & 0.9750 \\
12& 44& 0.9820 & 0.9715 & 0.9777 \\
16& 33& 0.9848 & 0.9758 & 0.9811 \\
20& 26& 0.9867 & 0.9789 & 0.9834 \\
$c_\infty$ & & 0.994 & 0.987 & 0.990 \\
\end{tabular}
\caption{The empirical value of the branching ratio $r_b$ using a range of window widths $W$ (in months). Second column shows the number of samples $m$ of width $W$ covering the available time series. }
\label{HawkesTab}
\end{table}

The length of this time series, in combination with the cadence with which it is measured, does not allow for as wide a range in window widths $W$ as is possible for \cite{HardBou2014}. The final row in \ref{HawkesTab} is determined by fitting a single power law to $c_\infty - r_b(W)$. That value for $c_\infty$ is shown where the $R^2$ of that fit is highest ($>0.995$ in all cases). This is a similar procedure to what is shown in \cite{HardBou2014} which shows this behaviour as a function of $W$.

This asymptotic value does differ slightly between the three versions of the time series, but it is clear that $r_b$ is very close to critical $\approx 0.99$. 

%\clearpage
\section{The SBI 2008}\label{subsec:SBIcodes}\noindent
The Standard Business Index (Dutch: Standaard Bedrijfsindeling), is a hierarchical classification that assigns businesses into specific sectors and sub-sectors. SBI labels consist of two to five digits: the first two indicate a broad division and each consecutive digit specifies a specific subdivision, nested up to five digits (cf. fig.~\ref{fig:SBItree} for an example). Not every division is subdivided at each depth; however, each sector uses at least two digits, which agree with international standards (ISIC), and the first four digits correspond to the European NACE classification, bar a few exceptions. The fifth level is specific to the Netherlands \citep{cbs_standard_2025, united_nations_statistics_division_isic_2025, european_commission_nace_2008}.

The SBI is revised approximately once every fifteen years. The bankruptcy microdata used here follow the SBI 2008 classification, with two exceptions; some older input-output tables, discussed in the next section, and the fact that a small number of revisions have been made over time (see \cite{kruiskamp_standaard_2022} or the \href{{https://sbi2008.cbs.nl/CBS.TypeerModule.TypeerServiceWebAPI/content/angular/app/\#/code?sbicode=46734}}{CBS SBI 2008 classification browser}). The SBI codes in the bankruptcy data are not all documented at the deepest sub-sector level, which means that the level of detail varies across observations. It is for this reason that only the first two digits are used in the analysis of the documented SBI sector (which are available for all observations in the data).
\begin{figure}
\centering
\begin{forest}
 for tree={
 font=\rmfamily,
 grow'=0,
 child anchor=west,
 parent anchor=south,
 anchor=west,
 calign=first,
 inner xsep=6pt,
 forked edges,
 edge path={
 \noexpand\path [draw, \forestoption{edge}]
 (!u.south west) +(12pt,0) |- (.child anchor) \forestoption{edge label};
 },
 before typesetting nodes={
 if n=1
 {insert before={[,phantom]}}
 {}
 },
 fit=band,
 before computing xy={l=15pt},
 } 
[52 $-$ Storage and services for transport
 [522 $-$ Transportation services
 [5224 $-$ Freight handling
 [52241 $-$ (un-)loading and transshipment ...
 ]
 ]
 ]
]
\end{forest}
\caption{Example of a full-depth sector hierarchy according to the SBI 2008.}
\label{fig:SBItree}
\end{figure}

\section{Input-output tables}\label{sec:IOtables}\noindent
SN publishes input-output (I/O) tables as part of the yearly National Accounts. These tables follow the framework introduced by Wassily Leontief in the late 1930s and summarise monetary flows between industries, distinguishing between intermediate transactions and quantities such as final demand and value added. Because I/O-tables categorise the connections between sectors, they form a natural basis for analysing industry-to-industry exposure. \cite{carvalho_micro_2014} shows that shocks can propagate through inter-sectoral linkages in an I/O matrix, while \cite{laeven_pandemics_2022} argue that exposure variables can be constructed from I/O-tables to analyse in- and out-exposure of linked sectors. Since here a key question is whether (bursts of) bankruptcies are associated across sectors, I/O-tables are used to construct exposure covariates. 

A practical issue in the use of the I/O-tables is the alignment of the sector definitions across different years. The sectors in the I/O-tables are reported using a classification consistent with the \textit{NACE}: the standard for European statistics established by the European Union, on which the SBI is based. However, the published I/O industries are not always available at the same level as the SBI sectors in the bankruptcy data. In particular, several industries appear only as groups ($13-15$, $37-39$ and $87-88$), whereas the SBI codes in the bankruptcy data can be more refined, despite being truncated at two digits (data are available for sector 38, for example, but this is combined into $37-39$ in the I/O-tables). 

The I/O-tables are available as three publications, covering $1995-2015$, $2015-2021$ and $2021-2024$. The earliest publication uses slightly different labels than the others, presumably an artefact of the revisions to the SBI in 2008. As of this writing, no table is published for $2025$.

To avoid making assumptions about the distribution of monetary flow within a group of sectors such as $13-15$, the sectors in the bankruptcy data are grouped on the same level of aggregation as the I/O-tables to preserve interpretability at the cost of losing some detail for those affected sectors. The grouped sectors are given in the appendix, along with their NACE names. For $13-15$, $37-39$ and $87-88$ the combined categories are already part of the later I/O-tables, so these are matched without losing detail. The main problems concern $59-60$ and $90-92$: these sectors appear in the later I/O-tables, but not in the $1995-2015$ publication. Therefore the sector groups are used and simply referred to as sectors.

The set of sectors used in the analysis is the intersection of the NACE sectors in the I/O-block and the (aggregated) sectors in the bankruptcy data. One final issue is that sector 09 is missing from the first year. To solve this, it is padded with zeros to keep the dimension of the matrices consistent across years.

\section{Construction of \texorpdfstring{$R$}{R}, \texorpdfstring{$\Delta$}{Delta} and the burst onset probability}\label{app:RDelCon}\noindent
Bankruptcies in the Netherlands are declared by courts on hearing-days, which are often (but not always) scheduled weekly and on Tuesdays. A naive approach is to construct equally spaced weekly intervals by choosing $\Delta = 7$ and indexing time on successive Tuesdays. However, since the hearing-days are not always equally spaced apart, it is inaccurate to treat $\Delta$ and $R$ as constants. 

A better approach is to construct intervals for each sector $s$, where $R$ is based on the time since the last burst in sector $s$, and $\Delta$ is based on the time since the last hearing-day. Formally, define $\Delta_t$ as the calendar-time length of the interval $(t-1,\,t]$ where $t-1$ and $t$ refer to successive hearing-days.

As $R_{s,t-1}$ should represent the elapsed time since the end of the last burst in sector $s$ measured just before the beginning of the interval $(t-1, t]$, it has to be constructed recursively. Analogous to the burst activity indicator, Eq.~(\ref{eq:buacind}), and burst onset indicator, Eq.~(\ref{eq:boaind}), let $F_{s,t}$ indicate the end of a burst:
\begin{equation}
F_{s,t} \equiv \mathds{1}\{A_{s,t-1}=1,\, A_{s,t}=0\}.
\end{equation}
Then define $R_{s,t-1}$ as
\begin{equation}
R_{s,t-1} = 
\begin{cases}
0 & \text{if } F_{s,t-1} = 1,\\
R_{s,t-2}+\Delta_{t-1} & \text{if } A_{s,t-1}=0 \text{ and } F_{s,t-1}=0,\\
\text{undefined} & \text{if } A_{s,t-1}=1 \text{ or if no previous}\nonumber\\
&\ \ \text{burst has ended yet}.
\end{cases}
\end{equation}
Thus, $R_{s,t-1}$ is the elapsed time since the end of the previous burst in sector $s$, measured at the start of the interval $(t-1,t]$. When a burst ends, it is set to zero.

To relate $\Delta_t$ and $R_{s,t-1}$ to the discrete-time hazard setup and construct the burst onset probabilities $p_{s,t}$, first define the sets
\begin{eqnarray}
\mathcal{I}_s &\equiv& \{t\in \{2,\dots, T\} \mid A_{s,t-1}=0\} \nonumber\\
\mathcal{I} &\equiv& \{(s,t)\mid t\in \mathcal{I}_s,\, R_{s,t-1} \text{ defined} \}.
\end{eqnarray}
Thus, $\mathcal{I}$ is defined as the set of sector-time pairs for which the previous hearing-day was inactive and the time since a previous burst is well-defined. Because the burst onset probability
\[
p_{s,t}=\mathbb{P}(B_{s,t}=1\mid A_{s,t-1}=0, w_{s,t-1})
\]
is conditioned on the previous hearing-day in sector $s$ being inactive, $p_{s,t}$ is only modelled on $\mathcal{I}$: if $A_{s,t-1}=1$, then $B_{s,t}=0$ which means $p_{s,t}$ is not defined. 

\section{Parameter estimation: burst onset}\noindent
The parameters $\gamma$, $k$ and $\lambda$ are estimated jointly by maximising the Bernoulli log-likelihood over the set of observations $\mathbb{I}$ on which $p_{s,t}$ is defined. First, the analytical log-likelihood is derived and the associated derivatives with respect to each parameter. The L-BFGS algorithm is used to maximise the log-likelihood. There are some complications regarding the identifiability and numerical stability of the parameters. Within the presented GLM, the burst onset indicators $B_{s,t}=\mathds{1}\{A_{s,t}=1, A_{s,t-1}=0\}$ are assumed to be ${\rm Bernoulli}(p_{s,t})$-distributed, where the success probability is given by the burst onset probability 
\begin{equation}
p_{s,t}=\mathbb{P}(B_{s,t}=1\mid A_{s,t-1}=0, w_{s,t-1}),
\end{equation}
i.e. the probability of a burst occurring in $(R_{s,t-1},\, R_{s,t-1}+\Delta_t]$, Therefore, the PMF of $B_{s,t}$ is
\begin{equation}
\mathbb{P}(B_{s,t}=m) = p_{s,t}^m\,(1-p_{s,t})^{1-m},\qquad m\in\{0,1\}.
\end{equation}
Assuming conditional independence across $(s,t)\in \mathbb{I}$, the corresponding likelihood can be written as
\begin{equation}
L(\gamma,k,\lambda) = \prod_{(s,t)\in \mathbb{I}} p_{s,t}^{B_{s,t}}(1-p_{s,t})^{1-B_{s,t}},
\end{equation}
hence the log-likelihood is
\begin{eqnarray}
\ell(\gamma, k, \lambda) &=& \sum_{(s,t)\in \mathbb{I}} \left[B_{s,t}\log(p_{s,t})+ \right.\\
&&\qquad\qquad\left. (1-B_{s,t})\log(1-p_{s,t})\right],\nonumber\label{eq:log-likelihoodActAnal}
\end{eqnarray}
where $p_{s,t}=1-\exp(-\exp(\eta_{s,t}))$, and
\begin{equation}
\eta_{s,t} = -k\log\lambda + \log[(R_{s,t-1}+\Delta_t)^k- R_{s,t-1}^k] + w_{s,t-1}^\top \gamma.
\end{equation}
Due to the complexity of the terms in the likelihood, directly differentiating with respect to the parameters yields non-linear score equations for which an analytical form is intractable. The maximum of $\ell$ is determined numerically, using the L-BFGS algorithm, which iteratively updates estimates of the parameters using the gradient of the log-likelihood (i.e. the score) to obtain an approximation of the maximum likelihood estimate $(\hat\gamma,\, \hat{k},\, \hat\lambda)$. 
%The L-BFGS algorithm is treated in detail in \ref{app:LBFGS}. 
L-BFGS is especially powerful if an analytical expression of the score is available, which is often more attainable than a direct solution of the score equations. As the scores have closed-form expressions for this GLM, the derivation of the analytical gradient is presented here.

%\pagebreak

\subsection{Deriving the gradient}\noindent
Recall that $k,\lambda\in\mathbb{R}_{>0}$ are the Weibull model parameters and that $\gamma\in\mathbb{R}^p$ is the covariate coefficient vector. Thus, the gradient $\nabla \ell$ of the log-likelihood is a $(p+2)$-dimensional vector. $\nabla\ell$ can be written as
\begin{equation}
\nabla \ell(\gamma, k, \lambda) = 
\begin{bmatrix}
\nabla_\gamma \ell & \dfrac{\partial\ell}{\partial k} & \dfrac{\partial\ell}{\partial \lambda}
\end{bmatrix}^\top,
\end{equation}
where $\nabla_\gamma \ell \equiv [
{{\partial\ell}/{\partial \gamma_1} \,\dots\, {\partial\ell}/{\partial \gamma_p}} ]^\top$. Now derive explicit expressions for the three score functions. First, write (\ref{eq:log-likelihoodActAnal}) as
\begin{eqnarray}
\ell(\gamma, k, \lambda) &=& \sum_{(s,t)\in \mathbb{I}} \ell_i, \\
\ell_i &\equiv& B_{s,t}\log p_{s,t}+ (1-B_{s,t})\log(1-p_{s,t}).\nonumber
\end{eqnarray}
To derive the score of $\ell$ with respect to a parameter $\theta$ where $\theta\in\{\gamma, k,\lambda\}$, it is convenient to first take the derivative with respect to $\eta_{s,t}$ and then apply the chain rule:
\begin{equation}
\frac{\partial}{\partial \theta} \ell = \sum_{(s,t)\in \mathbb{I}}\frac{\partial }{\partial \theta}\ell_i 
= \sum_{(s,t)\in \mathbb{I}} \frac{{\rm d} \ell_i}{{\rm d} \eta_{s,t}}\frac{\partial\eta_{s,t}}{\partial\theta}.\label{eq:logliktheta}
\end{equation}
To compute ${{\rm d} \ell_i}/{{\rm d} \eta_{s,t}}$, apply the chain rule again by differentiating with respect to $p_{s,t}$:
\begin{align}
\frac{{\rm d} \ell_i}{{\rm d} \eta_{s,t}} = \frac{{\rm d} \ell_i}{{\rm d} p_{s,t}}\frac{{\rm d} p_{s,t}}{{\rm d} \eta_{s,t}}.\label{eq:scoreeta}
\end{align}
For the first term:
\begin{equation}
\frac{{\rm d} \ell_i}{{\rm d} p_{s,t}} = \frac{B_{s,t}}{p_{s,t}} - \frac{1-B_{s,t}}{1-p_{s,t}}.
\end{equation}
For the second term, let $u_{s,t}=\exp(\eta_{s,t})$, then $p_{s,t} = 1-\exp(-u_{s,t})$. The second term then becomes
\begin{eqnarray}
\frac{{\rm d} p_{s,t}}{{\rm d} \eta_{s,t}} &=& \frac{{\rm d} p_{s,t}}{{\rm d} u_{s,t} }\frac{{\rm d} u_{s,t}}{{\rm d} \eta_{s,t}} \nonumber\\
&=& \exp(-u_{s,t})u_{s,t} \nonumber\\
&=& [1-(1-\exp(-u_{s,t}))]\exp(\eta_{s,t}) \nonumber\\
&=& (1-p_{s,t})\exp(\eta_{s,t}).
\end{eqnarray}
Substituting both terms into (\ref{eq:scoreeta}) yields
\begin{eqnarray}
 \frac{{\rm d}\ell_i}{{\rm d} \eta_{s,t}} &=& \frac{{\rm d} \ell_i}{{\rm d} p_{s,t}}\frac{{\rm d} p_{s,t}}{{\rm d} \eta_{s,t}} \nonumber\\
 &=& \left(\frac{B_{s,t}}{p_{s,t}} - \frac{1-B_{s,t}}{1-p_{s,t}}\right)\!(1-p_{s,t})\exp(\eta_{s,t}) \nonumber\\
 &=& \frac{B_{s,t}-p_{s,t}}{p_{s,t}}\exp(\eta_{s,t}).
\end{eqnarray}
Combining this result with (\ref{eq:logliktheta}) yields the full expression for $\partial \ell/\partial \theta$:
\begin{align}
\frac{\partial \ell}{\partial \theta} = \sum_{(s,t)\in \mathbb{I}} \frac{B_{s,t}-p_{s,t}}{p_{s,t}}\exp(\eta_{s,t}) \frac{\partial \eta_{s,t}}{\partial \theta}.\label{eq:sumtheta}
\end{align}
Using this expression, the partial derivatives $\partial \eta_{s,t}/\partial \theta$ for $\theta\in \{\gamma, k, \lambda\}$ can be calculated and used in the construction of the gradient. The derivative with respect to $\gamma$ is a straightforward calculation, while $k$ and $\lambda$ require more work.

\subsubsection{Reparametrisation}
In the Weibull setting, the parameters $k$ and $\lambda$ are assumed to be positive. Because the L-BFGS algorithm is unconstrained, it can propose negative values for $k$ and $\lambda$. Therefore, it is preferable to reparametrise these as 
\begin{equation}
\mu= \log k \,\,\,\text{and}\,\,\, \nu = \log \lambda,
\end{equation}
and so
\begin{equation}
k = e^\mu\,\,\, \text{and}\,\,\, \lambda=e^\nu.
\end{equation}
This ensures that $k,\lambda\in (0,\infty)$.
The linear predictor can be expressed in the new parameters as
\begin{eqnarray}
\eta_{s,t} &=& -k\log\lambda + \log\big[(R_{s,t-1}+\Delta_t)^k- R_{s,t-1}^k\big] \nonumber\\
&&\qquad\qquad + w_{s,t-1}^\top \gamma\nonumber\\
&=& -\exp(\mu)\nu + \log\left[(R_{s,t-1}+\Delta_t)^{\exp(\mu)}
\right.\nonumber\\
&&\qquad\left.- (R_{s,t-1})^{\exp(\mu)}\right] + w_{s,t-1}^\top \gamma, 
\end{eqnarray}
but for readability it is easier to operate on
\begin{equation}
\eta_{s,t}= -k\nu + \log\big[(R_{s,t-1}+\Delta_t)^k- R_{s,t-1}^k\big] + w_{s,t-1}^\top \gamma,
\end{equation}
and to remember that $k$ and $\lambda$ are functions of $\mu$ and $\nu$ when taking derivatives. The gradient of $\ell$ can be computed from (\ref{eq:sumtheta}) by first calculating $\partial \eta_{s,t}/\partial \theta$ with respect to the parameters $\theta\in \{\gamma, \mu,\nu\}$. For $\gamma$ and $\nu$:
\begin{equation}
\frac{\partial \eta_{s,t}}{\partial \gamma} = w_{s,t-1},\qquad\text{and}\qquad \frac{\partial \eta_{s,t}}{\partial \nu} = -k.
\end{equation}
To obtain $\partial \eta_{s,t}/\partial \mu$ apply the chain rule, which gives
\begin{eqnarray}
\frac{\partial \eta_{s,t}}{\partial \mu} &=& \frac{\partial \eta_{s,t}}{\partial k}\frac{\partial k}{\partial \mu}\\
&=& \left[-\nu + \frac{\frac{\partial}{\partial k}\!\left[(R_{s,t-1}+\Delta_t)^k-R_{s,t-1}^k\right]}{(R_{s,t-1}+\Delta_t)^k-R_{s,t-1}^k}\right] k\nonumber\\
= -k\nu&+&k\left[(R_{s,t-1}+\Delta_t)^k\log(R_{s,t-1}+\Delta_t)\right. \nonumber\\
&&\left.-R_{s,t-1}^k\log(R_{s,t-1})\right]/ \nonumber\\
&&\qquad\left[(R_{s,t-1}+\Delta_t)^k-R_{s,t-1}^k\right].
\nonumber
\end{eqnarray}

Substituting the three partial derivatives into (\ref{eq:sumtheta}) yields the scores: 
\begin{eqnarray}
\nabla_\gamma\ell &=& \sum_{(s,t)\in \mathbb{I}} v_{s,t} w_{s,t-1},\nonumber\\
\frac{\partial \ell(\gamma, \mu, \nu)}{\partial \nu} &=&\sum_{(s,t)\in \mathbb{I}} v_{s,t}(-k),
\end{eqnarray}
and
\begin{eqnarray}
\frac{\partial \ell(\gamma, \mu, \nu)}{\partial \mu}&=& \sum_{(s,t)\in \mathbb{I}} v_{s,t}\left[-k\nu + k\left[(R_{s,t-1}+\Delta_t)^k \right.\right.\nonumber\\
&&\log(R_{s,t-1}+\Delta_t)\left.-R_{s,t-1}^k\log(R_{s,t-1})\right]/ \nonumber\\
&&\left.\left[(R_{s,t-1}+\Delta_t)^k-R_{s,t-1}^k\right]\right].\label{eq:scoremu}
\end{eqnarray}
where $\nu = \log\lambda$, $\mu = \log k$, and
\begin{equation}
v_{s,t} = \frac{B_{s,t}-p_{s,t}}{p_{s,t}}\exp(\eta_{s,t}).
\end{equation}

Concerns about numerical problems may arise when implementing these expressions. The main numerical problems, which manifest inside the linear predictor $\eta_{s,t}$ and the score with respect to $\mu$, is resolved using the approach presented in appendix \ref{subsec:NumStab}. With the gradient expressed in the desired parametrisation, it can be supplied to the L-BFGS algorithm with an additional minus sign to convert it from a maximisation to a minimisation problem:
\begin{equation}
-\nabla \ell(\gamma, \mu, \nu) = 
\begin{bmatrix}
-\nabla_\gamma\ell &
\displaystyle-\frac{\partial \ell}{\partial \mu} &
\displaystyle-\frac{\partial \ell}{\partial \nu}
\end{bmatrix}^\top.
\end{equation}
To obtain estimates of $k$ and $\lambda$, use the invariance property of the MLE under one-to-one transformations: if $g$ is a one-to-one function and $\hat{\theta}$ is the MLE of $\theta$, then $g(\hat{\theta})$ is the MLE of $g(\theta)$. Since $\log(\cdot)$ is one-to-one, transforming back to the original parameters is done by inverting:
\begin{equation}
(\hat{\gamma},\, \hat{k},\, \hat{\lambda}) = (\hat{\gamma},\, \exp(\hat{\mu}),\,\exp(\hat{\nu})).
\end{equation}

\subsection{Parameter identifiability}\label{subsec:ident}\noindent
In statistical inference, it is crucial that parameter estimates correspond to the underlying model in a well-defined way. If two different parameter values lead to the same distribution, then the parameters cannot be unambiguously interpreted. Hence \textit{identifiability} of the parameters in the estimation is required.
%\begin{definition}
A parameter $\theta$ is called \textit{identifiable} if and only if the map $\theta\mapsto P_\theta$ is injective. Equivalently, 
\begin{equation}
P_{\theta_1}=P_{\theta_2} \implies \theta_1=\theta_2.
\label{def:identifiability}
\end{equation}%\parencite{kleijn_frequentist_2021}
%\end{definition}
If this property is violated, parameters cannot be interpreted unambiguously. As an example in our setting, this problem occurs within the linear predictor
\begin{equation}
\eta_{s,t} = -k\log\lambda + \log\big[(R_{s,t-1}+\Delta_t)^k- R_{s,t-1}^k\big] + w_{s,t-1}^\top \gamma.
\end{equation}
Specifically, $\eta_{s,t}$ depends on the data directly through $w_{s,t-1}$ and indirectly through $R_{s,t-1}$ and $\Delta_t$. In contrast, the first term $-k\log(\lambda)$ does not depend on the data. This is problematic, because the following could happen: Write $w_{s,t-1} = [1, z_{s,t}]$ and $\gamma = [\gamma_0, \beta]$, so that 
\begin{equation}
w_{s,t-1}^\top\gamma = \gamma_0 + z_{s,t}^\top\beta.
\end{equation}
Then $\eta_{s,t}$ becomes
\begin{equation}
\eta_{s,t} = \gamma_0-k\log\lambda + \log\big[(R_{s,t-1}+\Delta_t)^k- R_{s,t-1}^k\big] + z_{s,t}^\top \beta.
\end{equation}
Recognising that the constant part in this model as $\gamma_0-k\log \lambda$, this implies that $\gamma_0$ and $ \lambda$ are no longer identifiable separately: for any $c\in\mathbb{R}\setminus\{0\}$, let $\gamma_0\mapsto \gamma_0+c$ and $\lambda \mapsto \lambda\exp(c/k)$. Call this new expression $\eta_{s,t}'$. Then
\begin{eqnarray}
\eta_{s,t}' &=& (\gamma_0+c)- k\log(\smash{\lambda e^{c/k}}) +
\nonumber\\
&&\qquad\log\big[(R_{s,t-1}+\Delta_t)^k- R_{s,t-1}^k\big] + z_{s,t}^\top \beta \nonumber\\
&=& \gamma_0 + c - k\log\lambda - k{c}/{k}+ \nonumber\\
&&\log\big[(R_{s,t-1}+\Delta_t)^k- R_{s,t-1}^k\big] + z_{s,t}^\top \beta \nonumber\\
&=& \gamma_0 - k\log\lambda +\log\big[(R_{s,t-1}+\Delta_t)^k- R_{s,t-1}^k\big] \nonumber\\
&&\qquad + z_{s,t}^\top \beta\nonumber\\
&=& \eta_{s,t}.
\end{eqnarray}
But since $c\neq 0$, this implies $(\gamma_0,\lambda)\neq (\gamma_0 + c,\,\, \lambda e^{c/k})$, which violates (\ref{def:identifiability}), thus $\gamma_0$ and $\lambda$ are not identifiable separately. A simple solution is to remove the intercept column from the design matrix, so that the model no longer includes the intercept term $\gamma_0$. Then the term $-k\log\lambda$ functions as the intercept. Concretely, $w_{s,t-1}$ does not include the usual first column of 1's. This removal is necessary, but not sufficient: because categorical variables such as sectors and years are represented using indicator functions, the design-matrix still contains linear combinations. For example, in the year model where $\mathcal{S}$ is the set of years;
\begin{equation}
\sum_{y\in \mathcal{Y}}\mathds{1}\{Y(t) = y\} = 1,
\end{equation}
since for each $t$, one year applies. The most straightforward technique that ensures identifiability is to impose the \textit{sum-to-zero} constraints:
\begin{equation}
\sum_{y\in \mathcal{Y}} \gamma_y = 0 \qquad \text{and}\qquad \sum_{s\in \mathcal{S}} \gamma_{s} = 0,
\end{equation}
in the models where year ($y\in\mathcal{Y}$) and sector ($s\in\mathcal{S}$) covariates are present. It may also seem that the parameters $k$ and $\lambda$ are not separately identifiable because they appear together in the term $k\log \lambda$. However, because the term $\log\big[(R_{s,t-1}+\Delta_t)^k-R_{s,t-1}^k\big]$ in $\eta_{s,t}$ depends only on $k$ and not on $\lambda$, both $k$ and $\lambda$ can be estimated.

~\par

One may also question whether collinearity is present in the covariates. Given the modelling choices, this is unlikely. As an example: in the ``sector + year'' model $\text{BOA}_{SY}$, full collinearity only happens when all observations for a given sector occur within a single year. Some near-collinearity is to be expected, but in our setting this is not problematic.

\subsection{Numerical stability}\label{subsec:NumStab}\noindent
Additional technical issues that require careful handling arise when translating mathematics into a programming language. To improve numerical stability, terms are implemented on the log-scale and stable implementations of certain functions are used. 

\subsubsection{Improving the log-term in the linear predictor}
When implementing the term $\log[(R_{s,t-1}+\Delta_t)^k- R_{s,t-1}^k]$ in the linear predictor $\eta_{s,t}$, numerical instability may arise: for modest values of $k$ and large values of $R_{s,t-1}$ or $\Delta_t$, the terms within the logarithm can become large, and subtracting nearly equal large numbers can lead to loss of precision due to floating-point errors. Even though the errors might be small, repeated calculation of this term can lead to accumulation of errors. A more numerically stable way of writing $\log[(R_{s,t-1}+\Delta_t)^k- R_{s,t-1}^k]$ is as follows. Let
\begin{equation}
a=k\log(R_{s,t-1}+\Delta_t)\ \  \text{and}\ \  b= k\log R_{s,t-1},
\end{equation}
so that
\begin{eqnarray}
\exp(a) &=& (R_{s,t-1}+\Delta_t)^k, \nonumber\\
\exp(b) &=& R_{s,t-1}^k \nonumber\\ 
\exp(b-a) &=& \frac{R_{s,t-1}^k}{(R_{s,t-1}+\Delta_t)^k}.
\end{eqnarray}
Then can be written as
\begin{eqnarray}
&&(R_{s,t-1}+\Delta_t)^k - R_{s,t-1}^k = e^a - e^b = e^a(1-e^{b-a}) \qquad\nonumber\\
&&\qquad = (R_{s,t-1}+\Delta_t)^k\bigg(1-\frac{R_{s,t-1}^k}{(R_{s,t-1}+\Delta_t)^k}\bigg).
\end{eqnarray}

Taking the logarithm yields the more stable form
\begin{equation}
\log[(R_{s,t-1} + \Delta_t)^k-R_{s,t-1}^k] = k\log(R_{s,t-1} + \Delta_t) + \log(1-h) 
\end{equation}
where
\begin{equation}
h = \frac{R_{s,t-1}^k}{(R_{s,t-1} + \Delta_t)^k}.
\end{equation}
Since $R_{s,t-1}\geq 0$, $\Delta_t > 0$ and $k>0$, it follows that $h\in [0,1)$. 

\subsection{Improving terms in the gradient}\noindent
A second numerical instability issue arises when calculating the gradient of the log-likelihood with respect to $\mu = \log k$, given in (\ref{eq:scoremu}). specifically, direct evaluation of the term $-k\nu + kP$ is problematic, where $P$ is defined as
\begin{eqnarray}
P &\equiv& \left[(R_{s,t-1}+\Delta_t)^k\log(R_{s,t-1}+\Delta_t)
\right.\nonumber\\
&&\left.-R_{s,t-1}^k\log(R_{s,t-1})\right]/ \nonumber\\
&&\left[(R_{s,t-1}+\Delta_t)^k-R_{s,t-1}^k\right].
\end{eqnarray}
Calculating $P$ involves exponentiation, subtraction of possibly large similar numbers and division by small numbers. A more stable representation can be obtained by factoring. Again, let
\begin{eqnarray}
a &=& k\log(R_{s,t-1}+\Delta_t),\\
b &=& k\log R_{s,t-1} \nonumber\\ 
h &=& \exp(b-a)=\left(\frac{R_{s,t-1}}{{R_{s,t-1}+\Delta_t}}\right)^k\in[0,1).\nonumber
\end{eqnarray}
Dividing the numerator and denominator of $P$ by $(R_{s,t-1}+\Delta_t)^k = \exp(a)$ the improved 
form is
\begin{equation}
P = \frac{\log(R_{s,t-1}+\Delta_t) -h\log(R_{s,t-1})}{1-h}.
\end{equation}

\subsection{Improving ${h}$}\noindent
A further issue concerns the fact that $h$ is bounded between 0 and 1, when $R_{s,t-1} \gg \Delta_t$ it is possible that $h\approx1$. Then the denominator of $P$ can become $1-h\approx 0$ which can make the estimate of $P$ unstable. To reduce this risk, $h$ is written as
\begin{eqnarray}
h &=& \Big(\frac{R_{s,t-1}}{R_{s,t-1}+\Delta_t}\Big)^k \nonumber\\
&=& \Big(\frac{R_{s,t-1}+\Delta_t}{R_{s,t-1}}\Big)^{-k} \nonumber\\
&=& \exp\Big[-k\log\Big(1+\frac{\Delta_t}{R_{s,t-1}}\Big)\Big].
\end{eqnarray}
This representation removes the need to exponentiate both terms. Moreover, the stable \verb|log1p| (``log-plus-one") method in \verb|R| is applied, which is designed to accurately handle $\log(1+x)$ for $x$ close to zero. For $R_{s,t-1}=0$ set $h=0$ by definition. Hence, implement 
\begin{equation}
h=\exp\Big(-k\cdot{\rm log1p}\Big(\frac{\Delta_t}{R_{s,t-1}}\Big)\Big).
\end{equation}
The linear predictor $\eta_{s,t}$ can then be implemented as

\begin{eqnarray}
\eta_{s,t} &=& -k\log\lambda + \log[(R_{s,t-1}+\Delta_t)^k- R_{s,t-1}^k] \nonumber\\
&&\qquad + w_{s,t-1}^\top \gamma \nonumber\\
&=& -k\log\lambda + k\log(R_{s,t-1}+\Delta_t) + \log(1-h) 
\nonumber\\
&&\qquad + w_{s,t-1}^\top \gamma,\label{eq:neweta}
\end{eqnarray}
where $h=\exp(-k\cdot{\rm log1p}(\Delta_t/R_{s,t-1}))$, and 
(\ref{eq:neweta}) is implemented using the stable form for $h$ 
and \verb|log1p|. 
%One final safeguard against instability when $h\approx 1$ is given in \ref{app:AppendixH}. 

\section{Parameter estimation: burst size}\noindent
The parameters $\beta$ and $r$ are estimated by maximising the likelihood. A log link is used, which is a standard choice for NB regression, and has the added effect that the mean $m_{s,t}$ is always positive. Therefore, let
\begin{equation}
m_{s,t}(\beta) \equiv \exp(z_{s,t}^\top \beta),\qquad z_{s,t},\beta\in \mathbb{R}^p.
\end{equation}
While there is no a priori reason for the size component in the BSA to depend on the same set of covariates as the onset rate in the BOA, there is no strong reason to exclude any, which also means that the same nested model strategy is adopted as before. Using the hurdle model structure, the likelihood function over all observations can be written as
\begin{equation}
L(\beta, r) = \prod_{y_{s,t} = 0} (1-\pi_{s,t})\prod_{y_{s,t} > 0} \pi_{s,t}f_\text{NB}^+(y_{s,t}; r, m_{s,t}(\beta)),
\end{equation}
for all $(s,t)$. Thus, the full log-likelihood is
\begin{eqnarray}
\ell(\beta, r) &=& \sum_{y_{s,t} = 0} \log(1-\pi_{s,t}) + \sum_{y_{s,t} > 0} \left[\log(\pi_{s,t}) \right.\nonumber\\
&&\qquad\left.+\log f_\text{NB}^+(y_{s,t}; r, m_{s,t}(\beta))\right].
\end{eqnarray}
Because only the positive counts are fitted for, the first sum is constant by construction. Moreover, $\log(\pi_{s,t})$ in the second sum is an additive constant with respect to $r$ and $m_{s,t}$, hence the terms 
\begin{equation}
\sum_{y_{s,t}\hspace{1pt}=\hspace{1pt} 0} \log(1-\pi_{s,t}) \qquad\text{and}\qquad \sum_{y_{s,t}\hspace{1pt} >\hspace{1pt} 0} \log(\pi_{s,t})
\end{equation}
drop out in the maximisation, and the full log-likelihood simplifies to the following conditional likelihood:
\begin{equation}
\ell(\beta, r) = \sum_{y_{s,t}\hspace{1pt} >\hspace{1pt} 0} \log f_\text{NB}^+(y_{s,t};\hspace{1pt} r,\hspace{0.5pt} m_{s,t}(\beta)).\label{eq:loglikelihood1}
\end{equation}
Note that this quantity does not depend on the activation probability $\pi_{s,t}$. Taking the logarithm of the ZTNB PMF in (\ref{eq:ZTNBPMF}):
\begin{eqnarray}
\log f_\text{NB}^+(y; r, m) &=& \log\Gamma(r+y) - \log \Gamma(r) - \log(y!)
\nonumber\\
&&+ r\log r + y\log m \nonumber\\
&&-(r+y)\log({r+m}) \nonumber\\
&&-\log\left[1-\left(\frac{r}{r+m}\right)^r\, \right],
\end{eqnarray}
for $y\geq 1$. Lastly, write $m_{s,t}$ for $m_{s,t}(\beta)$ and let
%\opm{Do we need time offset? i.e. $m_{s,t}(\beta)=\exp(\log \Delta_t + z_{s,t}^\top\beta) =\Delta_t\exp(z_{s,t}^\top\beta)$? $->$ Check if counts depend on calendar time over the years. Include why or not as discussion part here.}
\begin{equation}
b_{s,t} \equiv \frac{r}{r+m_{s,t}}\qquad\text{and}\qquad
a_{s,t} \equiv b_{s,t}^{\,r}.\label{eq:definedquantities}
\end{equation}
The log-likelihood given in (\ref{eq:loglikelihood1}) can be expressed as
\begin{eqnarray}
\ell(\beta,r) &=& \sum_{y_{s,t} > 0}\Big[ \log\Gamma(r+y_{s,t}) - \log \Gamma(r) \\
&&- \log(y_{s,t}!) + r\log r + y_{s,t}\log m_{s,t} \nonumber\\
&&-(r+y_{s,t})\log({r+m_{s,t}})-\log(1-a_{s,t})\Big].\nonumber
\label{eq:loglikelihood}
\end{eqnarray}

\subsection{Score functions}\label{subsec:scoresBSA}\noindent
The MLE $(\hat{\beta}, \hat{r})$ can be approximated by minimising the analytical negative log-likelihood using the L-BFGS algorithm. As such, the gradient of $\ell$ is required.

\subsubsection{The score with respect to $\beta$}
The score with respect to $\beta$ can be written as
\begin{eqnarray}
\partb{\ell} &=& \sum_{y_{s,t}\hspace{1pt}>\hspace{1pt}0}z_{s,t}b_{s,t}[y_{s,t} - \mu_{s,t}^+], \nonumber\\ 
\mu_{s,t}^+ &=& \frac{r(1-b_{s,t})}{b_{s,t}(1-a_{s,t})},
\end{eqnarray}
and $\mu_{s,t}^+$ is the mean of the ZTNB random variable $Y_{s,t}\mid A_{s,t}=1$. Start by splitting the log-likelihood into one part that does not depend on $\beta$ and one part that does, by separating terms that include $m_{s,t}=\exp(z_{s,t}^\top \beta)$: 
\begin{eqnarray}
\ell(\beta,r) &=& \sum_{y_{s,t} > 0} \left[\log\Gamma(r+y_{s,t}) - \log \Gamma(r) - \log(y_{s,t}!)\right] \nonumber\\
&&+\sum_{y_{s,t} > 0} \left[r\log r + y_{s,t}\log m_{s,t} - \right.\nonumber\\
&&\left. (r+y_{s,t})\log({r+m_{s,t}})-\log(1-a_{s,t})\right].
\end{eqnarray}
Taking the derivative with respect to $\beta$ and applying the chain rule gives
\begin{eqnarray}
\partb{\ell} &=& \sum_{y_{s,t}\hspace{1pt}>\hspace{1pt}0}\partm{\ell}\partb{m_{s,t}} \nonumber\\
&=&\sum_{y_{s,t}\hspace{1pt}>\hspace{1pt}0}\partb{m_{s,t}}\left[\partm{}\big[y_{s,t}\log m_{s,t}\big] \right.\nonumber\\
&&\ \ -\partm{}\big[(r+y_{s,t})\log(r+m_{s,t})\big] \nonumber\\
&&\ \ \left.-\partm{}\log(1-a_{s,t})\right].
\end{eqnarray}
As $m_{s,t} = \exp(z_{s,t}^\top \beta)$, $\partial {m_{s,t}}/\partial \beta = z_{s,t}m_{s,t}$. Going term by term inside the brackets:
\begin{eqnarray}
\partm{}\big[y_{s,t}\log m_{s,t}\big] &=& \frac{y_{s,t}}{m_{s,t}}, \\
-\partm{}\big[(r+y_{s,t})\log(r+m_{s,t})\big] &=& -\frac{r+y_{s,t}} {r+m_{s,t}}, \nonumber
\end{eqnarray}
and by applying the chain rule,
\begin{equation}
\partm{}\big[-\log(1-a_{s,t})\big] = -\frac{ra_{s,t}}{\left(r+m_{s,t}\right)\left(1-a_{s,t}\right)}.
\end{equation}
The score with respect to $\beta$ can therefore be written as
\begin{eqnarray}
\partb{\ell} &=& \sum_{y_{s,t}\hspace{1pt}>\hspace{1pt}0}z_{s,t}m_{s,t} \left[\frac{y_{s,t}}{m_{s,t}}-\frac{r+y_{s,t}}{r+m_{s,t}} \right.\nonumber\\
&&\qquad\left.-\frac{ra_{s,t}}{\left(r+m_{s,t}\right)\left(1-a_{s,t}\right)} \right].
\end{eqnarray}
It is possible to further simplify this expression. First, distribute $m_{s,t}$ over the terms between brackets, which gives
\begin{eqnarray}
\partb{\ell} &=& \sum_{y_{s,t}>0}z_{s,t} \left[y_{s,t} - m_{s,t}\frac{(r+y_{s,t})}{r+m_{s,t}} \right.\nonumber\\
&&\left.\qquad- m_{s,t}\frac{ra_{s,t}}{(r+m_{s,t})(1-a_{s,t})}\right].\label{eq:betweenscorebeta}
\end{eqnarray}
Making use of (\ref{eq:definedquantities}):
\begin{equation}
m_{s,t} = r(1-b_{s,t})/b_{s,t}\quad \text{and}\quad r+m_{s,t}=r/b_{s,t}.
\end{equation}
The second term between the brackets of (\ref{eq:betweenscorebeta}) then becomes
\begin{eqnarray}
m_{s,t}\frac{(r+y_{s,t})}{r+m_{s,t}} &=& \frac{r(1-b_{s,t})} \nonumber\\ {b_{s,t}}\frac{r+y_{s,t}}{r/b_{s,t}} &=& (1-b_{s,t})(r+y_{s,t}). 
\end{eqnarray}
The third term between the brackets of (\ref{eq:betweenscorebeta}) is equal to
\begin{eqnarray}
m_{s,t}\frac{ra_{s,t}}{(r+m_{s,t})(1-a_{s,t})} &=& \frac{r(1-b_{s,t})}{b_{s,t}} \frac{ra_{s,t}}{r/b_{s,t}(1-a_{s,t})} \nonumber\\
&=& \frac{r(1-b_{s,t})a_{s,t}}{1-a_{s,t}}.
\end{eqnarray}
By substituting these results, (\ref{eq:betweenscorebeta}) can be written as
\begin{eqnarray}
\partb{\ell} &=& \sum_{y_{s,t}>0}z_{s,t} \left[y_{s,t} - (1-b_{s,t})(r+y_{s,t}) \right.\nonumber\\
&&\qquad\left. - \frac{r(1-b_{s,t})a_{s,t}}{1-a_{s,t}}\right].\label{eq:bracketterm}
\end{eqnarray}
By rearranging, the term between the brackets in (\ref{eq:bracketterm}) can be written as
\begin{eqnarray}
y_{s,t}&-&(1-b_{s,t})(r+y_{s,t}) - \frac{r(1-b_{s,t})a_{s,t}}{1-a_{s,t}}\nonumber\\
&=&
b_{s,t}y_{s,t} - r(1-b_{s,t})\Big(1+\frac{a_{s,t}}{1-a_{s,t}}\Big) \nonumber\\
&=& b_{s,t}y_{s,t} - \frac{r(1-b_{s,t})}{1-a_{s,t}},
\end{eqnarray}
so the score with respect to $\beta$ becomes
\begin{align}
\partb{\ell} = \sum_{y_{s,t}\hspace{1pt}>\hspace{1pt}0}z_{s,t}\left[b_{s,t}y_{s,t} - \frac{r(1-b_{s,t})}{1-a_{s,t}}\right].\label{eq:betabeta}
\end{align}
Finally, recall that the mean of $Y_{s,t}\mid A_{s,t}=1$ can be written as
\begin{equation}
\mu_{s,t}^+ = \cfrac{m_{s,t}}{1-\Big(\frac{r}{r+m_{s,t}}\Big)^r} = \frac{m_{s,t}}{1-a_{s,t}} = \frac{r(1-b_{s,t})}{b_{s,t}(1-a_{s,t})}.
\end{equation}
The last equation follows from the fact that $m_{s,t}=r(1-b_{s,t})/b_{s,t}$. Comparing this expression of $\mu_{s,t}^+$ to (\ref{eq:betabeta}), the second term between the brackets can be written as
\begin{equation}
\frac{r(1-b_{s,t})}{1-a_{s,t}} = \frac{b_{s,t}r(1-b_{s,t})}{b_{s,t}(1-a_{s,t})} = b_{s,t}\mu_{s,t}^+.
\end{equation}
Therefore the score with respect to $\beta$ simplifies to
\begin{eqnarray}
\partb{\ell} &=& \sum_{y_{s,t}\hspace{1pt}>\hspace{1pt}0}z_{s,t}b_{s,t}[y_{s,t} - \mu_{s,t}^+] \nonumber\\ 
\mu_{s,t}^+ &=& \frac{r(1-b_{s,t})}{b_{s,t}(1-a_{s,t})},\label{eq:gradb}
\end{eqnarray}
where $b_{s,t} = r/(r+m_{s,t})$ and $a_{s,t}=b_{s,t}^{r}$.

\subsubsection{The score with respect to $r$}\noindent
The size parameter $r$ of the NB distribution is assumed to be strictly positive. For this reason, reparametrise $r\equiv e^\rho$ to ensure positivity. 
The score with respect to $\rho$ is
\begin{eqnarray}
\partrh{\ell} &=& \sum_{y_{s,t}\hspace{1pt}>\hspace{1pt}0} \left[r[\psi(r+y_{s,t}) - \psi(r)] - b_{s,t}(y_{s,t} - \mu_{s,t}^+) 
\right.\nonumber\\
&&\qquad \left. + \frac{r\log(b_{s,t})}{1-a_{s,t}}\right].
\end{eqnarray}
$\partial \ell/\partial r$ is then calculated by the chain rule
\begin{align}
\partrh{\ell} = \partr{\ell}\partrh{r} = r\partr{\ell}.\label{eq:reparamr}
\end{align}
Recall that the log-likelihood is given by (\ref{eq:loglikelihood}) 
%\begin{align*}
%\ell(\beta,r) &= \sum_{y_{s,t}\hspace{1pt} >\hspace{1pt} 0}\Big[ %\log\Gamma(r+y_{s,t}) - \log \Gamma(r) - \log(y_{s,t}!)\notag\\
%&+ r\log r + y_{s,t}\log m_{s,t} -(r+y_{s,t})\log({r+m_{s,t}})-\log(1-%a_{s,t})\Big].
%\end{align*}
When taking the derivative with respect to $r$, the terms $-\log(y_{s,t}!)$ and $y_{s,t}\log m_{s,t}$ drop out, so $\partial \ell / \partial r$ becomes
\begin{eqnarray}
\partr{\ell} &=& \sum_{y_{s,t}>0}\bigg[\partr{} [\log\Gamma(r+y_{s,t}) - \log\Gamma(r)] + \partr{}[r\log r] \nonumber\\
&&\qquad - \partr{} [(r+y_{s,t})\log(r+m_{s,t})] \nonumber\\
&&\qquad - \partr{}\log(1-a_{s,t})\bigg].
\end{eqnarray}
The first term can be expressed in terms of the \textit{digamma function} $\psi$. As a consequence, the first term becomes
\begin{equation}
\partr{} \big[\log\Gamma(r+y_{s,t}) - \log\Gamma(r)\big] 
%= \frac{\Gamma'(r+y_{s,t})}{\Gamma(r+y_{s,t})} - \frac{\Gamma'(r)}{\Gamma(r)} = 
=\psi(r+y_{s,t}) - \psi(r).
\end{equation}
The other terms reduce to:
\begin{eqnarray}
\partr{} \big[r\log r\big] &=& 1+ \log r, \\ 
\partr{} \big[-(r+y_{s,t})\log(r+m_{s,t})\big] &=& -\log(r+m_{s,t})-\frac{r+y_{s,t}}{r+m_{s,t}},\nonumber
\end{eqnarray}
and
\begin{eqnarray}
\partr{}\big[-\log(1-a_{s,t})\big] &=& \frac{1}{1-a_{s,t}}\partr{a_{s,t}} 
\\
&=& \frac{a_{s,t}}{1-a_{s,t}}\left[\log(\frac{r}{r+m_{s,t}})+ \frac{m_{s,t}}{r+m_{s,t}}\right].\nonumber
\end{eqnarray}
Collecting terms produces:
\begin{eqnarray}
\partr{\ell} &=& \sum_{y_{s,t}>0} \bigg\{\psi(r+y_{s,t}) - \psi(r) + 1 + \log(r) \nonumber\\
&&- \log(r+m_{s,t})-\frac{r+y_{s,t}}{r+m_{s,t}} + \frac{a_{s,t}}{1-a_{s,t}} \nonumber\\
&&\qquad\times\left[\log(\frac{r}{r+m_{s,t}})+ \frac{m_{s,t}}{r+m_{s,t}}\right]\bigg\}.
\end{eqnarray}
Recall that $b_{s,t} = r/(r+m_{s,t})$. This implies that $1-b_{s,t} = m_{s,t}/(r+m_{s,t})$ and $\log(b_{s,t}) = \log(r)-\log(r+m_{s,t})$. Using this, the score with respect to $r$ can be written as
\begin{eqnarray}
\partr{\ell} &=& \sum_{y_{s,t}>0}\Big[ \psi(r+y_{s,t}) - \psi(r) +1 + \log(b_{s,t}) \nonumber\\
&&-\frac{r+y_{s,t}}{r+m_{s,t}} + \frac{a_{s,t}}{1-a_{s,t}}\big(\log(b_{s,t}) + 1-b_{s,t} \big)\Big] \nonumber\\
&=& \sum_{y_{s,t}>0}\Big[ \psi(r+y_{s,t}) - \psi(r) -\frac{y_{s,t}}{r+m_{s,t}} \\
&&+ \Big[1+\frac{a_{s,t}}{1-a_{s,t}}\Big]\big(\log(b_{s,t}) + 1-b_{s,t} \big)\Big].\nonumber
\end{eqnarray}
Using the chain rule expression in (\ref{eq:reparamr}), $\partial\ell/\partial\rho$ becomes
\begin{eqnarray}
\partrh{\ell} &=& \sum_{y_{s,t}>0}r\Big[ \psi(r+y_{s,t}) - \psi(r) -\frac{y_{s,t}}{r+m_{s,t}} \\
&&+\Big[1+\frac{a_{s,t}}{1-a_{s,t}}\Big]\big(\log(b_{s,t}) + 1-b_{s,t} \big)\Big]. \nonumber
\end{eqnarray}
This expression can be simplified: first, distribute $r$ across the terms and recall that $r+m_{s,t}=r/b_{s,t}$. The middle term in the summand can then be rewritten as
\begin{equation}
-r\frac{y_{s,t}}{r+m_{s,t}} = -r\frac{y_{s,t}b_{s,t}}{r} = -y_{s,t}b_{s,t},
\end{equation}
and the last term as
\begin{eqnarray}
r\hspace{5pt}\Big[1+\frac{a_{s,t}}{1-a_{s,t}}\Big]&(\log(b_{s,t})+1-b_{s,t}) \nonumber\\
&= \frac{r\log(b_{s,t})}{1-a_{s,t}} + \frac{r(1-b_{s,t})}{1-a_{s,t}} \nonumber\\
&= \frac{r\log(b_{s,t})}{1-a_{s,t}} + b_{s,t}\mu_{s,t}^+,
\end{eqnarray}
where $\mu_{s,t}^+$ is the mean of the ZTNB random variable.
Combining terms yields the score with respect to $\rho$:
\begin{eqnarray}
\partrh{\ell} &=&\sum_{y_{s,t}\hspace{1pt}>\hspace{1pt}0} \left[r[\psi(r+y_{s,t}) - \psi(r)] - y_{s,t}b_{s,t} + b_{s,t}\mu_{s,t}^+ 
\right.\nonumber\\
&&\left.+ \frac{r\log(b_{s,t})}{1-a_{s,t}}\right]\nonumber\\
&=& \sum_{y_{s,t}\hspace{1pt}>\hspace{1pt}0} \left[r[\psi(r+y_{s,t}) - \psi(r)] - b_{s,t}(y_{s,t} - \mu_{s,t}^+) \right.\nonumber\\
&&\left.+ \frac{r\log(b_{s,t})}{1-a_{s,t}}\right].\label{eq:gradrho}
\end{eqnarray}
The term $r\log(b_{s,t})/(1-a_{s,t})$ is a result of the zero-truncation. 

\subsection{Numerical stability}\noindent
To mitigate numerical over/underflow issues in the log-likelihood, for example when the L-BFGS line search chooses extreme values, some terms in the log-likelihood can be written in numerically stable forms. Recall that
\begin{eqnarray}
\ell(\beta,r) &= \sum_{y_{s,t} > 0}\Big[ \log\Gamma(r+y_{s,t}) - \log \Gamma(r) - \log(y_{s,t}!)\nonumber\\
&+ r\log r + y_{s,t}\log m_{s,t} -(r+y_{s,t})\log({r+m_{s,t}})- \nonumber\\
&\log(1-a_{s,t})\Big],
\end{eqnarray}
where
\begin{eqnarray}
m_{s,t} &=& \exp(z_{s,\hspace{1pt}t}^\top\beta), \qquad 
b_{s,t} = \frac{r}{r+m_{s,t}}\qquad\text{and} \nonumber\\
\qquad
a_{s,t} &=& b_{s,t}^{\,r}.
\end{eqnarray}
Note that $1-b_{s,t} = m_{s,t}/(r+m_{s,t})$, and that
\begin{eqnarray}
\log b_{s,t} &=& \log r - \log(r+m_{s,t}) \qquad\text{and}\nonumber\\ 
\log(1- b_{s,t}) &=& \log m_{s,t} - \log(r+m_{s,t}).
\end{eqnarray}
It follows that
\begin{eqnarray}
r \log b_{s,t} + &y\log(1-b_{s,t}) = r\log r + y\log m_{s,t} \nonumber\\
&- (r+y_{s,t})\log(r+m_{s,t}).
\end{eqnarray}
The factorial term is implemented using the gamma function: write $y_{s,t}!=\Gamma(y_{s,t}+1)$, so that $\log(y_{s,t}!) = \log\Gamma(y_{s,t}+1)$. Because $\log(1-x)$ loses numerical precision when $x$ is close to one, apply the \verb|R| function \verb|log1p|, defined as ${\rm log1p}(x)=\log(1+x)$, and also use the stable functions \verb|lgamma| and \verb|digamma| instead of $\log\Gamma(\cdot)$ and $\psi(\cdot)$. The numerically stable version of (\ref{eq:loglikelihood}) is given by
\begin{eqnarray}
\ell(\beta,r)&= \sum_{y_{s,t} > 0} \Big[\log\Gamma(r+y_{s,t}) -\log\Gamma(r)
\nonumber\\
&-\log\Gamma(y_{s,t}+1)+ r\log(b_{s,t}) \nonumber\\
&+ y_{s,t}{\rm log1p}(-b_{s,t})-{\rm log1p}(-a_{s,t}) \Big].
\end{eqnarray}

%\clearpage

\section{Burst onset tables}\label{app:appendixOnsetTable}\noindent
Tables \ref{tab:BOAseceffects}, \ref{tab:BOAsecyeareffects} and \ref{tab:BOAsecyearexp4effects} report the estimated covariate coefficients in the sector model ($\text{BOA}_S$), sector + year model ($\text{BOA}_{SY}$), and sector + year + full exposure model ($\text{BOA}_{SY}\text{E}_4$). The largest and smallest values are marked in bold within each category.

% -----------------------------------------------------

\begin{table}[h]
\centering
\begin{tabular}{lr|lr|lr}
\textbf{Sector} & \textbf{Estimate} &
  \textbf{Sector} & \textbf{Estimate} &
  \textbf{Sector} & \textbf{Estimate} \\
\hline
01 & $0.6654$ & 32 & $-0.01726$ & 66 & $1.267$ \\
02 & $-3.393$ & 33 & $0.5190$ & 68 & $1.736$ \\
03 & $-2.834$ & 35 & $-0.7885$ & 69 & $1.005$ \\
08 & $-4.266$ & 36 & $-4.232$ & 70 & $2.237$ \\
09 & $-3.517$ & $37$--$39$ & $0.03707$ & 71 & $1.708$ \\
10 & $0.7848$ & 41 & $2.203$ & 72 & $0.3105$ \\
11 & $-1.131$ & 42 & $0.6928$ & 73 & $1.354$ \\
12 & $-3.862$ & 43 & $2.203$ & 74 & $1.174$ \\
$13$--$15$ & $0.2779$ & 45 & $1.565$ & 75 & $-3.655$ \\
16 & $0.3605$ & 46 & $2.593$ & 77 & $0.9767$ \\
17 & $-1.432$ & 47 & $2.341$ & 78 & $1.890$ \\
18 & $0.9334$ & 49 & $1.918$ & 79 & $0.1099$ \\
{\bf 19} & {\bf $-$5.552} & 50 & $-0.5305$ & 80 & $0.08214$ \\
20 & $-0.9837$ & 51 & $-1.916$ & 81 & $1.254$ \\
21 & $-2.525$ & 52 & $1.088$ & 82 & $0.6761$ \\
22 & $-0.04367$ & 53 & $0.3717$ & 84 & $-4.844$ \\
23 & $-0.2395$ & 55 & $0.5173$ & 85 & $1.014$ \\
24 & $-1.085$ & 56 & $2.115$ & 86 & $1.131$ \\
25 & $1.368$ & 58 & $0.3859$ & $87$--$88$ & $1.533$ \\
26 & $-0.2008$ & $59$--$60$ & $0.1369$ & $90$--$92$ & $0.7033$ \\
27 & $-0.3876$ & 61 & $-0.553$ & 93 & $1.142$ \\
28 & $0.7964$ & 62 & $1.757$ & 94 & $0.4367$ \\
29 & $-0.5971$ & 63 & $0.3948$ & 95 & $-0.5469$ \\
30 & $-0.1536$ & {\bf 64} & {\bf $2.662$} & 96 & $0.8382$ \\
31 & $0.7967$ & 65 & $-2.777$ &  &  \\
\end{tabular}
\caption{Estimated coefficients in the BOA sector model ($\text{BOA}_S$).}
\label{tab:BOAseceffects}
\end{table}

% -----------------------------------------------------

\pagebreak
\begin{table*}[h]
\centering
\begin{tabular}{lr|lr|lr|lr}
\textbf{Sector} & \textbf{Estimate} &
  \textbf{Sector} & \textbf{Estimate} &
  \textbf{Sector} & \textbf{Estimate} &
  \textbf{Year} & \textbf{Estimate} \\
\hline
01 & $0.7339$ & 32 & $-0.03116$ & 66 & $1.396$ & 2009 & $0.3913$\\
02 & $-3.674$ & 33 & $0.5496$ & 68 & $1.901$ & 2010 & $0.2950$\\
03 & $-3.063$ & 35 & $-0.8750$ & 69 & $1.088$ & 2011 & $0.2642$\\
08 & $-4.491$ & 36 & $-4.564$ & 70 & $2.425$ & 2012 & $0.3694$\\
09 & $-3.753$ & $37$--$39$ & $0.01337$ & 71 & $1.848$ & {\bf 2013} & {\bf $0.4513$}\\
10 & $0.8343$ & 41 & $2.390$ & 72 & $0.3021$ & 2014 & $0.3362$\\
11 & $-1.287$ & 42 & $0.7481$ & 73 & $1.489$ & 2015 & $0.1629$\\
12 & $-4.040$ & 43 & $2.384$ & 74 & $1.261$ & 2016 & $0.09135$\\
$13$--$15$ & $0.2801$ & 45 & $1.698$ & 75 & $-3.965$ & 2017 & $-0.1244$\\
16 & $0.3968$ & 46 & $2.818$ & 77 & $1.043$ & 2018 & $-0.1631$\\
17 & $-1.560$ & 47 & $2.519$ & 78 & $2.048$ & 2019 & $-0.1597$\\
18 & $1.025$ & 49 & $2.081$ & 79 & $0.08826$ & 2020 & $-0.2969$\\
{\bf 19} & {\bf $-6.001$} & 50 & $-0.5943$ & 80 & $0.08532$ & {\bf 2021} & {\bf $-0.7376$}\\
20 & $-1.084$ & 51 & $-2.095$ & 81 & $1.350$ & 2022 & $-0.6194$\\
21 & $-2.752$ & 52 & $1.161$ & 82 & $0.7128$ & 2023 & $-0.2259$\\
22 & $-0.05693$ & 53 & $0.3723$ & 84 & $-4.964$ & 2024 & $-0.03472$\\
23 & $-0.2747$ & 55 & $0.5467$ & 85 & $1.083$ &  & \\
24 & $-1.177$ & 56 & $2.254$ & 86 & $1.199$ &  & \\
25 & $1.500$ & 58 & $0.4134$ & $87$--$88$ & $1.629$ &  & \\
26 & $-0.2304$ & $59$--$60$ & $0.1444$ & $90$--$92$ & $0.7592$ &  & \\
27 & $-0.4429$ & 61 & $-0.6057$ & 93 & $1.233$ &  & \\
28 & $0.8601$ & 62 & $1.893$ & 94 & $0.4640$ &  & \\
29 & $-0.6563$ & 63 & $0.4125$ & 95 & $-0.6006$ &  & \\
30 & $-0.1799$ & {\bf 64} & {\bf $2.863$} & 96 & $0.8800$ &  & \\
31 & $0.8715$ & 65 & $-3.032$ &  &  &  & \\
\end{tabular}
\caption{Estimated coefficients in the BOA sector + year model ($\text{BOA}_{SY}$).}
\label{tab:BOAsecyeareffects}
\end{table*}
\begin{table*}[h]
\centering
\begin{tabular}{lr|lr|lr|lr}
\textbf{Sector} & \textbf{Estimate} &
  \textbf{Sector} & \textbf{Estimate} &
  \textbf{Sector} & \textbf{Estimate} &
  \textbf{Year} & \textbf{Estimate} \\
\hline
01 & $0.7850$ & 32 & $0.2421$ & 66 & $1.180$ & {\bf 2009} & {\bf $0.7304$}\\
02 & $-2.737$ & 33 & $0.7370$ & 68 & $0.9413$ & 2010 & $0.1155$\\
03 & $-2.127$ & 35 & $-0.5003$ & 69 & $0.3800$ & 2011 & $0.1360$\\
08 & $-3.573$ & 36 & $-3.730$ & 70 & $2.306$ & 2012 & $0.3484$\\
09 & $-2.936$ & $37$--$39$ & $0.09898$ & 71 & $1.225$ & 2013 & $0.4542$\\
10 & $0.4991$ & 41 & $1.952$ & 72 & $0.06728$ & 2014 & $0.2565$\\
11 & $-1.2400$ & 42 & $0.2203$ & 73 & $1.457$ & 2015 & $0.09998$\\
12 & $-2.8500$ & 43 & $2.252$ & 74 & $0.6834$ & 2016 & $-0.1048$\\
$13$--$15$ & $0.3861$ & 45 & $1.321$ & 75 & $-3.479$ & 2017 & $-0.4118$\\
16 & $0.07771$ & 46 & $2.852$ & 77 & $0.5809$ & 2018 & $-0.2804$\\
17 & $-1.024$ & 47 & $2.488$ & 78 & $1.775$ & 2019 & $-0.1937$\\
18 & $0.9378$ & 49 & $1.890$ & 79 & $0.3895$ & 2020 & $-0.3137$\\
{\bf 19} & {\bf $-4.600$} & 50 & $-0.9715$ & 80 & $-0.2684$ & {\bf 2021} & {\bf $-0.5390$}\\
20 & $-0.4078$ & 51 & $-1.457$ & 81 & $1.070$ & 2022 & $-0.1454$\\
21 & $-2.332$ & 52 & $0.9592$ & 82 & $0.4234$ & 2023 & $-0.03927$\\
22 & $0.4059$ & 53 & $-0.2878$ & 84 & $-4.389$ & 2024 & $-0.1128$\\
23 & $-0.3673$ & 55 & $0.3900$ & 85 & $0.9741$ &  & \\
24 & $-0.9431$ & 56 & $2.398$ & 86 & $1.230$ &  & \\
25 & $1.504$ & 58 & $0.1448$ & $87$--$88$ & $1.827$ &  & \\
26 & $-0.5408$ & $59$--$60$ & $-0.1728$ & $90$--$92$ & $0.7549$ &  & \\
27 & $-0.8548$ & 61 & $-1.039$ & 93 & $1.380$ &  & \\
28 & $0.4509$ & 62 & $1.530$ & 94 & $0.1222$ &  & \\
29 & $-1.021$ & 63 & $-0.4034$ & 95 & $-0.6227$ &  & \\
30 & $-0.09340$ & {\bf 64} & {\bf $3.030$} & 96 & $0.7926$ &  & \\
31 & $0.8779$ & 65 & $-3.021$ &  &  &  & \\
\end{tabular}
\caption{Estimated coefficients in the BOA sector + year + full exposure model ($\text{BOA}_{SY}\text{E}_4$).}
\label{tab:BOAsecyearexp4effects}
\end{table*}

The estimated input and output exposure coefficients are
\begin{table}[ht]
\centering
\begin{tabular}{l|cccc}
$\hat{\gamma}$& $t-1$ & $l(30,t)$ & $l(91,t)$ & $l(182,t)$\\
\hline
in & 0.6821 & 0.2696 & 1.898 & 1.699\\
out & 0.3155 & $-$0.04931 & 1.229 & 0.7865\\
\end{tabular}
\caption{Estimated exposure coefficients at the four lags in the $\text{BOA}_{SY}\text{E}_4$ model.}
\end{table}

\begin{table*}
\centering
\scalebox{0.95}{%
\begin{tabular}{c p{12cm}}
\textbf{Sector} & \textbf{Full sector name}\\
\hline
\textbf{64} & Financiële instellingen (geen verzekeringen en pensioenfondsen)\\
\textbf{46} & Groothandel en handelsbemiddeling (niet in auto's en motorfietsen)\\
\textbf{47} & Detailhandel (niet in auto's)\\
\textbf{56} & Eet- en drinkgelegenheden\\
\textbf{70} & Holdings (geen financiële), concerndiensten binnen eigen concern en managementadvisering\\
\textbf{43} & Gespecialiseerde werkzaamheden in de bouw\\
\textbf{41} & Algemene burgerlijke en utiliteitsbouw en projectontwikkeling\\
\textbf{49} & Vervoer over land\\
\textbf{87}{$-$}\textbf{88} & \makecell[l]{Verpleging, verzorging en begeleiding met overnachting (87) \\[1pt]
Maatschappelijke dienstverlening zonder overnachting (88)}\\
\textbf{78} & Arbeidsbemiddeling, uitzendbureaus en personeelsbeheer\\
\end{tabular}%
}
\caption{Sectors with the largest estimated coefficients in the $\text{BOA}_{SY}\text{E}_4$ model with Dutch SBI 2008 sector names.}\label{tab:mostexposedBOA}

~\par~\par~\par

\scalebox{0.95}{%
\begin{tabular}{c p{12cm}}
\textbf{Sector} & \textbf{Full sector name}\\
\hline
\textbf{19} & Vervaardiging van cokesovenproducten en aardolieverwerking\\
\textbf{84} & Openbaar bestuur, overheidsdiensten en verplichte sociale verzekeringen\\
\textbf{36} & Winning en distributie van water\\
\textbf{08} & Winning van delfstoffen (geen olie en gas)\\
\textbf{75} & Veterinaire dienstverlening\\
\textbf{65} & Verzekeringen en pensioenfondsen (geen verplichte sociale verzekeringen)\\
\textbf{09} & Dienstverlening voor de winning van delfstoffen\\
\textbf{12} & Vervaardiging van tabaksproducten\\
\textbf{02} & Bosbouw, exploitatie van bossen en dienstverlening voor de bosbouw\\
\textbf{21} & Vervaardiging van farmaceutische grondstoffen en producten\\
\end{tabular}%
}
\caption{Sectors with the smallest estimated coefficients in the $\text{BOA}_{SY}\text{E}_4$ model with Dutch SBI 2008 sector names.}\label{tab:leastexposedBOA}
\end{table*}

\begin{table*}
\centering
\small
\begin{tabular}{@{}lcccc@{\hspace{0.75em}}|@{\hspace{0.75em}}lcccc@{}}
\toprule
$s$ & $\hat{p}^{\hspace{1pt}\text{avg}}_s$ & $B_s^{\hspace{0.5pt}\text{emp}}$ & $|\text{diff.}|$ &$|\mathcal{I}_s|$
&$s$ & $\hat{p}^{\hspace{1pt}\text{avg}}_s$ & $B_s^{\hspace{0.5pt}\text{emp}}$  & $|\text{diff.}|$ & $|\mathcal{I}_s|$\\
\midrule
01 & 0.09433 & 0.11027 & 0.01593 & 3156 & 50 & 0.03463 & 0.03642 & 0.00179 & 3377\\
02 & 0.00256 & 0.00261 & 0.00005 & 3444 & 51 & 0.00990 & 0.01009 & 0.00019 & 3469\\
03 & 0.00434 & 0.00445 & 0.00011 & 3146 & 52 & 0.14140 & 0.15321 & 0.01181 & 3022\\
08 & 0.00113 & 0.00118 & 0.00006 & 2534 & 53 & 0.07473 & 0.08058 & 0.00584 & 3264\\
09 & 0.00230 & 0.00230 & 0.00000 & 3041 & 55 & 0.08661 & 0.09241 & 0.00580 & 3214\\
10 & 0.09416 & 0.11665 & \cellcolor{red!30}{0.02249} & 3129 & 56 & 0.35539 & 0.35506 & 0.00033 & 2439\\
11 & 0.02067 & 0.02132 & 0.00064 & 3143 & 58 & 0.07790 & 0.08367 & 0.00576 & 3251\\
12 & 0.00165 & 0.00169 & 0.00003 & 2373 & $59$--$60$ & 0.06104 & 0.06622 & 0.00518 & 3307\\
$13$--$15$ & 0.07099 & 0.07481 & 0.00382 & 3275 & 61 & 0.03397 & 0.03563 & 0.00166 & 3396\\
16 & 0.07548 & 0.08168 & 0.00621 & 3232 & 62 & 0.25144 & 0.26726 & 0.01582 & 2679\\
17 & 0.01566 & 0.01583 & 0.00017 & 3475 & 63 & 0.07302 & 0.08367 & 0.01065 & 3215\\
18 & 0.13013 & 0.13739 & 0.00726 & 3057 & 64 & 0.51275 & 0.52880 & 0.01605 & 1719\\
19 & 0.00032 & 0.00029 & 0.00004 & 3459 & 65 & 0.00453 & 0.00457 & 0.00005 & 3500\\
20 & 0.02350 & 0.02389 & 0.00039 & 3433 & 66 & 0.16943 & 0.18160 & 0.01217 & 2924\\
21 & 0.00591 & 0.00587 & 0.00005 & 3410 & 68 & 0.23131 & 0.26617 & \cellcolor{red!60}{0.03486} & 2675\\
22 & 0.05509 & 0.05694 & 0.00185 & 3337 & 69 & 0.12900 & 0.14179 & 0.01279 & 3068\\
23 & 0.04474 & 0.04729 & 0.00255 & 3362 & 70 & 0.38046 & 0.39106 & 0.01060 & 2281\\
24 & 0.02119 & 0.02186 & 0.00067 & 3385 & 71 & 0.23172 & 0.25695 & \cellcolor{red!50}{0.02523} & 2697\\
25 & 0.18438 & 0.19610 & 0.01172 & 2871 & 72 & 0.07062 & 0.07662 & 0.00600 & 3263\\
26 & 0.04610 & 0.04884 & 0.00274 & 3358 & 73 & 0.18767 & 0.19440 & 0.00673 & 2891\\
27 & 0.03886 & 0.04133 & 0.00247 & 3387 & 74 & 0.14690 & 0.16377 & \cellcolor{red!20}{0.01687} & 2992\\
28 & 0.10420 & 0.11927 & 0.01507 & 3119 & 75 & 0.00202 & 0.00200 & 0.00002 & 3508\\
29 & 0.03231 & 0.03432 & 0.00201 & 3409 & 77 & 0.12537 & 0.13775 & 0.01238 & 3078\\
30 & 0.04782 & 0.05108 & 0.00325 & 3348 & 78 & 0.29075 & 0.29773 & 0.00698 & 2603\\
31 & 0.11447 & 0.11953 & 0.00506 & 3129 & 79 & 0.06286 & 0.06563 & 0.00277 & 3291\\
32 & 0.05481 & 0.05705 & 0.00223 & 3278 & 80 & 0.05869 & 0.06265 & 0.00397 & 3304\\
33 & 0.08632 & 0.09233 & 0.00601 & 3195 & 81 & 0.16320 & 0.17562 & 0.01242 & 2978\\
35 & 0.02794 & 0.02851 & 0.00058 & 3367 & 82 & 0.09757 & 0.10523 & 0.00766 & 3174\\
36 & 0.00114 & 0.00116 & 0.00002 & 3448 & 84 & 0.00067 & 0.00060 & 0.00007 & 1674\\
$37$--$39$ & 0.05781 & 0.06008 & 0.00227 & 3329 & 85 & 0.13500 & 0.14412 & 0.00912 & 3060\\
41 & 0.36756 & 0.38302 & 0.01546 & 2308 & 86 & 0.14641 & 0.15650 & 0.01009 & 3016\\
42 & 0.09965 & 0.10809 & 0.00844 & 3164 & $87$--$88$ & 0.21019 & 0.22147 & 0.01128 & 2804\\
43 & 0.37821 & 0.38157 & 0.00337 & 2301 & $90$--$92$ & 0.10227 & 0.10985 & 0.00758 & 3168\\
45 & 0.20625 & 0.23027 & \cellcolor{red!40}{0.02402} & 2775 & 93 & 0.15240 & 0.15973 & 0.00734 & 3005\\
46 & 0.50534 & 0.50908 & 0.00375 & 1817 & 94 & 0.08041 & 0.08630 & 0.00589 & 3233\\
47 & 0.42975 & 0.42650 & 0.00325 & 2136 & 95 & 0.03470 & 0.03644 & 0.00174 & 3403\\
49 & 0.29007 & 0.30440 & 0.01433 & 2546 & 96 & 0.11651 & 0.12204 & 0.00553 & 3130\\
\bottomrule
\end{tabular}
\caption{Table of burst onset probability estimates, averaged per sector (denoted $\hat{p}_{s}^{\hspace{1pt}\text{avg}}$), compared to the empirical average burst onset frequency $B_s^{\hspace{0.5pt}\text{emp}}$, where both quantities are averaged over all observations in $\mathcal{I}_s$. The absolute difference $|\text{diff.}|$  and the number of days in $\mathcal{I}_s$, i.e. $|\mathcal{I}_s|$, are also reported. The five largest differences are highlighted in shades of red.}
\label{tab:onsetavgs}
\end{table*}

\clearpage

%%%%%%%%%%%%%%%%%%%%%%%%%%%%%%%%%%%%%%%%%%%%%%%%%%%%%%%%%%%%%%%

\section{Burst size analysis tables}\label{app:appendixBSATable}\noindent
Tables \ref{tab:BSAseceffects}, \ref{tab:BSAsecyeareffects} and \ref{tab:BSAsecyearexp4effects} report the estimated covariate coefficients in the sector model ($\text{BSA}_S$), sector + year model ($\text{BSA}_{SY}$), and sector + year + full exposure model ($\text{BSA}_{SY}\text{E}_4$). The largest and smallest values are marked in bold within each category.

% -----------------------------------------------------

\begin{table}[h]
\centering
\begin{tabular}{lr|lr|lr}
\textbf{Sector} & \textbf{Estimate} &
  \textbf{Sector} & \textbf{Estimate} &
  \textbf{Sector} & \textbf{Estimate} \\
\hline
01 & $0.9500$ & 32 & $-0.2404$ & 66 & $1.065$ \\
02 & $0.6681$ & 33 & $-0.3486$ & 68 & $1.262$ \\
03 & $-1.722$ & 35 & $-0.5825$ & 69 & $0.1408$ \\
08 & $-3.354$ & 36 & $-3.747$ & 70 & $1.884$ \\
{\bf 09} & $\mathbf{-4.611}$ & $37$--$39$ & $-0.3692$ & 71 & $1.105$ \\
10 & $0.1474$ & 41 & $1.979$ & 72 & $-0.9067$ \\
11 & $-0.5679$ & 42 & $0.2544$ & 73 & $1.032$ \\
12 & $-0.6441$ & 43 & $2.152$ & 74 & $0.3138$ \\
$13$--$15$ & $-0.2937$ & 45 & $1.217$ & 75 & $-0.1372$ \\
16 & $-0.08076$ & {\bf 46} & $\mathbf{2.527}$ & 77 & $0.01107$ \\
17 & $-0.3577$ & 47 & $2.389$ & 78 & $1.624$ \\
18 & $0.8540$ & 49 & $1.417$ & 79 & $-0.2106$ \\
19 & $-2.262$ & 50 & $0.5055$ & 80 & $-0.477$ \\
20 & $-0.9904$ & 51 & $-1.509$ & 81 & $0.5108$ \\
21 & $-0.09669$ & 52 & $0.4429$ & 82 & $-0.09792$ \\
22 & $-0.2907$ & 53 & $-0.7312$ & 84 & $-2.262$ \\
23 & $-0.2511$ & 55 & $-0.1339$ & 85 & $0.3539$ \\
24 & $-0.8874$ & 56 & $1.812$ & 86 & $0.4223$ \\
25 & $0.9346$ & 58 & $-0.08076$ & $87$--$88$ & $0.8910$ \\
26 & $-0.4930$ & $59$--$60$ & $0.005592$ & $90$--$92$ & $0.03521$ \\
27 & $-0.5363$ & 61 & $-0.4678$ & 93 & $0.2730$ \\
28 & $0.1885$ & 62 & $1.143$ & 94 & $-0.2990$ \\
29 & $-0.4362$ & 63 & $-0.2775$ & 95 & $-0.7561$ \\
30 & $-0.4489$ & 64 & $2.368$ & 96 & $-0.04974$ \\
31 & $0.3631$ & 65 & $-1.234$ &  &  \\
\end{tabular}
\caption{Estimated coefficients in the BSA sector model ($\text{BSA}_{S}$).}
\label{tab:BSAseceffects}
\end{table}

% -----------------------------------------------------
\pagebreak

\begin{table*}[h]
\centering
\begin{tabular}{lr|lr|lr|lr}
\textbf{Sector} & \textbf{Estimate} &
  \textbf{Sector} & \textbf{Estimate} &
  \textbf{Sector} & \textbf{Estimate} &
  \textbf{Year} & \textbf{Estimate} \\
\hline
01 & $0.7248$ & 32 & $-0.2207$ & 66 & $0.9288$ & 2009 & $0.5292$\\
02 & $0.4837$ & 33 & $-0.3104$ & 68 & $1.197$ & 2010 & $0.5340$\\
03 & $-2.003$ & 35 & $-0.4911$ & 69 & $0.1349$ & 2011 & $0.4574$\\
08 & $-3.390$ & 36 & $-3.676$ & 70 & $1.887$ & 2012 & $0.6169$\\
{\bf 09} & {\bf $-4.455$} & $37$--$39$ & $-0.3246$ & 71 & $1.083$ & {\bf 2013} & {\bf $0.7036$}\\
10 & $0.1890$ & 41 & $1.936$ & 72 & $-0.7769$ & 2014 & $0.3551$\\
11 & $-0.01221$ & 42 & $0.1979$ & 73 & $0.9542$ & 2015 & $0.1511$\\
12 & $-0.2776$ & 43 & $2.061$ & 74 & $0.3663$ & 2016 & $0.04166$\\
$13$--$15$ & $-0.2379$ & 45 & $1.167$ & 75 & $0.1266$ & 2017 & $-0.1688$\\
16 & $-0.2140$ & {\bf 46} & {\bf $2.486$} & 77 & $0.05754$ & 2018 & $-0.2253$\\
17 & $-0.4907$ & 47 & $2.413$ & 78 & $1.579$ & 2019 & $-0.2256$\\
18 & $0.7159$ & 49 & $1.465$ & 79 & $-0.3088$ & 2020 & $-0.4341$\\
19 & $-2.431$ & 50 & $0.4633$ & 80 & $-0.5039$ & {\bf 2021} & {\bf $-1.126$}\\
20 & $-0.9649$ & 51 & $-1.550$ & 81 & $0.5533$ & 2022 & $-0.7552$\\
21 & $-0.2611$ & 52 & $0.4965$ & 82 & $-0.01663$ & 2023 & $-0.3863$\\
22 & $-0.4009$ & 53 & $-0.6094$ & 84 & $-1.645$ & 2024 & $-0.0676$\\
23 & $-0.3621$ & 55 & $-0.1347$ & 85 & $0.3900$ &  & \\
24 & $-0.9192$ & 56 & $1.945$ & 86 & $0.5364$ &  & \\
25 & $0.8277$ & 58 & $-0.1711$ & $87$--$88$ & $1.001$ &  & \\
26 & $-0.5499$ & $59$--$60$ & $-0.06067$ & $90$--$92$ & $0.01107$ &  & \\
27 & $-0.5805$ & 61 & $-0.5680$ & 93 & $0.3054$ &  & \\
28 & $0.1392$ & 62 & $1.176$ & 94 & $-0.2918$ &  & \\
29 & $-0.5529$ & 63 & $-0.2778$ & 95 & $-0.8044$ &  & \\
30 & $-0.468$ & 64 & $2.411$ & 96 & $0.05874$ &  & \\
31 & $0.2553$ & 65 & $-1.412$ &  &  &  & \\
\end{tabular}
\caption{Estimated coefficients in the BSA sector + year model ($\text{BSA}_{SY}$).}
\label{tab:BSAsecyeareffects}
\end{table*}

% -----------------------------------------------------
\pagebreak

\begin{table*}[h]
\centering
\begin{tabular}{lr|lr|lr|lr}
\textbf{Sector} & \textbf{Estimate} &
  \textbf{Sector} & \textbf{Estimate} &
  \textbf{Sector} & \textbf{Estimate} &
  \textbf{Year} & \textbf{Estimate} \\
\hline
01 & $0.8746$ & 32 & $-0.1409$ & 66 & $0.9038$ & {\bf 2009} & {\bf $0.7360$}\\
02 & $0.7598$ & 33 & $-0.1729$ & 68 & $0.9627$ & 2010 & $0.2968$\\
03 & $-1.455$ & 35 & $-0.3478$ & 69 & $-0.1111$ & 2011 & $0.2682$\\
08 & $-3.143$ & 36 & $-3.141$ & 70 & $1.722$ & 2012 & $0.4651$\\
{\bf 09} & {\bf $-3.711$} & $37$--$39$ & $-0.2805$ & 71 & $0.8223$ & 2013 & $0.6218$\\
10 & $0.1803$ & 41 & $1.675$ & 72 & $-0.8130$ & 2014 & $0.2749$\\
11 & $-0.006664$ & 42 & $0.08502$ & 73 & $0.9151$ & 2015 & $0.1131$\\
12 & $-0.1783$ & 43 & $1.863$ & 74 & $0.1728$ & 2016 & $-0.03290$\\
$13$--$15$ & $-0.1246$ & 45 & $1.056$ & 75 & $0.1247$ & 2017 & $-0.2585$\\
16 & $-0.2242$ & {\bf 46} & {\bf $2.331$} & 77 & $-0.1193$ & 2018 & $-0.2297$\\
17 & $-0.2118$ & 47 & $2.226$ & 78 & $1.355$ & 2019 & $-0.2165$\\
18 & $0.7627$ & 49 & $1.370$ & 79 & $-0.1246$ & 2020 & $-0.4097$\\
19 & $-2.209$ & 50 & $0.3794$ & 80 & $-0.7306$ & {\bf 2021} & {\bf $-0.8600$}\\
20 & $-0.6718$ & 51 & $-1.279$ & 81 & $0.3556$ & 2022 & $-0.4832$\\
21 & $-0.2137$ & 52 & $0.4221$ & 82 & $-0.1478$ & 2023 & $-0.2526$\\
22 & $-0.1214$ & 53 & $-0.8677$ & 84 & $-1.670$ & 2024 & $-0.0328$\\
23 & $-0.2110$ & 55 & $-0.1483$ & 85 & $0.2940$ &  & \\
24 & $-0.7492$ & 56 & $1.823$ & 86 & $0.5645$ &  & \\
25 & $0.9258$ & 58 & $-0.2879$ & $87$--$88$ & $1.064$ &  & \\
26 & $-0.6536$ & $59$--$60$ & $-0.2631$ & $90$--$92$ & $0.00009721$ &  & \\
27 & $-0.5819$ & 61 & $-0.6153$ & 93 & $0.3891$ &  & \\
28 & $0.1407$ & 62 & $0.9873$ & 94 & $-0.4525$ &  & \\
29 & $-0.5969$ & 63 & $-0.5107$ & 95 & $-0.8572$ &  & \\
30 & $-0.3927$ & 64 & $2.310$ & 96 & $-0.04347$ &  & \\
31 & $0.3459$ & 65 & $-1.584$ &  &  &  & \\
\end{tabular}
\caption{Estimated coefficients in the BSA sector + year + full exposure model ($\text{BSA}_{SY}\text{E}_4$).}
\label{tab:BSAsecyearexp4effects}
\end{table*}

The input and output exposure coefficient estimates are
\begin{table}[ht]
\centering
\begin{tabular}{l|cccc}
$\hat{\beta}$ & $t-1$ & $l(30,t)$ & $l(91,t)$ & $l(182,t)$\\
\hline
in & $-$0.5175 & $-$0.6582 & 0.7841 & 0.7852 \\
out & $-$0.2765 & $-$0.1043 & 0.4306 & 0.2329\\
\end{tabular}
\caption{Estimated exposure coefficients at the four lags in the $\text{BSA}_{SY}\text{E}_4$ model.}
\end{table}

\begin{table*}[ht]
\centering
\scalebox{0.95}{%
\begin{tabular}{c p{12cm}}
\textbf{Sector} & \textbf{Full sector name}\\
\hline
\textbf{46} & Groothandel en handelsbemiddeling (niet in auto's en motorfietsen)\\
\textbf{64} & Financiële instellingen (geen verzekeringen en pensioenfondsen)\\
\textbf{47} & Detailhandel (niet in auto's)\\
\textbf{43} & Gespecialiseerde werkzaamheden in de bouw\\
\textbf{56} & Eet- en drinkgelegenheden\\
\textbf{70} & Holdings (geen financiële), concerndiensten binnen eigen concern en managementadvisering\\
\textbf{41} & Algemene burgerlijke en utiliteitsbouw en projectontwikkeling\\
\textbf{49} & Vervoer over land\\
\textbf{78} & Arbeidsbemiddeling, uitzendbureaus en personeelsbeheer\\
\textbf{87}{$-$}\textbf{88} & \makecell[l]{Verpleging, verzorging en begeleiding met overnachting (87)\\[1pt]
Maatschappelijke dienstverlening zonder overnachting (88)}\\
\end{tabular}%
}
\caption{Sectors with the largest estimated coefficients in the $\text{BSA}_{SY}\text{E}_4$ model with Dutch SBI 2008 sector names.}\label{tab:mostexposedBSA}

~\par~\par~\par

\scalebox{0.95}{%
\begin{tabular}{c p{12cm}}
\textbf{Sector} & \textbf{Full sector name}\\
\hline
\textbf{09} & Dienstverlening voor de winning van delfstoffen\\
\textbf{08} & Winning van delfstoffen (geen olie en gas)\\
\textbf{36} & Winning en distributie van water\\
\textbf{19} & Vervaardiging van cokesovenproducten en aardolieverwerking\\
\textbf{84} & Openbaar bestuur, overheidsdiensten en verplichte sociale verzekeringen\\
\textbf{65} & Verzekeringen en pensioenfondsen (geen verplichte sociale verzekeringen)\\
\textbf{03} & Visserij en kweken van vis en schaaldieren\\
\textbf{51} & Luchtvaart\\
\textbf{53} & Post en koeriers\\
\textbf{95} & Reparatie van computers en consumentenartikelen\\
\end{tabular}%
}
\caption{Sectors with the smallest estimated coefficients in the $\text{BSA}_{SY}\text{E}_4$ model with Dutch SBI 2008 sector names.}\label{tab:leastexposedBSA}
\end{table*}

\begin{table*}[ht]
\centering
\small
\begin{tabular}{@{}lcccc@{\hspace{0.75em}}|@{\hspace{0.75em}}lcccc@{}}
\toprule
$s$ & $\hat{\mu}^{+}_{s}$ & $\mu_s^{+,\,\text{emp}}$ & $|\text{diff.}|$ & $n_s^\text{pos}$ &
$s$ & $\hat{\mu}^{+}_{s}$ & $\mu_s^{+,\,\text{emp}}$ & $|\text{diff.}|$ & $n_s^\text{pos}$\\
\midrule
01 & 1.96069 & 1.90526 & 0.05543 & 380 & 50 & 1.68052 & 1.60000 & 0.08052 & 130\\
02 & 1.83175 & 1.70000 & \cellcolor{red!40}{0.13175} & 10 & 51 & 1.08654 & 1.08333 & 0.00321 & 36\\
03 & 1.07024 & 1.06667 & 0.00357 & 15 & 52 & 1.58272 & 1.56471 & 0.01802 & 510\\
08 & 1.01475 & 1.00000 & 0.01475 & 4 & 53 & 1.18708 & 1.18382 & 0.00325 & 272\\
09 & 1.00594 & 1.00000 & 0.00594 & 8 & 55 & 1.32882 & 1.32609 & 0.00274 & 322\\
10 & 1.47539 & 1.42748 & 0.04791 & 393 & 56 & 3.05457 & 2.98907 & 0.06550 & 1098\\
11 & 1.23168 & 1.21429 & 0.01740 & 70 & 58 & 1.35903 & 1.34386 & 0.01517 & 285\\
12 & 1.24658 & 1.20000 & 0.04658 & 5 & $59$--$60$ & 1.39414 & 1.37333 & 0.02080 & 225\\
$13$--$15$ & 1.28281 & 1.28016 & 0.00266 & 257 & 61 & 1.25408 & 1.23664 & 0.01744 & 131\\
16 & 1.35726 & 1.34386 & 0.01340 & 285 & 62 & 2.09597 & 2.08159 & 0.01438 & 858\\
17 & 1.26507 & 1.26316 & 0.00191 & 57 & 63 & 1.29048 & 1.28472 & 0.00575 & 288\\
18 & 1.89033 & 1.82881 & 0.06152 & 479 & 64 & 4.28711 & 4.27573 & 0.01138 & 1817\\
19 & 1.03793 & 1.00000 & 0.03793 & 2 & 65 & 1.11191 & 1.11765 & 0.00574 & 17\\
20 & 1.14437 & 1.14286 & 0.00151 & 84 & 66 & 2.01363 & 2.00654 & 0.00709 & 612\\
21 & 1.34346 & 1.33333 & 0.01012 & 21 & 68 & 2.31713 & 2.20674 & \cellcolor{red!30}{0.11039} & 861\\
22 & 1.28305 & 1.28141 & 0.00165 & 199 & 69 & 1.43852 & 1.42521 & 0.01331 & 468\\
23 & 1.30032 & 1.29167 & 0.00866 & 168 & 70 & 3.06666 & 3.12102 & 0.05436 & 1256\\
24 & 1.16035 & 1.15789 & 0.00246 & 76 & 71 & 2.05195 & 2.04458 & 0.00737 & 830\\
25 & 1.90096 & 1.89291 & 0.00805 & 663 & 72 & 1.15649 & 1.15444 & 0.00205 & 259\\
26 & 1.23499 & 1.23077 & 0.00422 & 169 & 73 & 1.98093 & 1.97674 & 0.00419 & 645\\
27 & 1.22258 & 1.22069 & 0.00189 & 145 & 74 & 1.52238 & 1.50000 & 0.02238 & 544\\
28 & 1.47460 & 1.44365 & 0.03096 & 417 & 75 & 1.42601 & 1.33333 & \cellcolor{red!20}{0.09267} & 9\\
29 & 1.25082 & 1.24409 & 0.00673 & 127 & 77 & 1.39310 & 1.37473 & 0.01837 & 459\\
30 & 1.26142 & 1.24138 & 0.02004 & 174 & 78 & 2.68556 & 2.67773 & 0.00783 & 934\\
31 & 1.52826 & 1.52334 & 0.00492 & 407 & 79 & 1.32688 & 1.30303 & 0.02385 & 231\\
32 & 1.29590 & 1.29534 & 0.00057 & 193 & 80 & 1.23243 & 1.23474 & 0.00232 & 213\\
33 & 1.26944 & 1.26603 & 0.00342 & 312 & 81 & 1.61349 & 1.60183 & 0.01165 & 545\\
35 & 1.21112 & 1.21212 & 0.00100 & 99 & 82 & 1.34276 & 1.33799 & 0.00477 & 358\\
36 & 1.01177 & 1.00000 & 0.01177 & 5 & 84 & 1.03992 & 1.00000 & 0.03992 & 2\\
$37$--$39$ & 1.26261 & 1.26087 & 0.00174 & 207 & 85 & 1.53462 & 1.51891 & 0.01571 & 476\\
41 & 3.38922 & 3.31107 & 0.07814 & 1228 & 86 & 1.58297 & 1.55385 & 0.02912 & 520\\
42 & 1.49017 & 1.47283 & 0.01734 & 368 & $87$--$88$ & 1.88998 & 1.85734 & 0.03264 & 729\\
43 & 3.50051 & 3.69879 & \cellcolor{red!50}{0.19828} & 1235 & 90$-$92 & 1.38713 & 1.38315 & 0.00398 & 368\\
45 & 2.19100 & 2.15769 & 0.03331 & 761 & 93 & 1.48773 & 1.48106 & 0.00667 & 528\\
46 & 4.33966 & 4.78010 & \cellcolor{red!60}{0.44044} & 1719 & 94 & 1.28046 & 1.27891 & 0.00155 & 294\\
47 & 4.33880 & 4.33833 & 0.00047 & 1401 & 95 & 1.18231 & 1.17829 & 0.00402 & 129\\
49 & 2.42081 & 2.38990 & 0.03091 & 990 & 96 & 1.36345 & 1.35381 & 0.00964 & 407\\
\bottomrule
\end{tabular}
\caption{Table of estimated conditional positive means per sector $\hat{\mu}_{s}^+$, compared to the empirical average conditional positive mean, denoted $\mu_s^{+,\hspace{1pt}\text{emp}}$. The difference between these quantities is given as $|\text{diff.}|$, and $n_s^\text{pos}$ denotes the number of active hearing-days in the corresponding sector. The five largest differences are highlighted in shades of red.}
\label{tab:BSAavgs}
\end{table*}

%%%%%%%%%%%%%%%%%%%%%%%%%%%%%%%%%%%%%%%%%%%%%%%%%%%%%%%%%%%

\printcredits

\bibliographystyle{tmp-model2-names}

\bibliography{tmp-refs}

\begin{thebibliography}{28}
\expandafter\ifx\csname natexlab\endcsname\relax\def\natexlab#1{#1}\fi
\providecommand{\url}[1]{\texttt{#1}}
\providecommand{\href}[2]{#2}
\providecommand{\path}[1]{#1}
\providecommand{\DOIprefix}{doi:}
\providecommand{\ArXivprefix}{arXiv:}
\providecommand{\URLprefix}{URL: }
\providecommand{\Pubmedprefix}{pmid:}
\providecommand{\doi}[1]{\href{http://dx.doi.org/#1}{\path{#1}}}
\providecommand{\Pubmed}[1]{\href{pmid:#1}{\path{#1}}}
\providecommand{\bibinfo}[2]{#2}
\ifx\xfnm\relax \def\xfnm[#1]{\unskip,\space#1}\fi
%Type = Article
\bibitem[{Allison(1982)}]{allison_discrete-time_1982}
\bibinfo{author}{Allison, P.D.}, \bibinfo{year}{1982}.
\newblock \bibinfo{title}{Discrete-{Time} {Methods} for the {Analysis} of
  {Event} {Histories}}.
\newblock \bibinfo{journal}{Sociological Methodology} \bibinfo{volume}{13},
  \bibinfo{pages}{61}.
\newblock \DOIprefix\doi{10.2307/270718}.
%Type = Book
\bibitem[{Box-Steffensmeier and Jones(2004)}]{box-steffensmeier_event_2004}
\bibinfo{author}{Box-Steffensmeier, J.M.}, \bibinfo{author}{Jones, B.S.},
  \bibinfo{year}{2004}.
\newblock \bibinfo{title}{Event {History} {Modeling}: {A} {Guide} for {Social}
  {Scientists}}.
\newblock \bibinfo{edition}{1} ed., \bibinfo{publisher}{Cambridge University
  Press}.
\newblock \DOIprefix\doi{10.1017/CBO9780511790874}.
%Type = Article
\bibitem[{Carroll(2003)}]{carroll_use_2003}
\bibinfo{author}{Carroll, K.J.}, \bibinfo{year}{2003}.
\newblock \bibinfo{title}{On the use and utility of the {Weibull} model in the
  analysis of survival data}.
\newblock \bibinfo{journal}{Controlled Clinical Trials} \bibinfo{volume}{24},
  \bibinfo{pages}{682--701}.
\newblock \DOIprefix\doi{10.1016/S0197-2456(03)00072-2}.
%Type = Article
\bibitem[{Carvalho(2014)}]{carvalho_micro_2014}
\bibinfo{author}{Carvalho, V.M.}, \bibinfo{year}{2014}.
\newblock \bibinfo{title}{From {Micro} to {Macro} via {Production} {Networks}}.
\newblock \bibinfo{journal}{Journal of Economic Perspectives}
  \bibinfo{volume}{28}, \bibinfo{pages}{23--48}.
\newblock \URLprefix \url{https://pubs.aeaweb.org/doi/10.1257/jep.28.4.23},
  \DOIprefix\doi{10.1257/jep.28.4.23}.
%Type = Misc
\bibitem[{CBS(2025)}]{cbs_standard_2025}
\bibinfo{author}{CBS}, \bibinfo{year}{2025}.
\newblock \bibinfo{title}{Standard {Industrial} {Classifications}}.
%Type = Misc
\bibitem[{CBS(2026a)}]{CBS_faillissementen_2026}
\bibinfo{author}{CBS}, \bibinfo{year}{2026}a.
\newblock \bibinfo{title}{Faillissementen, zittingsdaggecorrigeerd}.
\newblock \URLprefix \url{https://www.cbs.nl/nl-nl/cijfers/detail/83085NED}.
%Type = Techreport
\bibitem[{CBS(2026b)}]{cbs_uitgesproken_2026}
\bibinfo{author}{CBS}, \bibinfo{year}{2026}b.
\newblock \bibinfo{title}{Uitgesproken faillissementen door {Nederlandse}
  rechtbanken naar type gefailleerde, activiteit en regio}.
\newblock \bibinfo{type}{Technical Report}. CBS.
\newblock \URLprefix
  \url{https://www.cbs.nl/nl-nl/onze-diensten/maatwerk-en-microdata/microdata-zelf-onderzoek-doen/microdatabestanden/faillissementen-uitgesproken-faillissementen}.
%Type = Article
\bibitem[{Cox(1972)}]{cox_regression_1972}
\bibinfo{author}{Cox, D.R.}, \bibinfo{year}{1972}.
\newblock \bibinfo{title}{Regression models and life-tables}.
\newblock \bibinfo{journal}{Journal of the Royal Statistical Society: Series B}
  \bibinfo{volume}{34}, \bibinfo{pages}{187--220}.
%Type = Article
\bibitem[{Diamoutene et~al.(2016)Diamoutene, Barro, Somda, Noureddine and
  Foguem}]{diamoutene_survival_2016}
\bibinfo{author}{Diamoutene, A.}, \bibinfo{author}{Barro, D.},
  \bibinfo{author}{Somda, M.A.S.}, \bibinfo{author}{Noureddine, F.},
  \bibinfo{author}{Foguem, B.K.}, \bibinfo{year}{2016}.
\newblock \bibinfo{title}{Survival analysis in living and engineering
  sciences}.
\newblock \bibinfo{journal}{JP Journal of Biostatistics} \bibinfo{volume}{13},
  \bibinfo{pages}{223--238}.
\newblock \DOIprefix\doi{10.17654/BO013020223}.
%Type = Book
\bibitem[{{European Commission}(2008)}]{european_commission_nace_2008}
\bibinfo{editor}{{European Commission}} (Ed.), \bibinfo{year}{2008}.
\newblock \bibinfo{title}{{NACE} {Rev}. 2: statistical classification of
  economic activities in the {European} {Community}}.
\newblock \bibinfo{publisher}{Publications Office},
  \bibinfo{address}{Luxembourg}.
%Type = Article
\bibitem[{Hardiman and Bouchaud(2014)}]{HardBou2014}
\bibinfo{author}{Hardiman, S.J.}, \bibinfo{author}{Bouchaud, J.P.},
  \bibinfo{year}{2014}.
\newblock \bibinfo{title}{Branching-ratio approximation for the self-exciting
  hawkes process}.
\newblock \bibinfo{journal}{Phys. Rev. E} \bibinfo{volume}{90},
  \bibinfo{pages}{062807}.
\newblock \URLprefix \url{https://link.aps.org/doi/10.1103/PhysRevE.90.062807},
  \DOIprefix\doi{10.1103/PhysRevE.90.062807}.
%Type = Article
\bibitem[{Hawkes(1970)}]{Hawkes1970}
\bibinfo{author}{Hawkes, A.}, \bibinfo{year}{1970}.
\newblock \bibinfo{title}{{S}pectra of some mutually exciting point processes}.
\newblock \bibinfo{journal}{Biometrika} \bibinfo{volume}{58},
  \bibinfo{pages}{83}.
%Type = Article
\bibitem[{Hawkes(1971)}]{Hawkes1971}
\bibinfo{author}{Hawkes, A.}, \bibinfo{year}{1971}.
\newblock \bibinfo{title}{{P}oint spectra of some mutually exciting point
  processes}.
\newblock \bibinfo{journal}{Journal of the Royal Statistical Society: Series B}
  \bibinfo{volume}{33}, \bibinfo{pages}{438--443}.
%Type = Article
\bibitem[{Holford(1976)}]{holford_life_1976}
\bibinfo{author}{Holford, T.R.}, \bibinfo{year}{1976}.
\newblock \bibinfo{title}{Life {Tables} with {Concomitant} {Information}}.
\newblock \bibinfo{journal}{Biometrics} \bibinfo{volume}{32},
  \bibinfo{pages}{587}.
\newblock \DOIprefix\doi{10.2307/2529747}.
%Type = Article
\bibitem[{Hong et~al.(2007)Hong, Lee and Lee}]{hong_power_2007}
\bibinfo{author}{Hong, B.H.}, \bibinfo{author}{Lee, K.E.},
  \bibinfo{author}{Lee, J.W.}, \bibinfo{year}{2007}.
\newblock \bibinfo{title}{Power law in firms bankruptcy}.
\newblock \bibinfo{journal}{Physics Letters A} \bibinfo{volume}{361},
  \bibinfo{pages}{6--8}.
\newblock \URLprefix
  \url{https://linkinghub.elsevier.com/retrieve/pii/S0375960106014046},
  \DOIprefix\doi{10.1016/j.physleta.2006.09.034}.
%Type = Book
\bibitem[{Klein and Moeschberger(2005)}]{klein_survival_2005}
\bibinfo{author}{Klein, J.P.}, \bibinfo{author}{Moeschberger, M.L.},
  \bibinfo{year}{2005}.
\newblock \bibinfo{title}{Survival analysis: techniques for censored and
  truncated data}.
\newblock Statistics for biology and health. \bibinfo{edition}{second edition,
  corrected third printing} ed., \bibinfo{publisher}{Springer},
  \bibinfo{address}{New York Berlin Heidelberg}.
%Type = Article
\bibitem[{Kruiskamp(2022)}]{kruiskamp_standaard_2022}
\bibinfo{author}{Kruiskamp, P.}, \bibinfo{year}{2022}.
\newblock \bibinfo{title}{Standaard {Bedrijfs} {Indeling} 2008 {Versie} 2018
  {Update} 2022}.
\newblock \bibinfo{journal}{CBS} .
%Type = Article
\bibitem[{Kumar and Klefsjö(1994)}]{kumar_proportional_1994}
\bibinfo{author}{Kumar, D.}, \bibinfo{author}{Klefsjö, B.},
  \bibinfo{year}{1994}.
\newblock \bibinfo{title}{Proportional hazards model: a review}.
\newblock \bibinfo{journal}{Reliability Engineering \& System Safety}
  \bibinfo{volume}{44}, \bibinfo{pages}{177--188}.
\newblock \DOIprefix\doi{10.1016/0951-8320(94)90010-8}.
%Type = Article
\bibitem[{Laeven(2022)}]{laeven_pandemics_2022}
\bibinfo{author}{Laeven, L.}, \bibinfo{year}{2022}.
\newblock \bibinfo{title}{Pandemics, intermediate goods, and corporate
  valuation}.
\newblock \bibinfo{journal}{Journal of International Money and Finance}
  \bibinfo{volume}{120}, \bibinfo{pages}{102505}.
\newblock \URLprefix
  \url{https://linkinghub.elsevier.com/retrieve/pii/S026156062100156X},
  \DOIprefix\doi{10.1016/j.jimonfin.2021.102505}.
%Type = Techreport
\bibitem[{Lopes~Cardozo(2024)}]{lopes_cardozo_internship_2024}
\bibinfo{author}{Lopes~Cardozo, M.}, \bibinfo{year}{2024}.
\newblock \bibinfo{title}{Internship report}.
\newblock \bibinfo{type}{Technical Report}. Centraal Bureau voor de Statistiek.
%Type = Article
\bibitem[{Moran et~al.(2025)Moran, Pijpers, Weitzel, Bouchaud and
  Panja}]{moran_critical_2025}
\bibinfo{author}{Moran, J.}, \bibinfo{author}{Pijpers, F.P.},
  \bibinfo{author}{Weitzel, U.}, \bibinfo{author}{Bouchaud, J.P.},
  \bibinfo{author}{Panja, D.}, \bibinfo{year}{2025}.
\newblock \bibinfo{title}{Critical fragility in sociotechnical systems}.
\newblock \bibinfo{journal}{Proceedings of the National Academy of Sciences}
  \bibinfo{volume}{122}, \bibinfo{pages}{e2415139122}.
\newblock \DOIprefix\doi{10.1073/pnas.2415139122}.
%Type = Article
\bibitem[{Moran et~al.(2024)Moran, Romeijnders, Doussal, Pijpers, Weitzel,
  Panja and Bouchaud}]{moran_timeliness_2024}
\bibinfo{author}{Moran, J.}, \bibinfo{author}{Romeijnders, M.},
  \bibinfo{author}{Doussal, P.L.}, \bibinfo{author}{Pijpers, F.P.},
  \bibinfo{author}{Weitzel, U.}, \bibinfo{author}{Panja, D.},
  \bibinfo{author}{Bouchaud, J.P.}, \bibinfo{year}{2024}.
\newblock \bibinfo{title}{Timeliness criticality in complex systems}.
\newblock \bibinfo{journal}{Nature Physics} \bibinfo{volume}{20},
  \bibinfo{pages}{1352--1358}.
\newblock \DOIprefix\doi{10.1038/s41567-024-02525-w}.
%Type = Article
\bibitem[{Ott and Sommerer(1994)}]{ott_blowout_1994}
\bibinfo{author}{Ott, E.}, \bibinfo{author}{Sommerer, J.C.},
  \bibinfo{year}{1994}.
\newblock \bibinfo{title}{Blowout bifurcations: the occurrence of riddled
  basins and on-off intermittency}.
\newblock \bibinfo{journal}{Physics Letters A} \bibinfo{volume}{188},
  \bibinfo{pages}{39--47}.
\newblock \DOIprefix\doi{10.1016/0375-9601(94)90114-7}.
%Type = Article
\bibitem[{Peng(2003)}]{peng_estimating_2003}
\bibinfo{author}{Peng, Y.}, \bibinfo{year}{2003}.
\newblock \bibinfo{title}{Estimating baseline distribution in proportional
  hazards cure models}.
\newblock \bibinfo{journal}{Computational Statistics \& Data Analysis}
  \bibinfo{volume}{42}, \bibinfo{pages}{187--201}.
\newblock \DOIprefix\doi{10.1016/S0167-9473(02)00158-5}.
%Type = Article
\bibitem[{Pomeau and Manneville(1980)}]{pomeau_intermittent_1980}
\bibinfo{author}{Pomeau, Y.}, \bibinfo{author}{Manneville, P.},
  \bibinfo{year}{1980}.
\newblock \bibinfo{title}{Intermittent transition to turbulence in dissipative
  dynamical systems}.
\newblock \bibinfo{journal}{Communications in Mathematical Physics} .
%Type = Article
\bibitem[{Prentice and Gloeckler(1978)}]{prentice_regression_1978}
\bibinfo{author}{Prentice, R.L.}, \bibinfo{author}{Gloeckler, L.A.},
  \bibinfo{year}{1978}.
\newblock \bibinfo{title}{Regression {Analysis} of {Grouped} {Survival} {Data}
  with {Application} to {Breast} {Cancer} {Data}}.
\newblock \bibinfo{journal}{Biometrics} \bibinfo{volume}{34},
  \bibinfo{pages}{57}.
\newblock \DOIprefix\doi{10.2307/2529588}.
%Type = Article
\bibitem[{Shumway(2001)}]{Shumway01}
\bibinfo{author}{Shumway, T.}, \bibinfo{year}{2001}.
\newblock \bibinfo{title}{Forecasting bankruptcy more accurately: A simple
  hazard model}.
\newblock \bibinfo{journal}{The Journal of Business} \bibinfo{volume}{74},
  \bibinfo{pages}{101--124}.
%Type = Misc
\bibitem[{{United Nations Statistics
  Division}(2025)}]{united_nations_statistics_division_isic_2025}
\bibinfo{author}{{United Nations Statistics Division}}, \bibinfo{year}{2025}.
\newblock \bibinfo{title}{{ISIC}}.

\end{thebibliography}



\end{document}